\documentclass{pasj00}
\newcommand{\simgt}{\lower.5ex\hbox{$\; \buildrel > \over \sim \;$}}
\newcommand{\simlt}{\lower.5ex\hbox{$\; \buildrel < \over \sim \;$}}

\newcommand{\bmf}[1]{\mbox{\boldmath$#1$}}

\def\btheta{\mbox{\boldmath $\theta$}}

\begin{document}
\SetRunningHead{Okabe, Takada, Umetsu et al.}{}
\Received{}%{yyyy/mm/dd}
\Accepted{}%{yyyy/mm/dd}

\title{LoCuSS: Subaru Weak Lensing Study of 30 Galaxy Clusters 
\thanks{Based in part on data collected at Subaru Telescope and obtained from
the SMOKA, which is operated by the Astronomy Data Center, National
Astronomical Observatory of Japan. 
}}

%%% begin:list of authors
\author{%
Nobuhiro \textsc{Okabe}\altaffilmark{1,3},
Masahiro \textsc{Takada}\altaffilmark{2}, 
Keiichi \textsc{Umetsu}\altaffilmark{3,4},
Toshifumi \textsc{Futamase}\altaffilmark{1},
Graham P. \textsc{Smith}\altaffilmark{5}
}
\altaffiltext{1}{Astronomical Institute, Tohoku University,
Aramaki, Aoba-ku, Sendai, 980-8578, Japan}
\email{okabe@astr.tohoku.ac.jp}

\altaffiltext{2}{Institute for the Physics and Mathematics of the
Universe (IPMU)\\ The University of Tokyo,
 5-1-5 Kashiwa-no-Ha, Kashiwa
City, Chiba 277-8582, Japan}
%\email{masahiro.takada@ipmu.jp}

\altaffiltext{3}{Academia Sinica Institute of Astronomy and Astrophysics
(ASIAA), P.O. Box 23-141, Taipei 10617, Taiwan} 

\altaffiltext{4} {Leung
Center for Cosmology and Particle Astrophysics, National Taiwan
University, Taipei 10617, Taiwan}

\altaffiltext{5}{School of Physics and Astronomy,  
University of Birmingham, Edgbaston, Birmingham, B15 2TT, UK}

%%% end:list of authors

%%% Please use the following style in case that sorting by 
%%% affilation is impossible. 
%
% \author{%
%   D-Firstname \textsc{D-Familyname}\altaffilmark{1}
%   E-Firstname \textsc{E-Familyname}\altaffilmark{1,2}
%   and
%   F-Firstname \textsc{F-Familyname}\altaffilmark{2}}
% \altaffiltext{1}{Address of Institute}
% \email{ddddd@xxx.xxx.xx.xx}
% \email{eeeee@xxx.xxx.xx.xx}
% \altaffiltext{2}{Address of Institute}

%% `\KeyWords{}' always has to be placed before `\maketitle'.
\maketitle

\begin{abstract}
We use high-quality Subaru/Suprime-Cam imaging data to conduct a
detailed weak-lensing study of the distribution of dark matter in a
sample of 30 $X$-ray luminous galaxy clusters at $0.15\le z\le0.3$.  A
weak-lensing signal is detected at high statistical significance in
each cluster, the total signal-to-noise ratio of the detections
ranging from $5$ to $13$.  
%MT
%We concentrate on fitting
%spherical models to the tangential distortion profiles of the clusters.
%When the models are fitted to the clusters individually, 
Comparing spherical models to the tangential distortion profiles of the
 clusters individually,
we are unable
to discriminate statistically between singular isothermal sphere (SIS)
and Navarro Frenk \& White (NFW) models.  However when the tangential
distortion profiles 
are combined and then models are
fitted to the stacked profile, the SIS model is rejected at $6\sigma$
and $11\sigma$, respectively, for low ($M_{\rm
vir}<6\times10^{14}h^{-1}M_\odot$) and high ($M_{\rm
vir}>6\times10^{14}h^{-1}M_\odot$) mass bins.  We also use the
individual cluster NFW model fits to investigate the relationship
between cluster mass and concentration, finding that concentration
($c_{\rm vir}$) decreases with increasing cluster mass ($M_{\rm
vir}$).  The best-fit $c_{\rm vir}-M_{\rm vir}$ relation is: 
$c_{\rm vir}(M_{\rm vir})=8.75^{+4.13}_{-2.89}\times (M_{\rm vir}/10^{14}
      h^{-1}M_\odot)^{\alpha}$ with $\alpha\approx -0.40\pm 0.19$: 
i.e.\ a non-zero slope is detected at $2\sigma$ significance.  
%MT
This relation gives a concentration of 
$c_{\rm vir}=3.48^{+1.65}_{-1.15}$ for clusters
   with $M_{\rm vir}=10^{15}
h^{-1}M_\odot$, which is inconsistent at $4\sigma$ significance
   with the values of 
$c_{\rm vir}\sim 10$ reported for strong-lensing-selected clusters.
%We then
%investigate the optimal radius within which to measure cluster mass,
%finding 
We find
that the measurement error on cluster mass is smaller at
higher over-densities $\Delta\simeq500-2000$, than at the virial
over-density $\Delta_{\rm vir}\simeq 110$; typical fractional errors
at $\Delta\simeq 500-2000$
are improved to $\sigma(M_\Delta)/M_{\Delta }\simeq 0.1-0.2$ 
compared with $0.2$--$0.3$ at $\Delta_{\rm vir}$. 
%from the
% virial ones  $0.2-0.3$.  
Furthermore, 
comparing the 3D spherical mass with the 2D cylinder mass, obtained from 
the aperture mass method at a given aperture radius $\theta_\Delta$, reveals
 $M_{\rm 2D}(<\theta_{\Delta})/M_{\rm
3D}(<r_\Delta=D_l\theta_{\Delta})\simeq 1.46$ and $1.32$ for $\Delta
=500$ and $\Delta_{\rm vir}$, respectively.  The amplitude of this
offset agrees well with that predicted by integrating an NFW model of
 cluster-scale halos
along the line-of-sight.  
%Overall, our results demonstrate the power of
%high-quality imaging data
%for making detailed weak-lensing studies of the
%matter distribution on both individual cluster and statistical bases. 
\end{abstract}

\section{INTRODUCTION}

 The mass and internal structure of galaxy clusters reflect the
properties of primordial density perturbations and the nature of dark
matter.  A most striking prediction from numerical simulations based
on the cold dark matter (CDM) model of structure formation is that
dark matter halos can be described by a universal mass density profile
as found by Navarro, Frenk \& White (1996, 1997, hereafter NFW; also
see Moore et al.\ 1999; Fukushige \& Makino 2001; most recently
Navarro et al.\ 2008 and references therein).  These results have
shown that cluster-scale halos should have relatively shallow,
low-concentration mass profiles, where the power-law slope of density
profile becomes more negative with increasing radius, approaching an
asymptotic slope of $-3$ around the virial radius.  The dark matter
halo mass function, of which galaxy clusters represent the high mass
tail, is also sensitive to cosmological parameters, including for
example the dark energy equation of state parameter $w$ (e.g.\ White
et al.\ 1993; Kitayama \& Suto 1997; Haiman et al. 2001; Vikhlinin et
al.\ 2008).  Testing predictions from numerical dark matter
simulations and probing dark energy require precise measurements of
galaxy cluster masses, however it is non-trivial to define what is
meant by the mass of a cluster because clusters do not have any clear
boundary between themselves and the surrounding large-scale structure.
By convention cluster mass is therefore defined as the mass enclosed
within a three-dimensional sphere of a given radius with respect to
the halo center such as the virial mass (e.g., White 2002).  Given a
working definition of mass, a method of mass measurement must be
chosen, each of which suffers a number of advantages and
disadvantages, as discussed below.

The deep potential well of a galaxy cluster causes weak shape
distortions of background galaxy images due to differential deflection
of light rays, resulting in a coherent distortion pattern around the
cluster center, known as weak gravitational lensing (Narayan \&
Bartelmann 1996; Bartelmann \& Schneider 2001; Schneider 2006).
Measuring this coherent distortion pattern allows us to map directly
the mass distribution in a  cluster without requiring any
assumptions on the dynamical/physical state of the system (e.g.,
Kaiser \& Squires 1993; Fahlman et al. 1994).  Other methods rest on
some assumptions: the velocity dispersion of member galaxies invoke
assumptions on the velocity anisotropies, dynamical equilibrium of the
cluster, and the geometry of the system.  Methods based on
observations of $X$-ray and the Sunyaev-Zel'dovich (SZ) effect usually
rest on assumptions of hydrostatic equilibrium and spherical symmetry.
However, lensing-based methods also suffer several limitations.
First, 
lensing observables are sensitive to the total mass 
distribution
projected along the line of sight from an observer to source
galaxies.
Therefore mass concentrations along the line of sight
through a given cluster, which are not physically associated with the
cluster, increase the uncertainty on cluster mass measurements from
the lensing observables (Metzler et al. 2000; White et al. 2001;
Hoekstra 2001; Hamana et al. 2004).  Second, exhaustive spectroscopic
redshift information is not available for cluster lensing
observations.  The limited information on source galaxy redshifts
derived from the available broad-band photometry results in
degeneracies between cluster parameters and the estimated source
galaxy redshifts.  Furthermore, in practice it is not straightforward
to isolate background, therefore lensed, galaxies based on the
photometric data alone.  In fact including unlensed galaxies (mostly
cluster members for low-$z$ clusters of interest) into the lensing
analysis appears to cause a significant dilution of the lensing
distortion signals, thereby yielding biased estimations on cluster
parameters (e.g., Broadhurst et al. 2005; Medezinski et al. 2007;
Limousin et al. 2007; Hoekstra 2007; Umetsu \& Broadhurst 2008).

Recently a possible tension between the CDM model predictions and the
lensing observations has been reported: anomalously high concentration
parameter estimates have been obtained for A1689, Cl0024 and
MS2137 (Gavazzi et al.\ 2003; Kneib et al.\ 2003; Broadhurst et al.\
2005; Broadhurst et al.\ 2008; Oguri et al.\ 2009).  However, before a
serious problem with CDM may be claimed, it is important (among other
things) to address carefully the selection bias inherent in studying
strong lensing clusters.  Specifically, strong lensing clusters are
likely biased towards clusters with high concentrations and/or
significantly non-spherical mass distribution (e.g., Oguri et al.\
2005; Hennawi et al.\ 2007; Corless et al.\ 2009).  A systematic weak
lensing study of a large cluster sample is therefore an essential step
towards resolving, or confirming, this tension.  An important aspect
of such a study is to minimize possible selection biases towards
clusters with simpler (presumably ``relaxed'') or more complex
(presumably ``unrelaxed'') gravitational potentials.  A selection
function that is blind to such factors would support increased
understanding of possible biases in cluster mass estimates as a
function of the dynamical state and shape of clusters (Dahle et al.\
2002; Smith et al.\ 2005; Clowe et al.\ 2006; Bardeau et al.\ 2007;
Hoekstra 2007; Okabe \& Umetsu 2008).  

Such a systematic weak lensing
study would also be invaluable as the foundation for a careful
comparison of lensing-based mass estimates with those from other
methods, en route to measuring precisely the shape, scatter and
normalization of mass-observable scaling relations and to calibrating
the systematic errors inherent in each mass measurement method (Smith
et al.\ 2003, 2005; Hicks et al.\ 2006; Zhang et al.\ 2007, 2008;
Mahdavi et al.\ 2008; Miyazaki et al.\ 2007; Hamana et al.\ 2008;
Vikhlinin et al.\ 2008a; Henry et al.\ 2008; Berg\'e et al.\ 2008,
Umetsu et al.\ 2008).  In particular, well-calibrated mass-observable
scaling relations are critically important for the use of cluster
counting experiments to constrain the nature of dark energy (e.g.,
Lima \& Hu 2005).  Such studies are complementary to the
cross-correlation method of background galaxy shapes around clusters
binned on cluster richness, $X$-ray luminosity, etc.\ -- the so-called
{\em stacked lens} 
(Mandelbaum et al.\ 2008; Sheldon et al.\ 2007a,b; Johnston et al.\
2007), where the average properties of cluster mass
profile as well as the average mass-observable relation can be
obtained, but the information on individual clusters is lost.

In this paper, we use Subaru/Suprime-Cam observations of 30 galaxy
clusters at $0.15\le z\le0.3$ to study in detail the dark matter density
profile of the clusters.  These clusters are a sub-set of those
studied by the Local Cluster Substructure Survey (LoCuSS)
project\footnote{\sf http://www.sr.bham.ac.uk/locuss} (PI:
G.~P.~Smith; also see Smith et al.\ in preparation), and have therefore
been selected in a manner blind to the dynamical status and cluster
morphology -- see \S\ref{sec:locuss} for more details.  The superb
image quality, wide field capability, and 8-m aperture of
Subaru/Suprime-Cam (Miyazaki et al.\ 2002) allow us to investigate in
detail the accuracy achievable on cluster mass measurements with
ground-based weak lensing data.  It is also important to note that the
redshift range of the clusters in this study is well-matched to the
field-of-view of Suprime-Cam (about one-quarter square degrees) -- one
pointing spans the entire virialized region of each cluster
(cluster-centric radii of $\sim1-2r_{\rm vir}$), which is essential to
achieve robust constraints on cluster virial masses.  We explore
different methods of cluster mass estimation, specifically, fitting of
several different parametric mass profiles and the model-independent
lensing aperture mass method.  We discuss the pros and cons of each
method, and compare the results quantitatively.  We also identify the
optimal radial scale at which to measure cluster masses with weak
lensing data.  In studying all of these issues, we pay particular
attention to possible systematic errors inherent in the lensing
methods.  Most importantly, we demonstrate the importance of
correcting for dilution of the lensing signal by faint cluster
galaxies when seeking to measure robustly cluster mass and
concentration.

The structure of this paper is as follows. We describe the details of
our cluster sample and lensing analysis in Section~\ref{sec:locuss},
and define background galaxy samples to use for the lensing analysis
in Section~\ref{sec:sample}.  After describing our methods to estimate
cluster parameters from the lensing observables in
Section~\ref{sec:model}, we present the main results in
Section~\ref{sec:results}.  Section~\ref{sec:discussion} is devoted to
summary and discussion of our findings.  To improve the readability of
the paper for the non-lensing expert, several technical discussions
are presented in the appendices, for example, Appendix~\ref{app:samp}
describes details of how the background galaxy samples are defined based
on the available broad-band photometry.
We also present the two-dimensional mass maps and distortion profiles of
all the clusters in Appendix~\ref{sec:massmap}. Throughout this paper
we will assume the concordance $\Lambda$CDM model that is specified by
$\Omega_{m0}=0.27$, $\Omega_{\Lambda}=0.73$ and $H_0=72.0 \ {\rm km
s}^{-1} {\rm Mpc}^{-1}$ (Komatsu et al. 2009).

\section{CLUSTER SAMPLE AND DATA ANALYSIS}
\label{sec:locuss}

\subsection{LoCuSS}

The Local Cluster Substructure Survey (LoCuSS; Smith et al.\ in prep.;
also see Zhang et al.\ 2008) is a systematic multi-wavelength survey
of 
%more than 100 
$X$-ray luminous clusters ($L_X[0.1-2.4{\rm
keV}]\simgt2\times10^{44}~$erg/s) at redshifts of $0.15\le z\le0.3$
and declinations of $-70^\circ\le\delta\le+70^\circ$, selected from
the ROSAT All Sky Survey (RASS; Ebeling et al.\ 1998, 2000;
B\"ohringer et al.\ 2004).  The LoCuSS selection function is
deliberately blind to the physical properties of clusters, other
than the requirement to be bright enough in the X-ray band to lie
above the RASS flux limit.  The sample is therefore expected to span a
broad range of dynamical stages of cluster evolution, including
extreme merger and extreme ``relaxed'' systems.  One of the main goals
of the survey is to calibrate mass-observable scaling relations, and
to identify the main astrophysical systematic uncertainties in the use
of these relations for cluster cosmology, in a similar vein to Smith
et al.'s (2003) preliminary results on $\sigma_8$.  The Subaru data
presented in this paper form the backbone of the scaling relation
aspects of the survey.  More generally, studies of the LoCuSS sample
are in progress combining data from a wide range of ground-based
(Gemini, Keck, VLT, Subaru, SZA, Palomar, MMT, NOAO and UKIRT) and
space-based (HST, GALEX, Spitzer, Chandra and XMM-Newton) facilities.

The Subaru prime focus camera, Suprime-Cam (Miyazaki et al.\ 2002),
has the widest field-of-view (FoV) ($27'\times34'$) among 8-m class
telescopes, and can cover the entire region of a cluster at low
redshift $z\simeq0.2$ (up to a few Mpc in radius) with one pointing.
The large telescope aperture, wide FoV, and superb image quality of
Subaru/Suprime-Cam therefore mean this is a uniquely powerful facility
for an efficient ground-based weak-lensing study of a large sample of
low redshift galaxy clusters.

\subsection{Sample and Observations}

\begin{figure}
\begin{center}
%%\FigureFile(80mm,80mm){f1.eps}
%\FigureFile(90mm,90mm){lx-z_v3.eps}
\FigureFile(90mm,90mm){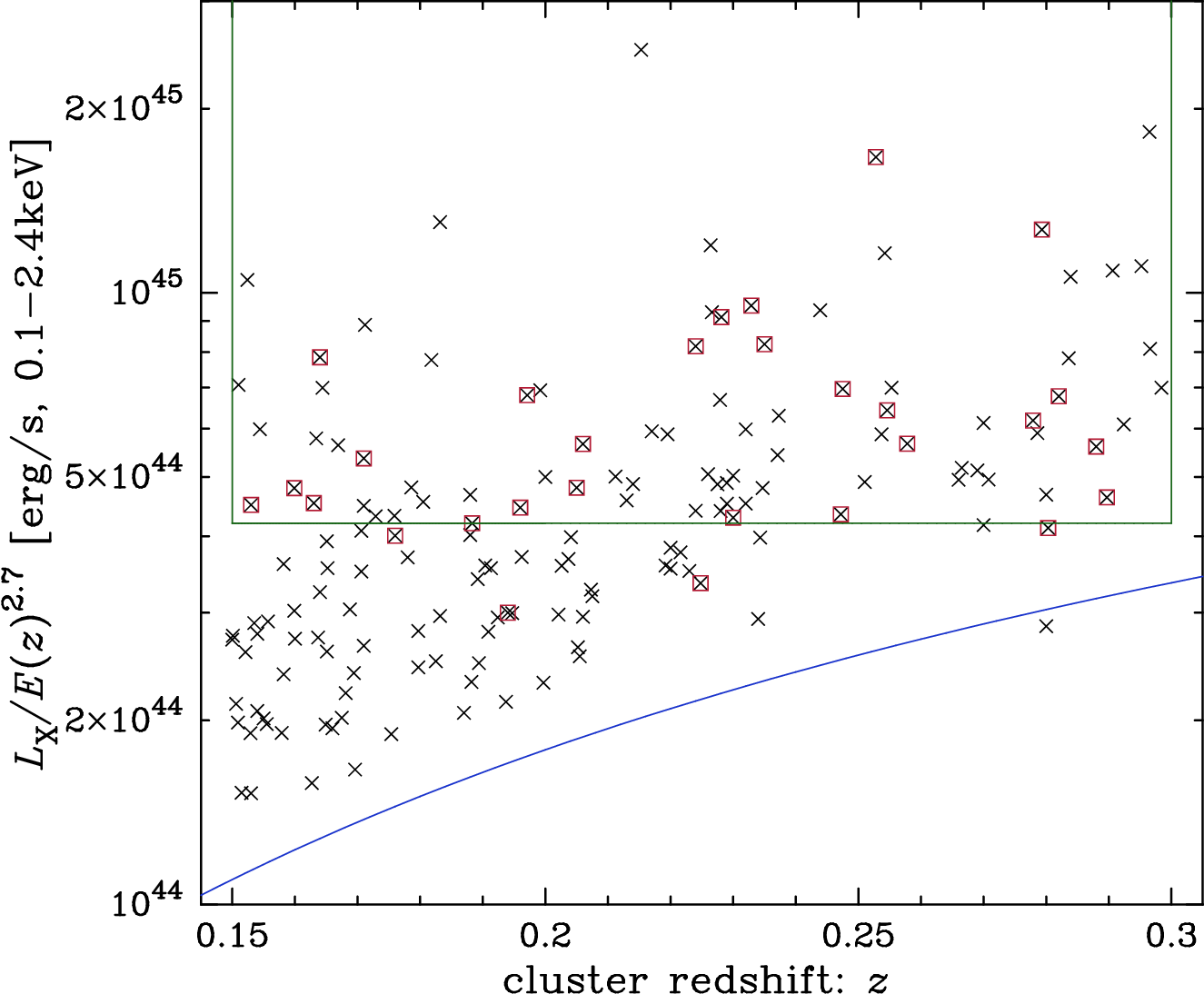}
\end{center}
\caption{The open square symbols denote the cluster sample studied in
this paper in 2D plane of the ROSAT $X$-ray luminosity and cluster
redshift, while the cross symbols show all sample of the LoCuSS
clusters. 
%MT
The box denotes our 
%``Low-$L_X$''  and 
``High-$L_X$'' sample
%: the Low-$L_X$ sample is defined by
% $2.2\times 10^{44}
%<L_X/E(z)^{2.7}<4.2\times 10^{44}$~erg~s$^{-1}$ and $0.15<z<0.2$, while
% the High-$L_X$ sample is 
defined by
 $L_X/E(z)^{2.7}>4.2\times 10^{44}$~erg~s$^{-1}$ as
 $0.15<z<0.3$, where $E(z)$ is the redshift evolution of
 the Hubble expansion rate. 
Also for comparison the solid curve denotes the luminosity with
 constant flux $f_{\rm X}=3\times 10^{-12}\mbox{erg/s/cm}^2$ assuming our
 fiducial cosmological model. 
}
 \label{fig:lx_vs_z}
\end{figure}

Clusters at declinations of $-20^\circ\le\delta\le+60^\circ$ are
are observable with Subaru
  at sufficiently high elevations 
($\simgt50^\circ$) 
 to ensure that high quality
 data suitable for faint galaxy shape measurements can be obtained.
%Clusters at declinations of $-20^\circ\le\delta\le+60^\circ$ are
%observable with the Subaru Telescope.  We further restricted our
%selection from the full LoCuSS sample to clusters at sufficiently high
%elevation ($\simgt50^\circ$) 
%to ensure that high-quality data suitable
%for faint galaxy shape measurements would be obtained.  
Through the
open use program (S05B, S06A and S07A; PI: T. Futamase) we collected
Suprime-Cam data for 20 clusters, selected solely on the observability
on the allocated observing nights.  We also added 10 cluster data from
the Subaru archive (SMOKA; \cite{bab02}). In total we study 30
clusters in this paper.

Figure~\ref{fig:lx_vs_z} shows our cluster sample in the
$L_X/E(z)^{2.7}-{\rm redshift}$ plane, where the vertical axis is
plotted in the unit of $L_X/E(z)^{2.7}$ ($E(z)$ is the normalized Hubble
expansion rate at redshift $z$: $E(z)\equiv H(z)/H_0$).  Note that the
vertical axis roughly scales with cluster mass as $L_X/E(z)^{2.7}\propto
M$ as studied in Popesso et al. (2005).
The $L_X$ distribution of our cluster sample appears to be similar to  
that of an $X$-ray luminosity limited sample with $L_{X}/ 
E(z)^{2.7}\simgt4.2\times10^{44}\mbox{erg s}^{-1}$ (Fig~\ref{fig:lx_vs_z}).  To test  
this quantitatively, we drew 3,000 random samples of 29 
clusters\footnote{All 30 clusters minus ZwCl0740 - see below for the  
details).} from the parent sample (cross symbols) defined by $L_{ X}/ 
E(z)^{2.7}\ge 4.2\times 10^{44}{\rm erg~s}^{-1}$.
The average luminosity distribution of these samples are 
statistically indistinguishable from our representative sample of 29  
clusters:
the 3000 samples have $\langle L_X/E(z)^{2.7}\rangle=6.22\times 10^{44} 
{\rm erg~s}^{-1}$ and  $\sigma({L_X/E(z)^{2.7}})=0.53\times 10^{44} 
{\rm erg~s}^{-1}$ for the
average and the standard deviation, respectively, which contain the  
average of Subaru sample, $\langle L_X/E(z)^{2.7}\rangle=6.12\times  
10^{44}{\rm
erg~s}^{-1}$ within the $1\sigma$ range. Our sample is therefore  
statistically indistinguishable from a volume-limited sample.  The  
observed clusters and the basic parameters of the observations are  
listed in Table~\ref{tab:cog}.
%Most of our clusters
%appear to be close to an
%$X$-ray luminosity limited sample with $L_{ X}/E(z)^{2.7}\simgt 4.2\times
%10^{44} \mbox{erg s}^{-1}$ rather than a flux limited sample.  To be
%more quantitative, we checked that, if 29
%clusters\footnote{All 30 clusters minus ZwCl0740 which is a cluster, but
%not included in the LoCuSS sample (see below for the details).}, 
%as in our sample, are randomly
%selected from the whole LoCuSS clusters (cross symbols) by imposing
%the luminosity threshold $L_{ X}/E(z)^{2.7}\ge 4.2\times 10^{44}{\rm
%  erg~s}^{-1}$, such samples are statistically indistinguishable from
%our cluster sample based on the chi-square test described in \S~14.3
%in Press et al. 1992. We found that most of the tests yield probabilities
%greater than 10\% ($Q(\chi^2)>0.1$) 
%that  the Subaru sample and the randomly generated
%samples look drawn from the same LoCuSS sample. 
%Hence our sample is statistically
%indistinguishable from a volume-limited sample.
%The observed clusters and the basic
%parameters of the observations are listed in Table~\ref{tab:cog}.

As discussed in detail in
\S\ref{sec:sample}~\&~Appendix~\ref{app:samp}, we use two filter data,
if available, in order to minimize, via color selection, dilution of
the weak-lensing signal caused by contamination of the background
galaxy catalog by faint cluster and foreground galaxies.

%%% 30 clusters
\begin{table*}
  \caption{Cluster Sample}\label{tab:cog}
\begin{center}
\begin{tabular}{cccccccc}
\hline
\hline
 Name         %&  Alt Name
                 &  RA
                 &  Dec
                 &  Redshift
                 &  $L_X$
                 &  $i'$
                 &  $V$
                 &  seeing  \\
                 %&
                 &  (J2000)
                 &  (J2000)
                 &    $z$
                 &  ($10^{44} {\rm erg s^{-1}}$)
                 &  (min)
                 &  (min)
                 &  (arcsec) \\
(1)              %&   (2)
                 &   (2)
                 &   (3)
                 &   (4)
                 &   (5)
                 &   (6)
                 &   (7)
                 &   (8) \\
\hline 
A68
                 %&
                 &  00 37 05.28
                 &  $+09~09~10.8$
                 &  0.2546
                 &  8.81 %8.81065                   %14.89 BCS
                 &  $16.0$\altaffilmark{a}
                 &  $30.0$\altaffilmark{a}
                 &  0.83  \\
A115
                 %&  
                 &  00 55 59.76 
                 &  $+26~22~40.8$ 
                 &  0.1971
                 &  8.63 % 8.63314
                 &  25.0\altaffilmark{a}
                 &  9.0\altaffilmark{d} %%9.0\altaffilmark{d} no useable
                 &  0.71  \\
ZwCl0104.4+0048
                 %&  
                 &  01 06 48.48
                 &  $+01~02~42.0$
                 &  0.2545
                 &  5.80 % 5.79882
                 &  35.0\altaffilmark{a}
                 &   -
                 &  0.65 \\
A209             %&  RXCJ0131.8-1336
                 &  01 31 53.00
                 &  $-13~36~34.0$
                 &  0.2060 
                 &  7.27 %7.27136
                 &  22.0\altaffilmark{d} %%with AG probe && Foucss test
                 &  30.0\altaffilmark{d}
                 &  0.63 \\
RXJ0142.0+2131   %&
                 &  01 42  02.64
                 &  $+21~31~19.2$
                 &  0.2803
                 &  5.86 %5.85799
                 &  40.0\altaffilmark{a}
                 &  30.0\altaffilmark{a}
                 &  0.67 \\ 
A267
                 %&  RXCJ0152.7+0100
                 &  01 52 48.72
                 &  $+01~01~08.4$
                 &  0.2300
                 &  8.11 %8.11243
                 &  40.0\altaffilmark{a}
                 &  30.0\altaffilmark{a}
                 &  0.61 \\
A291             %&  RXCJ0201.7-0212
                 &  02 01 44.20
                 &  $-02~12~03.0$
                 &  0.1960
                 &  5.65 %5.64574
                 &  36.0\altaffilmark{a}
                 &  30.0\altaffilmark{a}
                 &  0.71 \\
A383             %&  RXCJ0248.0-0332
                 &  02 48 02.00
                 &  $-03~32~15.0$ 
                 &  0.1883
                 &  5.27 %5.27113
                 &  36.0\altaffilmark{a}
                 &  30.0\altaffilmark{a}
                 &  0.67 \\
A521             %& RXCJ0454.1-1014       
                 &  04 54  6.88
                 &  $-10~13~24.6$
                 &  0.2475
                 &  9.46 %9.45543
                 &  22.0\altaffilmark{d,g} %%with AG probe && without Foucss test
                 &  22.0\altaffilmark{d}
                 &  0.61\\       
A586             %&  
                 &  07 32 22.32
                 &  $+31~38~02.4$
                 &  0.1710 
                 &  6.58 %6.57988
                 &  35.0\altaffilmark{c}
                 &  20.0\altaffilmark{b}
                 &  0.83\\
ZwCl0740.4+1740\altaffilmark{1}            %& Z1432
                 &  07 43 23.16
                 &  $+17~33~40.0$
                 &  0.1114 %photo-z
                 &  - %3.11834
                 &  25.0\altaffilmark{c}
                 &  20.0\altaffilmark{c}
                 &  0.83\\
ZwCl0823.2+0425            %&  
                 &  08 25 57.84
                 &  $+04~14~47.5$ 
                 &  0.2248
                 &  4.41 %4.41420
                 &  35.0\altaffilmark{c}
                 &  16.0\altaffilmark{c}
                 &  0.71\\
ZwCl0839.9+2937            %&  MS 0839.9+2938
                 &  08 42 56.07
                 &  $+29~27~25.7$
                 &  0.1940
                 &  3.79 %3.79290
                 &  35.0\altaffilmark{c}
                 &  -
                 &  0.77\\
A611             %&
                 &  08  00  55.92
                 &  $+36~03~39.6$
                 &  0.2880
                 &  8.05 %8.04734
                 &  30.0\altaffilmark{c}
                 &  16.0\altaffilmark{c}
                 &  0.79\\
A689             %&  
                 &  08  37  25.44
                 &  $+14~58~58.8$
                 &  0.2793
                 &  17.99 %17.99408
                 &  40.0\altaffilmark{c}
                 &  20.0\altaffilmark{c}
                 &  0.69\\
A697             %&
                 &  08  42  57.84
                 &  $+36~21~54.0$
                 &  0.2820
                 &  9.64 %9.64497
                 &  40.0\altaffilmark{c}
                 &  16.0\altaffilmark{c}
                 &  0.73\\
A750             %&
                 &  09  09  11.76
                 &  $+10~59~20.4$
                 &  0.1630
                 &  5.50 %5.50296
                 &  28.0\altaffilmark{d}  %%with AG probe && Foucss test
                 &  32.0\altaffilmark{d}
                 &  0.71\\
A963             %&
                 &  10  17 01.20
                 &  $+39~01~44.4$
                 &  0.2060
                 &  6.16 %6.15976
                 &  $I_{\rm c}$, 50.0\altaffilmark{d,f} %%without AG probe && with Foucss test
                 &  -
                 &  0.75 \\
A1835
                 %&
                 &  14  01 02.40
                 &  $+02~52~55.2$
                 &  0.2528
                 &  22.80 %22.79882
                 &  20.0\altaffilmark{b}
                 &  20.0\altaffilmark{b}
                 &  0.89 \\
ZwCl1454.8+2233
                 %&  MS1455.0+2232
                 &  14 57 14.40
                 &  $+22~20~38.4$ 
                 &  0.2578
                 &  7.80 %7.80473
                 &  36.0\altaffilmark{b}
                 &  15.0\altaffilmark{b}
                 &  0.81 \\
A2009            %&  
                 &  15 00 20.40
                 &  $+21~21~43.2$
                 &  0.1530
                 &  5.40 %5.39645
                 &  $R_{\rm c}$, 26.0\altaffilmark{d,e,g} %%with AG probe && without Foucss test
                 &  - %%25.6\altaffilmark{d,e}
                 &  0.75\\
ZwCl1459.4+4240            %&  RXCJ1501.3+4220
                 &  15 01 23.13
                 &  $+42~20~39.6$ 
                 &  0.2897
                 &  6.66 %6.66272
                 &  ${R_{\rm c}}$, 27.0\altaffilmark{d} 
                 &  18.0\altaffilmark{d}
                 &  0.57\\
A2219 		 &  16 40 22.56
                 &  $+46~42~21.6$	
                 &  0.2281
                 &  12.07
                 &  ${R_{\rm c}}$, 24.0\altaffilmark{d} 
                 &  18.0\altaffilmark{d}
                 &  0.99 \\
RXJ1720.1+2638   %& 
                 &  17 20 08.88
                 &  $+26~38~06.0$
                 &  0.1640
                 &  9.54 %9.53846
                 &  32.0\altaffilmark{b} 
                 &  20.0\altaffilmark{b} 
                 &  0.71 \\
A2261            %&  
                 &  17  22  27.60
                 &  $+32~07~37.2$
                 &  0.2240 
                 &  10.76 %10.75740
                 &  ${R_{\rm c}}$, 27.0\altaffilmark{d}
                 &  18.0\altaffilmark{d}
                 &  0.63\\
A2345            %&  RXCJ2127.1-1209
                 &  21 27 11.00
                 &  $-12~09~33.0$
                 &  0.1760
                 &  4.95 %4.95086
                 &  30.0\altaffilmark{a}
                 &   -
                 &  0.77 \\
RXJ2129.6+0005   %&  RBS1748
                 &  21 29 37.92 
                 &  $+00~05~38.4$
                 &  0.2350
                 &  11.00 %11.00000
                 &  44.0\altaffilmark{a}  %%Rc, 27min, 0.71arcsec archive
                 &  30.0\altaffilmark{a}
                 &  0.85 \\
A2390            
                 &  21  53  36.72
                 &  $+17~41~31.2$
                 &  0.2329
                 &  12.69 %12.68639
                 &  ${R_{\rm c}}$, 38.0\altaffilmark{d} 
                 &  12.0\altaffilmark{d}
                 &  0.65\\ %% both
A2485            %&  RXCJ2248.5-1606
                 &  22 48 31.13
                 &  $-16~06~25.6$
                 &  0.2472
                 &  5.90 %5.89664
                 &  40.0\altaffilmark{a}
                 &  30.0\altaffilmark{a}
                 &  0.67 \\
A2631            %&  RXCJ2337.6+0016
                 &  23 37 40.08
                 &  $+00~16~33.6$
                 &  0.2780
                 &  7.85 %7.84615
                 &  ${R_{\rm c}}$, 24.0\altaffilmark{d}
                 &  12.0\altaffilmark{d}
                 &  0.65\\
\hline
\end{tabular}
\end{center}
\textrm{ NOTES $\singlebond$
 Column (1): cluster name; 
 Column (2), (3): right ascension (RA) and declination (Dec) (J2000.0); 
 Column (4): redshift;  
 Column (5): the ROSAT $X$-ray luminosity 
in the 0.1-2.4keV band;
 Column (6), (7): exposure times in the $i'$ and $V$, respectively. Note
 that a case such as $I_{\rm c},50.0$ means $50$ min exposures for the
 $I_{\rm c}$ filter. 
 Column (8): the seeing FWHMs for either of $i'$, $R_{\rm c}$ and
 $I_{\rm c}$ filters that are used for our weak lensing analysis. \\
 {\altaffilmark{1} ZwCl0740.4+1740 was observed by an incorrect
 pointing, and is nevertheless added to our sample because a
 cluster exists in the field. The redshift is taken from NED.} \\
 \altaffilmark{a} Observed in the semester S05B\\
 \altaffilmark{b} Observed in the semester S06A\\
 \altaffilmark{c} Observed in the semester S07A\\
 \altaffilmark{d} Data retrieved from SMOKA} \\
 \altaffilmark{e} Data of w67c1 chip is not usable \\
 \altaffilmark{f} Data taken without AG (acquisition and guide) probe
 for guide stars \\
 \altaffilmark{g} Data taken without performing focus test before taking
 images
\end{table*}

\subsection{Image Processing and Photometry}

The data were reduced using the same algorithm as that described by
Okabe \& Umetsu (2008).  Briefly, the standard pipeline reduction
software for Suprime-Cam, SDFRED (\cite{yag02,ouc04}), was used for
flat-fielding, instrumental distortion correction, differential
refraction, PSF matching, sky subtraction and stacking.  The size of the
seeing disk in the final reduced data is very important for successful
weak-lensing measurements.  The full width half maximum (FWHM) of point
sources in the reduced data is listed for each cluster in
Table~\ref{tab:cog}.  A small seeing disk of ${\rm FWHM}\simeq
0.\!\!^{\prime\prime}7$ was achieved.  An astrometric solution was
obtained for each cluster observations by comparing the final mosaiced
image with the 2MASS catalog (\cite{skr06}).  The typical residuals on
these fits were less than the CCD pixel size ($0.\!\!^{\prime\prime}22$)
-- i.e.\ sufficient for our lensing study.  Photometric calibration was
achieved by reference to standard star observations that were
interspersed between the science observations, and SDSS photometry where
available (Adelman-McCarthy et al. 2008).  Uncertainties on the
photometric calibration were typically $\simlt0.1{\rm mag}$.
Photometric catalogs were constructed from the mosaic
images using SExtractor (\cite{ber96}).

\subsection{Weak Lensing Distortion Analysis}\label{subsec:wlana}

Our weak lensing analysis was done using the IMCAT package kindly
provided by N. Kaiser\footnote{http://www.ifa.hawaii/kaiser/IMCAT},
which was developed based on the formalism described in Kaiser,
Squires \& Broadhurst (1995; hereafter KSB).  We also incorporated
modifications by \citet{erb01} into the analysis pipeline (see Okabe
\& Umetsu 2008 for the details).

In order to obtain accurate lensing measurements, it is of critical
importance to correct for atmospheric distortion effects due to seeing
smearing and PSF anisotropy.  After constructing object catalogs of
galaxies and stars that are detected with significant signal-to-noise
($>6\sigma$ for our study), yet unsaturated, we first measure the
image ellipticity of individual galaxies by computing the weighted
quadrupole moments of the surface brightness with respect to the
galaxy center. Then, according to the KSB method, the PSF anisotropy
is corrected for as
\begin{equation} 
e'_{\alpha} = e_{\alpha} - P_{\rm sm}^{\alpha \beta} q^*_{\beta}, 
\label{eq:qstar}
\end{equation}
where $P_{\rm sm}^{\alpha\beta}$ is the {smear polarisability} matrix
being close to diagonal, and $q^*_{\alpha} = (P^*_{{\rm
sm}})^{-1}_{\alpha \beta}e_*^{\beta}$ is the stellar anisotropy kernel
(hereafter quantities with asterisk denote the quantities for stellar
objects). We select bright unsaturated stars identified in the
half-light radius, $r_h$, vs. (either of $i'$, $R_{\rm c}$ and $I_{\rm
c}$) magnitude diagram to estimate $q^*_\alpha$ for individual stellar
objects. To obtain an estimate on $q^*_\alpha$ at each galaxy position
in Eq.~(\ref{eq:qstar}), we need to construct a map of $q^*_\alpha$
that smoothly varies with angular position.  We therefore divide the
frame into several chunks the sizes of which are determined based on
the typical coherent scale of the measured PSF anisotropy pattern.  We
then fit the discrete distribution of $q^*$ in each chunk
independently to second-order bi-polynomials of vector $\btheta$
to obtain $q_*^{\alpha}(\btheta)$ at each galaxy position, in conjunction
with iterative $\sigma$-clipping rejection on each component of the
residual, $e_{\alpha}^{*{\rm (res)}} = e^*_{\alpha}-P_{*{\rm
sm}}^{\alpha\beta}q^*_{\beta}(\btheta)$.

\begin{table*}
  \caption{PSF Anisotropy Correction}\label{tab:e_star}
\begin{center}
\begin{tabular}{ccccccccc}
\hline
\hline
 Cluster         &  $\bar{e}_1^{*}$
                 &  $\bar{e}_2^{*}$
                 &  $\sigma (e^{*})$
                 &  $\bar{e}_1^{\rm * res}$
                 &  $\bar{e}_2^{\rm * res}$
                 &  $\sigma (e^{\rm * res})$
                 &  $N^*$
                 &  $\bar{r}_h^{*}$ \\
                 &  $10^{-2}$
                 &  $10^{-2}$
                 &  $10^{-2}$
                 &  $10^{-4}$
                 &  $10^{-4}$
                 &  $10^{-3}$
                 &  
                 &  (arcsec)\\
(1)              &  (2)
                 &  
                 &   
                 &  (3)
                 &  
                 &  
                 &  (4)
                 &  (5)   \\
\hline
A68              & $+1.59$
                 & $+1.10$
                 & $2.11$
                 & $-0.49\pm2.69$
                 & $-0.80\pm1.81$
                 & $6.49$
                 & $402$
                 & $0.44$  \\
A115             & $+0.86$
                 & $-1.98$
                 & $3.52$
                 & $+2.73\pm2.46$
                 & $+1.01\pm1.51$
                 & $6.80$
                 & $554$
                 & $0.37$  \\
%ZwCl0104.4+0048
ZwCl0104
             & $-4.28$
                 & $+0.79$
                 & $3.07$
                 & $+3.48\pm3.34$
                 & $-1.38\pm1.77$
                 & $8.13$
                 & $463$
                 & $0.32$  \\
A209             & $+0.85$
                 & $-4.74$
                 & $3.29$
                 & $+0.33\pm2.33$
                 & $+6.05\pm2.26$
                 & $6.64$
                 & $418$
                 & $0.33$  \\
%RXJ0142.0+2131   
RXJ0142
& $+1.80$
                 & $-2.07$
                 & $2.91$
                 & $+0.21\pm1.80$
                 & $+2.34\pm1.26$
                 & $5.35$
                 & $594$
                 & $0.35$  \\
A267             & $-1.98$
                 & $-0.85$
                 & $2.91$
                 & $-0.25\pm2.17$
                 & $+0.90\pm1.63$
                 & $5.59$
                 & $425$
                 & $0.32$  \\
A291             & $+0.61$
                 & $+1.94$
                 & $2.52$
                 & $-1.45\pm1.51$
                 & $-0.28\pm1.37$
                 & $4.14$
                 & $412$
                 & $0.38$  \\
A383             & $-0.66$
                 & $+2.19$
                 & $2.18$
                 & $+0.27\pm1.86$
                 & $-1.21\pm1.51$
                 & $4.25$
                 & $316$
                 & $0.35$  \\
A521             & $-0.29$
                 & $+1.47$
                 & $2.53$
                 & $-1.25\pm1.68$
                 & $-1.37\pm1.01$
                 & $6.34$
                 & $1046$
                 & $0.33$  \\
A586             & $-0.99$
                 & $+0.53$
                 & $2.22$
                 & $-0.09\pm0.76$
                 & $+0.91\pm0.56$
                 & $3.28$
                 & $1196$
                 & $0.47$  \\
%ZwCl0740.4+1740
ZwCl0740
            & $-3.11$
                 & $-1.50$
                 & $1.90$
                 & $+1.24\pm1.14$
                 & $+0.69\pm0.61$
                 & $4.35$
                 & $1126$
                 & $0.44$  \\
%ZwCl0823.2+0425 
ZwCl0823
           & $+1.24$
                 & $-0.36$
                 & $2.68$
                 & $+0.10\pm1.01$
                 & $+0.30\pm0.68$
                 & $4.79$
                 & $1543$
                 & $0.39$  \\
%ZwCl0839.9+2937
ZwCl0839
            & $-1.86$
                 & $+3.32$
                 & $3.01$
                 & $+0.14\pm1.42$
                 & $-2.97\pm1.38$
                 & $4.63$
                 & $544$
                 & $0.41$  \\
A611             & $-2.06$
                 & $-2.56$
                 & $2.81$
                 & $-0.41\pm1.01$
                 & $+2.14\pm1.08$
                 & $3.62$
                 & $596$
                 & $0.42$  \\
A689             & $-2.55$
                 & $-1.77$
                 & $2.51$
                 & $+2.46\pm1.71$
                 & $+0.78\pm1.02$
                 & $4.66$
                 & $549$
                 & $0.36$  \\
A697             & $-0.74$
                 & $-0.13$
                 & $4.28$
                 & $+0.10\pm0.86$
                 & $+0.07\pm0.66$
                 & $1.95$
                 & $325$
                 & $0.38$  \\
A750             & $+0.15$
                 & $+0.81$
                 & $2.04$
                 & $+1.11\pm1.67$
                 & $-1.88\pm1.07$
                 & $5.63$
                 & $806$
                 & $0.38$  \\
A963             & $-1.02$
                 & $-0.17$
                 & $2.50$
                 & $-2.65\pm2.03$
                 & $+0.86\pm1.56$
                 & $4.42$
                 & $298$
                 & $0.39$  \\
A1835            & $-1.52$
                 & $+1.88$
                 & $1.67$
                 & $+0.26\pm1.24$
                 & $-0.24\pm0.90$
                 & $3.57$
                 & $547$
                 & $0.47$  \\
%ZwCl1454.8+2233
ZwCl1454
            & $-1.28$
                 & $-0.30$
                 & $1.99$
                 & $-0.13\pm1.40$
                 & $+0.92\pm0.87$
                 & $3.83$
                 & $538$
                 & $0.43$  \\
A2009            & $+2.10$
                 & $-0.48$
                 & $1.82$
                 & $-0.56\pm2.46$
                 & $+0.05\pm1.10$
                 & $6.30$
                 & $546$
                 & $0.40$  \\
%ZwCl1459.4+4240
ZwCl1459
            & $-0.94$
                 & $-1.73$
                 & $2.25$
                 & $+4.27\pm2.97$
                 & $+3.95\pm1.73$
                 & $7.83$
                 & $518$
                 & $0.31$  \\
A2219     & $1.83$
          & $0.27$
          & $0.97$
          & $-0.94\pm0.85$
          & $0.32\pm0.53$
          & $2.58$
          & $657$
          & $0.51$ \\
%RXJ1720.1+2638
RXJ1720
   & $+1.16$
                 & $+0.12$
                 & $2.60$
                 & $-0.96\pm1.15$
                 & $-0.39\pm0.61$
                 & $4.11$
                 & $998$
                 & $0.39$  \\
A2261            & $+0.85$
                 & $-1.36$
                 & $1.53$
                 & $-0.20\pm1.21$
                 & $+1.32\pm0.61$
                 & $4.08$
                 & $911$
                 & $0.34$\\
A2345            & $+2.92$
                 & $+0.89$
                 & $3.23$
                 & $+1.10\pm1.79$
                 & $+1.36\pm1.22$
                 & $6.86$
                 & $994$
                 & $0.39$  \\
%RXJ2129.6+0005  
RXJ2129
 & $+1.20$
                 & $+4.92$
                 & $2.50$
                 & $-0.55\pm0.91$
                 & $-2.95\pm0.84$
                 & $4.72$
                 & $1446$
                 & $0.45$  \\
A2390            & $-2.39$
                 & $-1.41$
                 & $2.17$
                 & $+1.13\pm1.34$
                 & $+0.79\pm0.70$
                 & $6.44$
                 & $1811$
                 & $0.36$  \\
A2485            & $-2.03$
                 & $+2.21$
                 & $2.36$
                 & $+0.32\pm2.11$
                 & $-0.96\pm1.34$
                 & $5.46$
                 & $476$
                 & $0.35$  \\
A2631            & $-2.35$
                 & $+1.02$
                 & $2.58$
                 & $+0.63\pm1.73$
                 & $-0.18\pm1.01$
                 & $4.78$
                 & $571$
                 & $0.34$  \\
\hline
\end{tabular}
\end{center}
\textrm{ NOTES $\singlebond$ Column (1): cluster name (we used the
 abbreviation for some clusters' names); Column (2): mean and standard
 deviation for two components of stellar ellipticities before PSF
 correction; Column (3): mean and standard deviation after the PSF
 anisotropy correction; Column (4): number of stellar objects used in
 the analysis; Column (5): median stellar half-light radius in unit of
 arcseconds }
\end{table*}
Table~\ref{tab:e_star} summarizes the results of the PSF anisotropy
correction in each cluster field.  While the mean and rms of the
original stellar ellipticities are typically both a few per cent, the
correction described above reduces the residual ellipticities to
$|\overline{e}_{\alpha}^{*{\rm res}}| \simlt 10^{-4}$ and the rms of
the residuals, $\sigma(e^{*{\rm res}})$ to less than $10^{-2}$ for all
clusters, even down to a few times $10^{-3}$ for a few clusters.
Measurements of cluster distortion signals of $>10^{-2}$ should therefore
be robust, based on this PSF anisotropy correction.

Next we correct for the isotropic smearing of galaxy images caused by
seeing and the Gaussian window function used for the shape
measurements.  An estimate on the pre-seeing reduced distortion signal,
$g_\alpha$, for each galaxy can be obtained from
\begin{eqnarray}
g_\alpha&=& (P_g^{-1})_{\alpha\beta} e'_{\beta}, \label{eq:raw_g}
\end{eqnarray}
where $P^g_{\alpha\beta}$ is the pre-seeing shear polarizability
tensor.  To reduce noise, we employ the following procedures.  First
the tensor $P^g_{\alpha\beta}$ for each galaxy is estimated based on
the scalar correction approximation (\cite{hud98,hoe98,erb01}) as
\begin{equation}
(P_{g})_{\alpha\beta}=\frac{1}{2}{\rm tr}[P_g]\delta_{\alpha\beta}\equiv
P_g^{\rm s}\delta_{\alpha\beta}. 
\label{eq:Pgs}
\end{equation}
The tensor $P_{g}^{\rm s}$ for individual galaxies is still noisy,
especially for small and faint galaxies, and therefore we employ a
practically useful procedure developed by van Waerbeke et al.\ (2000;
see also \cite{erb01,ham03}).  We first identify 30 neighboring
galaxies around each galaxy in the magnitude-$r_g$ plane where $r_g$
is the Gaussian smoothing radius used in the KSB method, and then
compute, over the defined neighboring sample, the median value
$\langle P_g^{\rm s}\rangle$ as an estimate on $P_g$ used in
Eq.~(\ref{eq:Pgs}). We thus use the following estimator for the
reduced distortion signal of each galaxy:
\begin{equation}
g_{\alpha} = \frac{e'_{\alpha}}{\left< P_g^{\rm s}\right>}.
\label{eq:e_rshear}
\end{equation}
Using this equation we also compute the variance $\sigma^2_g(\equiv
\sigma^2_{g_1}+\sigma^2_{g_2})$ for each galaxy over the neighboring
sample.  We will below use the dispersion $\sigma_g^2$ to estimate a
statistical error in measuring the lensing distortion signals as
described around Eq.~(\ref{eq:sig_g+}).  Typically $\sigma_g\sim 0.4$
for our galaxy samples.

\subsection{Tangential Distortion Profile}

As indicated in Eq.~(\ref{eq:e_rshear}), the reduced distortion is
given by two components reflecting the spin-2 field nature (e.g.\ see
Bartelmann \& Schneider 2000): $\bmf{g}=(g_1,g_2)$. For cluster
lensing it is useful to define, for each galaxy, the tangential
distortion component, $g_+$, and the $45$ degree rotated component,
$g_\times$, with respect to the cluster center:
\begin{eqnarray}
g_{+(i)}&=&-(g_{1(i)}\cos2\varphi+g_{2(i)}\sin2\varphi), \nonumber\\
g_{\times (i)}&=&-g_{1(i)}\sin2\varphi+g_{2(i)}\cos2\varphi, 
\label{eq:gti}
\end{eqnarray}
where subscript $(i)$ denotes the $i$-th galaxy, and $\varphi$ is the
position angle between the first coordinate axis on the sky and the
vector connecting the cluster center and the galaxy position.  The
minus sign for the definition of $g_+$ is introduced so that $g_+$
becomes positive or negative when a background galaxy shape is
tangentially or radially deformed with respect to the cluster center,
respectively.  The tangential distortion contains the full information
on the lensing signals if the lensing mass distribution 
is
axisymmetric on the sky.  Clusters can be considered as gravitationally
bound spherical objects to zeroth order, therefore studying the
tangential distortion profile is a sensible first step.  The position
of the brightest cluster galaxy (BCG) is adopted as the cluster center
for this analysis (see also \S~\ref{sec:member}).

The weak lensing signal of a cluster is typically $0.01-0.1$ in
ellipticities and cannot be distinguished from the intrinsic
ellipticity of individual galaxies.  The coherent weak-lensing
distortion pattern is therefore only measurable at high significance
when averaged over sufficient background galaxies, thus beating down
the ``shape noise'' attributable to the (assumed) random intrinsic
ellipticity distribution of background galaxies.  Note that the
assumption of random intrinsic galaxy ellipticities and orientations
is safe because the majority of galaxies considered are separated by
cosmological distances and therefore are not physically associated
with each other.  The tangential distortion profile is estimated as
\begin{equation}
\langle{g_+}\rangle(\theta_n)=\frac{\sum_{i}w_{(i)} g_{+(i)}}
{\sum_i w_{(i)}},
\label{eq:1d_gt}
\end{equation}
with 
The weight function $w_{(i)}$ 
being given by 
\begin{equation}
w_{(i)}\equiv \frac{1}{\alpha^2+\sigma_{g(i)}^2}. 
\end{equation}
In Eq.~(\ref{eq:1d_gt}) the summation runs over all the galaxies
residing in the $n$-th radial bin $\theta_n$ with a given bin
width. The weighting $w_{(i)}$ is used to down-weight galaxies whose
shapes are less reliably measured, based on the uncertainty in the
shape measurement, $\sigma_{g(i)}$, estimated for the $i$-th galaxy
(see \S~\ref{subsec:wlana}), following Van Waerbeke et al.\ (2000). We
use $\alpha=0.4$ throughout this paper.

The other distortion component, $g_\times$ (see Eq.~[\ref{eq:gti}]),
should vanish after the azimuthal average in the weak lensing regime.
Therefore, the measured $\langle g_\times\rangle(\theta_n)$ in each
radial bin serves as a monitor of systematics errors, most likely
arising from imperfect PSF correction.

The statistical uncertainty on the tangential distortion profile in
each radial bin can be estimated as
\begin{equation}
\sigma_{g_+}^2(\theta_n) =\frac{1}{2}\frac{\sum_i w_{(i)}^2 \sigma_{g(i)}^2}
{\left(\sum_i w_{(i)} \right)^2},
\label{eq:sig_g+}
\end{equation}
where the prefactor $1/2$ accounts for the fact that $\sigma_{g(i)}$
is the r.m.s. for the sum of two distortion components (see below
Eq.~\ref{eq:e_rshear}).  The statistical error in $\langle g_\times
\rangle(\theta_n)$ is the same as that given in
Eq.~(\ref{eq:sig_g+}).  Here we have assumed that dominant source of
the measurement errors is the intrinsic ellipticities that are
uncorrelated between different radial bins.  We thus, for simplicity,
ignore the error contribution arising from cosmic shearing effects on
galaxy images caused by large-scale structures along the line-of-sight
through a cluster.  As discussed in Hoekstra (2003; also see Dodelson
2004), the cosmic shear contribution may reduce an accuracy of cluster
mass estimation from the distortion profile.
However we checked that the cosmic shear contamination is insignificant,
because the shot noise turns out to be more significant than assumed in
Hoekstra (2003), due to a 
%MT
smaller
%fewer 
number of galaxies 
used in the lensing analysis
after our background galaxy selection (see Oguri et
al. 2010 for the detail). 

\begin{table*}
  \caption{Background Galaxy Samples and the Lensing Distortion Signals}
\label{tab:wl_para}
\begin{center}
\begin{tabular}{c|cc|ccc|ccc}
\hline
\hline
 Cluster         &  $r_h$
                 &  mag 
                 &  $n_{\rm g}^{\rm all}$
                 &  $\langle\langle g_{+}\rangle\rangle$
                 &  $(S/N)$ 
                 &  $n_{\rm g}^{\rm red+blue}$
                 &  $\langle\langle g_{+}\rangle\rangle$
                 &  $(S/N)$ \\
                 &  (pix. scale)
                 &  (AB mag)
                 &  (arcmin$^{-2}$)
                 &  ($10^{-2}$)
		 &
                 &  (arcmin$^{-2}$)
                 &  ($10^{-2}$)
                 &  \\
                 %&  Morphology
(1)              &  (2)
                 &  (3)
                 &  (4) 
                 &  (5)
                 &  (6)
                 &  (7)
                 &  (8)
		 &  (9) \\
\hline 
A68        & $ [ 2.25 - 10.00 ] $
           & $ [ 22.0 - 26.0 ] $
           & $19.85$
           & $2.60\pm0.53$
           & $6.20$
           & $9.61$
           & $3.65\pm0.94$
           & $5.83$ \\
A115        & $ [ 1.96 - 4.00 ] $
           & $ [ 22.0 - 26.0 ] $
           & $14.23$
           & $1.80\pm0.60$
           & $5.99$
           & $5.94$
           & $3.79\pm1.05$
           & $5.39$ \\
%ZwCl0104.4+0048
ZwCl0104
        & $ [ 1.70 - 10.00 ] $
           & $ [ 22.0 - 26.0 ] $
           & $42.88$
           & $2.44\pm0.42$
           & $7.60$
           & -
           & -
           & - \\
A209        & $ [ 1.79 - 10.00 ] $
           & $ [ 22.0 - 25.8 ] $
           & $37.13$
           & $3.78\pm0.66$
           & $14.30$
           & $20.90$
           & $5.83\pm1.10$
           & $12.85$ \\
%RXJ0142.0+2131
RXJ0142
        & $ [ 1.85 - 10.00 ] $
           & $ [ 22.0 - 26.0 ] $
           & $36.92$
           & $3.56\pm0.68$
           & $9.78$
           & $20.71$
           & $6.60\pm1.18$
           & $9.47$ \\
A267        & $ [ 1.66 - 10.00 ] $
           & $ [ 22.0 - 26.0 ] $
           & $42.82$
           & $4.61\pm0.64$
           & $11.90$
           & $24.10$
           & $5.48\pm0.85$
           & $9.30$ \\
A291        & $ [ 2.04 - 10.00 ] $
           & $ [ 21.5 - 26.3 ] $
           & $36.84$
           & $2.23\pm0.44$
           & $9.50$
           & $18.06$
           & $3.17\pm0.68$
           & $8.26$ \\
A383        & $ [ 1.81 - 10.00 ] $
           & $ [ 22.0 - 26.2 ] $
           & $48.15$
           & $4.01\pm0.40$
           & $12.57$
           & $33.81$
           & $4.95\pm0.52$
           & $12.00$ \\
A521        & $ [ 1.74 - 10.00 ] $
           & $ [ 22.0 - 26.2 ] $
           & $43.65$
           & $3.02\pm0.54$
           & $11.74$
           & $27.26$
           & $3.98\pm0.78$
           & $9.81$ \\
A586        & $ [ 2.50 - 10.00 ] $
           & $ [ 22.0 - 26.0 ] $
           & $22.02$
           & $6.73\pm0.86$
           & $11.31$
           & $7.48$
           & $11.20\pm1.99$
           & $9.08$ \\
%ZwCl0740.4+1740
ZwCl0740
        & $ [ 2.25 - 10.00 ] $
           & $ [ 21.5 - 25.5 ] $
           & $21.55$
           & $1.89\pm0.48$
           & $7.19$
           & $16.73$
           & $2.23\pm0.60$
           & $6.56$ \\
%ZwCl0823.2+0425
ZwCl0823
        & $ [ 2.11 - 10.00 ] $
           & $ [ 22.0 - 25.9 ] $
           & $26.61$
           & $3.64\pm0.51$
           & $11.24$
           & $16.92$
           & $4.08\pm0.60$
           & $10.38$ \\
%ZwCl0839.9+2937
ZwCl0839
        & $ [ 2.23 - 10.00 ] $
           & $ [ 22.1 - 26.0 ] $
           & $26.82$
           & $3.67\pm0.61$
           & $7.67$
           & -
           & -
           & - \\
A611        & $ [ 2.15 - 10.00 ] $
           & $ [ 22.0 - 26.0 ] $
           & $31.23$
           & $3.04\pm0.47$
           & $9.63$
           & $21.00$
           & $4.08\pm0.59$
           & $9.81$ \\
A689        & $ [ 1.86 - 10.00 ] $
           & $ [ 22.0 - 26.0 ] $
           & $39.80$
           & $0.52\pm0.42$
           & $6.38$
           & $22.08$
           & $1.49\pm0.58$
           & $5.29$ \\
A697        & $ [ 2.04 - 10.00 ] $
           & $ [ 22.0 - 26.2 ] $
           & $39.10$
           & $3.01\pm0.49$
           & $12.53$
           & $20.58$
           & $5.21\pm0.76$
           & $12.06$ \\
A750        & $ [ 2.06 - 10.00 ] $
           & $ [ 22.0 - 26.0 ] $
           & $31.67$
           & $2.80\pm0.38$
           & $13.70$
           & $13.59$
           & $4.51\pm0.67$
           & $10.30$ \\
A963        & $ [ 2.08 - 10.00 ] $
           & $ [ 21.5 - 26.1 ] $
           & $43.57$
           & $3.15\pm0.44$
           & $11.45$
           & -
           & -
           & - \\
A1835        & $ [ 2.44 - 10.00 ] $
           & $ [ 20.0 - 24.5 ] $
           & $19.76$
           & $4.16\pm0.52$
           & $11.79$
           & $14.93$
           & $4.60\pm0.66$
           & $11.11$ \\
%ZwCl1454.8+2233
ZwCl1454
        & $ [ 2.21 - 10.00 ] $
           & $ [ 21.8 - 24.8 ] $
           & $20.85$
           & $3.90\pm0.88$
           & $7.39$
           & $9.78$
           & $4.05\pm1.12$
           & $5.84$ \\
A2009        & $ [ 2.14 - 10.00 ] $
           & $ [ 22.4 - 26.1 ] $
           & $25.47$
           & $3.85\pm0.56$
           & $9.42$
           & -
           & -
           & - \\
%ZwCl1459.4+4240
ZwCl1459
        & $ [ 1.64 - 10.00 ] $
           & $ [ 22.0 - 26.3 ] $
           & $56.26$
           & $3.13\pm0.56$
           & $9.80$
           & $11.18$
           & $5.93\pm1.66$
           & $7.45$ \\
A2219        & $ [ 2.65 - 10.00 ] $
           & $ [ 22.0 - 26.0 ] $
           & $25.77$
           & $4.13\pm0.52$
           & $12.13$
           & $10.33$
           & $9.26\pm1.10$
           & $11.27$ \\
%RXJ1720.1+2638
RXJ1720
        & $ [ 2.03 - 10.00 ] $
           & $ [ 20.0 - 24.2 ] $
           & $20.12$
           & $2.92\pm0.52$
           & $6.70$
           & $10.72$
           & $4.77\pm0.77$
           & $7.75$ \\
A2261        & $ [ 1.76 - 10.00 ] $
           & $ [ 22.0 - 26.0 ] $
           & $44.77$
           & $4.13\pm0.37$
           & $16.49$
           & $16.76$
           & $6.85\pm0.71$
           & $12.90$ \\
A2345        & $ [ 2.01 - 4.00 ] $
           & $ [ 22.0 - 25.8 ] $
           & $15.68$
           & $2.26\pm0.63$
           & $6.78$
           & -
           & -
           & - \\
%RXJ2129.6+0005
RXJ2129
        & $ [ 2.34 - 10.00 ] $
           & $ [ 22.0 - 25.5 ] $
           & $26.51$
           & $2.80\pm0.50$
           & $8.42$
           & $15.49$
           & $3.97\pm0.71$
           & $7.89$ \\
A2390        & $ [ 1.89 - 10.00 ] $
           & $ [ 22.0 - 26.2 ] $
           & $33.58$
           & $5.91\pm0.67$
           & $14.73$
           & $10.70$
           & $9.55\pm1.31$
           & $10.66$ \\
A2485        & $ [ 1.81 - 10.00 ] $
           & $ [ 22.0 - 25.9 ] $
           & $38.35$
           & $2.79\pm0.49$
           & $9.52$
           & $15.06$
           & $3.50\pm0.82$
           & $6.78$ \\
A2631        & $ [ 1.79 - 10.00 ] $
           & $ [ 22.0 - 26.3 ] $
           & $47.11$
           & $3.25\pm0.43$
           & $10.55$
           & $30.89$
           & $5.47\pm0.61$
           & $11.30$ \\
\hline
\end{tabular}
\end{center}
\textrm{ NOTES $\singlebond$ Column (1): cluster name; Column (2): the
range of half light radius used in selecting the background galaxy
sample; Column (3): the magnitude range of the background galaxy sample;
Column (4): the angular number density of background galaxies for the
faint galaxy sample (see text for the details); Column (5): the mean
strength of the tangential distortion profile averaged over radii from
$\simeq 1'$ up to the outermost radius, measured for the faint galaxy
sample (see Eq.[\ref{eqn:aagt}] for the definition); Column (6): the
total signal-to-noise ratio for the tangential distortion signal (see
Eq.~[\ref{eqn:sn}]); Column (7): the number density for the red plus
blue background galaxy sample (the row marked ``--'' denotes the cluster
where no color information on galaxies is available); Columns (8) and
(9): similar to Columns (5) and (6), but for the red+blue galaxy sample.
}\\
\end{table*}

\section{GALAXY SAMPLE SELECTION AND SOURCE REDSHIFT ESTIMATION}
\label{sec:sample}

\begin{figure}
\begin{center}
\FigureFile(80mm,80mm){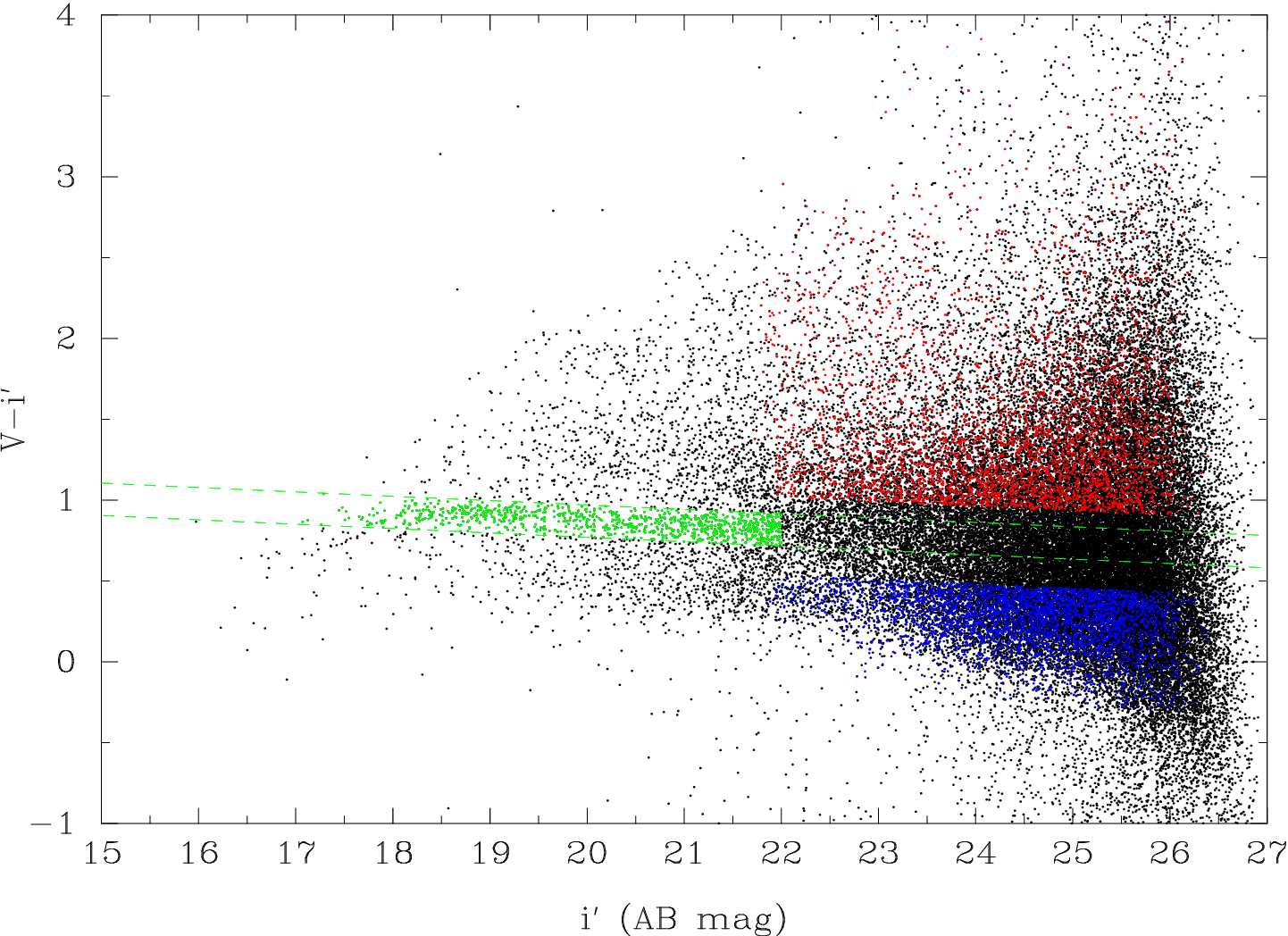}
\end{center}
\caption{The color-magnitude diagram and our galaxy samples for A68 as
 one representative example (see text for details).  Green points
 are the member galaxy sample for E/S0 galaxies of this cluster, where
 the two dashed curves denote the width of the red sequence. The red
 and blue points are the background galaxy samples redder and blue
 than the red-sequence, respectively, used for the lensing distortion
 analysis. Note that these shear catalogs are chosen imposing another
 condition that galaxies are well resolved to make reliable shape
 measurements, so do not include all the red/blue galaxies in the
 diagram.} \label{fig:cmr}
\end{figure}

\subsection{Galaxy Sample Selection}

Including unlensed galaxies, mainly cluster members in the case of our
low-$z$ clusters, into the background galaxy catalog dilutes the
measured lensing strengths.  Therefore it is of vital importance to
minimize contamination of the background galaxy catalogs in order to
obtain robust lensing measurements (Broadhurst et al.\ 2005; Limousin
et al.\ 2007).  For the clusters for which data in two filters are
available (see Table~\ref{tab:cog}), we therefore define the following
four galaxy samples:
\begin{itemize}
\item {\em Member galaxy sample}: brightest cluster galaxy (BCG)
      plus galaxies that are contained in the cluster red sequence and
      brighter than $22$ mag (AB) in the red-band magnitude (either of
      $i'$, $R_c$, and $I_c$).
\item {\em Faint galaxy sample}: galaxies that are in the magnitude
      range listed in Table~\ref{tab:wl_para} (typically fainter than
      $22$ mag) and are well resolved so as to be usable for weak
      lensing analysis.
\item {\em Red galaxy sample}: galaxies contained in the faint galaxy
      sample, but redder than the cluster red sequence at least by a
      finite color offset that is chosen to reduce the monitored
      dilution effect on the lensing distortion signal.
\item {\em Blue galaxy sample}: galaxies contained in the faint galaxy
      sample, but bluer than the red sequence at least by a
      finite color offset.
\end{itemize}
As one example, Figure~\ref{fig:cmr} shows these galaxy samples in the
color-magnitude diagram of A68.  For the clusters for which single
filter data are available, we use the faint galaxy sample for our
lensing analysis.  In Appendix~\ref{app:samp} we describe in detail
how the galaxy samples are defined based on the color-magnitude
information of galaxies, constructed from two filter data.  We
briefly summarize the method below.

Broadhurst et al.\ (2005) showed that selecting galaxies redder than
the cluster red sequence yields galaxy samples dominated by background
galaxies because the red galaxy colors are caused by larger
$k$-corrections than for lower redshift objects.  These photometric
results have also been confirmed spectroscopically by Rines \& Geller
(2008).  However, using solely red galaxies for weak-lensing analysis
generally leads to low signal-to-noise ratios because of the
relatively small number density of red galaxies.  We therefore use
combined red plus blue galaxy samples as our fiducial sample of
background galaxies in order to obtain a higher statistical precision
for our lensing measurements.  

In brief the method for determining the appropriate color cuts with
which to define the red and blue galaxy samples -- i.e.\ how much
redder or bluer than the cluster red sequence needs to 
%be qualified
qualify as 
a red or blue galaxy respectively -- consists of plotting the mean
distortion profile strength $\langle \langle g_+\rangle \rangle$ (see
Eq.~[\ref{eqn:aagt}] for the definition) as a function of color offset
from the cluster red sequence.  In regions of high contamination
(i.e.\ small color offsets) the value of $\langle\langle
g_+\rangle\rangle$ is depressed due to the stronger dilution of
distortion signals.  The optimal color offsets are thus chosen to
minimize this effect 
%MT
(e.g., see Figs.~12) 
 --
the typical color cuts  are ${\rm |\Delta color|}\simeq [0.1,
  0.4]$ for the redder and bluer galaxies than the cluster
red-sequence.  

\begin{figure}
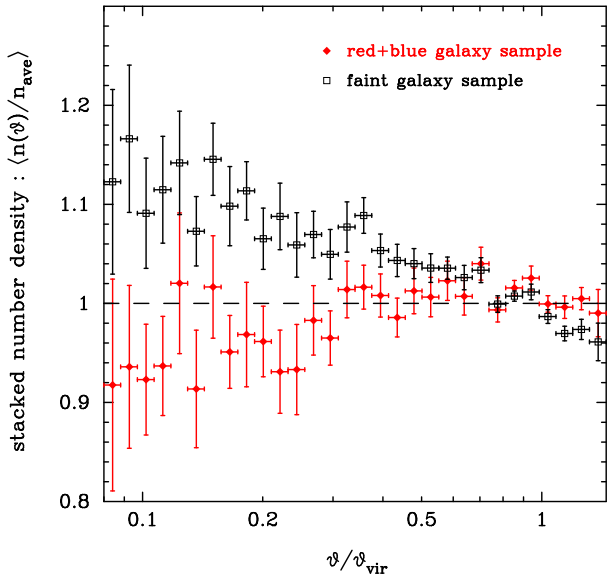

\begin{center}%%NO updated using new theta_vir
%\FigureFile(80mm,80mm){Nprofile.ps}
\FigureFile(80mm,80mm){f3.ps}
\end{center}
\caption{The radial profiles of galaxy number density stacked over 21
 clusters with two passband data, for the faint background galaxy
 samples (square symbols) and the red+blue galaxy samples (circle),
 respectively. The profiles are computed by stacking the number
 densities in each radial bins normalized by the virial radius of each
 cluster for the galaxy samples of 21 clusters. The profile for the
 faint galaxy samples shows increasing number densities with decreasing
 radius, indicating contamination of member galaxies. On the other hand,
 the red+blue galaxy sample does not show any excess at the inner radii.}
 \label{fig:Nprofile}
\end{figure}
A justification on 
 the effectiveness of the red and
blue color selection method is given by Figure~\ref{fig:Nprofile}
showing the stacked number density profiles
as a function of cluster-centric radius
for the faint and red$+$blue galaxy samples, respectively.  
These stacked profiles
are constructed from the 21 clusters for which data are available in
two filters, excluding ZwCl0740 (the lowest redshift cluster).  
The number
density profile of the faint galaxy sample shows 
increasing densities at smaller cluster-centric radii as expected for a
catalog that is contaminated by faint cluster members.  
On the other hand, 
the number density profile of red+blue galaxies does not
show any evidence of excess in the number densities at the small
radii, and consistent with a constant density within the Poisson
errors. One may notice that a slight depression in the number density
at small radii, for the red+blue galaxy sample. This is probably due to 
an overestimation in the solid angle in computing the number
density.  Since the red+blue galaxy sample is defined from the faint
galaxy sample by excluding  galaxies around the red-sequence, we have to
take into account the masking effect of the excluded galaxies on the
solid angle on the sky, and this masking contamination is more
significant at smaller radii due to the increased contribution of member
galaxies. However we ignored this effect, and this likely causes to
underestimate the number densities at the small radii
for the red+blue galaxy sample. 
%MT
Furthermore the number densities at such small radii may be affected by
the magnification effect on background galaxies that causes the galaxies
to be included or excluded in the sample within a given magnitude range. 

Table~\ref{tab:wl_para} summarizes the 
background galaxy samples obtained based on the methods outlined above,
the number density of galaxies in the faint and red$+$blue galaxy
samples, the mean strength of tangential distortion profile and the
total signal-to-noise ratio measured for each galaxy sample.  Here the
signal-to-noise ratios are defined from Eqs.~(\ref{eq:1d_gt}) and
(\ref{eq:sig_g+}) as
\begin{equation}
\left(\frac{S}{N}\right)^2\equiv \sum_{n=1}^{N_{\rm bin}}
\frac{\left[\langle g_+\rangle(\theta_n)\right]^2}
{\sigma_{g_+}^2(\theta_n)}.
\label{eqn:sn}
\end{equation}
We emphasize that the Subaru data allow us to achieve a significant
detection of the tangential distortion at $(S/N)\simgt5$ for all the
cluster fields.  Comparison of the 5th and 8th columns in
Table~\ref{tab:wl_para} also confirms quantitatively the impact of
dilution on weak lensing signal -- the mean distortion signal is larger for
the red$+$blue galaxy sample than for the faint galaxy sample in every
cluster for which two filter data are available.  However, the number
density of galaxies in the red$+$blue samples is a factor of
$\sim1.5-3$ lower than in the faint galaxy samples (see 4th and 7th
columns of Table~\ref{tab:wl_para}).  Nevertheless, as a result of the
balance between these competing effects, the total signal-to-noise
ratios are only degraded by $\simlt20\%$ in most cases, and by
$\sim30\%$ for a few clusters (see 6th and 9th columns).

%MT
\subsection{Source Redshift Estimation}
\label{ssec:zs}

\begin{table}
  \caption{Distance ratio averaged over the redshift distribution of 
background galaxy sample 
%{\bf ZZZ have changed this to single column ZZZ}
}
\begin{center}
\begin{tabular}{l|cc}
\hline
\hline \\
Name  & $\langle D_{\rm ls}/D_{\rm s} \rangle_{{\rm faint}}$
      & $\langle D_{\rm ls}/D_{\rm s} \rangle_{{\rm red+blue}}$ \\
\hline
A68  &  $0.655$
            &  $0.660$ \\
A115  &  $0.701$
            &  $0.715$ \\
ZwCl0104  &  $0.657$
            &  $-$ \\
A209  &  $0.709$
            &  $0.714$ \\
RXJ0142  &  $0.628$
            &  $0.635$ \\
A267  &  $0.683$
            &  $0.687$ \\
A291  &  $0.723$
            &  $0.744$ \\
A383  &  $0.732$
            &  $0.733$ \\
A521  &  $0.667$
            &  $0.668$ \\
A586  &  $0.728$
            &  $0.738$ \\
ZwCl0740  &  $0.820$
            &  $0.827$ \\
ZwCl0839  &  $0.709$
            &  $-$ \\
A611  &  $0.615$
            &  $0.626$ \\
A697  &  $0.623$
            &  $0.637$ \\
A963  &  $0.686$
            &  $-$ \\
A1835  &  $0.584$
            &  $0.603$ \\
ZwCl1454  &  $0.622$
            &  $0.633$ \\
A2009  &  $0.779$
            &  $-$ \\
ZwCl1459  &  $0.637$
            &  $0.706$ \\
RXJ1720  &  $0.697$
            &  $0.727$ \\
A2219  &  $0.684$
            &  $0.762$ \\
A2261  &  $0.697$
            &  $0.733$ \\
RXJ2129  &  $0.667$
            &  $0.672$ \\
A2390  &  $0.691$
            &  $0.736$ \\
A2485  &  $0.670$
            &  $0.688$ \\
A2631  &  $0.639$
            &  $0.655$ \\
\hline
\label{tab:zs}
\end{tabular}
\end{center}
\textrm{ NOTES $\singlebond$
 Column (1): cluster name; 
 Column (2): distance ratio averaged over the redshift distribution of
 {\it faint} galaxy sample, calibrated based on the COSMOS photometric
 redshift catalog
 Column (3): distance ratio averaged over the redshift distribution of
 {\it red+blue} galaxy sample
}
\end{table}

The overall normalization of lensing distortion signals depends on the
redshift distribution of background galaxies. An uncertainty in the
source redshift causes biases in parameter estimations.

The lensing amplitude for a given cluster of known redshift
scales with the mean distance ratio averaged over the population of
source galaxies:
\begin{equation}
\left\langle\frac{D_{\rm ls}}{D_{\rm s}}\right\rangle=
\int\!dz~ \frac{dp_{\rm wl}}{dz} \frac{D_{\rm ls}}{D_{\rm s}}, 
\label{eq:ave_beta}
\end{equation}
where $D_{\rm ls}$ and $D_{\rm s}$ are the angular diameter distances
from the lens to source and from the observer to source. The probability
distribution function
$dp_{\rm wl}/dz$ is the redshift distribution of source galaxies used in
the lensing analysis.

Since redshifts of our imaging galaxies are not available, we instead
employ a statistical approach as follows. In order to estimate $dp_{\rm
wl}/dz$ we used the COSMOS photometric redshift catalog given in Ilbert
et al. (2008). The photometric redshifts were estimated by combining 30
broad, intermediate and narrow bands covering a wide range of
wavelengths from UV, optical to mid infrared, and also calibrated using
the spectroscopic subsample. Hence the catalog provides currently the
most reliable redshift distribution, for magnitude limited galaxy sample
selected with $i<25$ in the Subaru $i$-band data. In addition the
catalog is constructed from the sufficiently large survey area of about
2 square degrees, therefore it can be considered as a fairly
representative sample of distant galaxies.

To estimate redshifts of our galaxy samples, we first construct a
subsample of galaxies from the COSMOS catalog by imposing the same
color cut used in our weak lensing analysis for each cluster field
(see Table~\ref{tab:wl_para} for the cuts). Then we compute the
average distance ratio (\ref{eq:ave_beta}) using the redshift
distribution of the COSMOS subsample based on the available photo-$z$
information. Although our source galaxies contain galaxies fainter
than $i=25$ as listed in Table~\ref{tab:wl_para}, we ignored the
contribution because the fraction of such faint galaxies in our source
galaxies is not large, and the redshift distribution does not so much
change for the range of limiting magnitudes, $25<i<26$.
Table~\ref{tab:zs} shows the estimated distance ratio for each cluster
field. Note that the distance ratio for a faint galaxy sample varies
with clusters because of the different ranges of
magnitudes used to define the faint galaxy sample as well as the
differences of cluster redshifts.  A typical error inferred from the
statistical errors of photometric redshifts is found to be, at most, a
few per cent in $\langle D_{\rm ls}/D_{\rm s}\rangle$.  We will come
back to a possible residual uncertainty in source redshifts in
\S~\ref{subsec:zs}, and it turns out the uncertainty, even if it
exists, does not cause any significant changes on cluster parameters
because our clusters are all at low redshifts, 
%MT
$[0.15,0.3]$.
% {\bf ZZ do  you mean 0.15-0.3? ZZ}.

\section{MODELING OF LENSING DISTORTION SIGNALS}
\label{sec:model}

In this section we describe the modeling methods that will be applied
to the data to constrain cluster mass and density
profile shape presented in \S\ref{sec:results}.

\subsection{Model-Dependent Estimate of 3D Cluster Mass}

The definition of mass most often used in the literature is the
three-dimensional mass enclosed within a spherical region of a given
radius $r_{\Delta}$ inside of which the mean interior density is
$\Delta$ times the critical mass density, $\rho_{\rm cr}(z)$, at
redshift of a cluster:
\begin{equation}
M_\Delta=\frac{4\pi}{3}r_\Delta^3\rho_{\rm cr}(z)
\Delta. 
\label{eq:M_Delta}
\end{equation}
Conventionally either a constant over-density such as
$\Delta\simeq200$ or the virial over-density $\Delta=\Delta_{\rm
vir}(z)$ (e.g., Nakamura \& Suto 1997 for the definition of
$\Delta_{\rm vir}$; see also Tomita 1969; Gunn \& Gott 1972) are used.
%MT
Note that $\Delta_{\rm vir}\simeq 112$ for halos at
redshift $z=0.2$ for our fiducial cosmological model. 
This spherical over-density mass is very useful from a theoretical
viewpoint because the dark matter halo mass function derived from
numerical simulations is well fitted by a simple analytical formula
such as the Press-Schechter function (Press \& Schechter 1974; see
also White 2002) if halo masses are computed using the spherical
top-hat average of mass distribution in each halo regions in
simulations.

However weak-lensing observables do not provide direct estimates of
the three-dimensional masses of clusters 
because the
lensing signal probes the two-dimensional projected mass distribution.  We
therefore estimate $M_\Delta$ by fitting a three-dimensional model to
the data.  In short this consists of projecting the three-dimensional
model to predict the observables based on a given set of model
parameters, and then varying those model parameters to find the
best-fit model and associated parameter uncertainties:  $M_\Delta$ is
then estimated by marginalizing over the other parameters.  The
tangential distortion profile (\ref{eq:1d_gt}) is one-dimensional, and
expressed as a function of cluster-centric radius.  If we simply
assume a spherically symmetric mass distribution that corresponds to a
circularly symmetric mass distribution on the sky after projection,
the model distortion profile can be expressed, in the absence of
noise, as
\begin{equation}
g_+(\theta)=\frac{\gamma(\theta)}{1-\kappa(\theta)}, 
\label{eq:model_gt}
\end{equation}
where $\kappa(\theta)$ and $\gamma(\theta)$ are the convergence and
shear profiles of the cluster (note that the shear has the tangential
component alone for a circularly symmetric lens). It should be also
noted that, exactly speaking, the equation above is valid for a single
source redshift, and needs to be modified when source galaxies have
redshift distribution (e.g.\ see \S~4.3.2 in Bartelmann \& Schneider
2000), but this effect is 
very small for low-redshift clusters
at $z\simeq0.2$, given the deep Subaru data.

We now discuss briefly the choice of parametric form for the cluster
mass models.  The NFW model is a theoretically well-motivated mass
model based on dark matter only numerical simulations.  NFW found that
the mass density profile of ``equilibrium'' CDM halos is well fitted
by the following analytic function over a wide range of halo masses:
\begin{equation}
\rho_{\rm NFW}(r)=\frac{\rho_s}{(r/r_s)(1+r/r_s)^2},
\label{eq:rho_nfw}
\end{equation}
where $\rho_s$ is the central density parameter and $r_s$ is the scale
radius to divide the two distinct regimes of asymptotic mass density
slopes $\rho\propto r^{-1}$ and $r^{-3}$.  The NFW profile is thus
specified by two parameters.  The enclosed mass within a sphere of
radius $r_\Delta$ can be obtained by integrating the NFW profile up to
$r_\Delta$:
\begin{equation}
M_{{\rm NFW}, \Delta}=\frac{4\pi \rho_s r_\Delta^3}{c_\Delta^3}\left[
\ln(1+c_\Delta)-\frac{c_\Delta}{1+c_\Delta}\right], 
\label{eq:M_nfw}
\end{equation}
where we have introduced the concentration parameter, the ratio of
$r_\Delta$ relative to the scale radius, $c_\Delta\equiv
r_\Delta/r_s$. By equating Eqs.~(\ref{eq:M_Delta}) and
(\ref{eq:M_nfw}), the NFW profile can be specified in terms of the two
parameters $M_\Delta$ and $c_\Delta$, instead of $\rho_s$ and $r_s$,
once cosmological parameters and the spherical top-hat over-density
$\Delta$ are specified. We will use this parametrization of the NFW
profile throughout the rest of this paper.

It is then straightforward to compute the lensing profiles,
$\kappa(\theta; M_\Delta, c_\Delta)$ and $\gamma(\theta; M_\Delta,
c_\Delta)$, given the NFW profile (Bartelmann 1996; Wright \& Brainerd
2000; Takada \& Jain 2003). Inserting these profiles into
Eq.~(\ref{eq:model_gt}) gives the NFW prediction for the tangential
distortion profile to be compared with the measurement. In doing this,
note that the lensing fields are dimension-less and given in the units
of the critical projected mass density defined as
\begin{equation}
\Sigma_{\rm cr}\equiv \frac{c^2}{4\pi G}
 D_{\rm l}^{-1}
\left\langle\frac{D_{\rm ls}}{D_{\rm s}}\right\rangle^{-1}, 
\end{equation}
where $D_{\rm l}$ is the angular diameter distance to a given cluster,
and the average distance ratio $\langle D_{\rm ls}/D_{\rm s}\rangle$ is
estimated for source galaxy samples in each cluster field as described
in \S~\ref{ssec:zs}.  

An alternative simpler model often used in the literature is a
singular isothermal sphere (SIS) model. This model is specified by one
parameter, the one-dimensional velocity dispersion $\sigma_v^2$, and
the density profile is given by:
\begin{equation}
\rho_{\rm SIS}(r)=\frac{\sigma_{v}^2}{2\pi G}\frac{1}{r^2}. 
\label{eq:sis}
\end{equation}
Integrating this profile over a spherical region of radius $r_\Delta$
gives the enclosed mass:
\begin{equation}
M_{{\rm SIS},\Delta}=\frac{2\sigma_v^2}{G}r_\Delta. 
\label{eq:M_sis}
\end{equation}
Again by equating Eqs.~(\ref{eq:M_Delta}) and (\ref{eq:M_sis}), the SIS
model is fully specified by either of $\sigma_v^2$ or the over-density
mass $M_\Delta$. The lensing fields, obtained by integrating the profile
above along the line-of-sight, are found to be
\begin{equation}
\kappa(\theta)=\gamma(\theta)=\frac{\theta_E}{2\theta},
\label{eq:sis_kappa}
\end{equation}
where $\theta_E$ is the Einstein radius defined as $\theta_E\equiv
4\pi (\sigma_v/c)^2D_{\rm ls}/D_{\rm s}$ (e.g.\ see Bartelmann \&
Schneider 2001 for further details).

We also consider a cored isothermal sphere (CIS) model that is
obtained by introducing a softening ``core'' into an SIS model in an
empirical manner. We use the CIS model given as
\begin{equation}
\kappa_{\rm CIS}(\theta)=\frac{\theta_E}{2(\theta+\theta_c)},
\label{eq:cis}
\end{equation}
where $\theta_E$ is not exactly same as that for the SIS model given by
Eq.~(\ref{eq:sis_kappa}), so should be considered as a model parameter,
and $\theta_c$ is the core radius parameter. Note that, for the limit
$\theta_c\rightarrow 0$, the CIS model becomes equivalent to the SIS
model. The CIS model above is given by two parameters, similarly to the
NFW model. 

By comparing the goodness-of-fit of each model to the measured
distortion profile for each cluster, we will discuss which of these
mass models are preferred for real clusters.

\subsection{Model-Independent Estimate of 2D Cluster Mass}
\label{sec:model-independ-mass}

It is also possible and very useful to derive a model-independent
estimate of cluster mass from weak-lensing data.  In the weak lensing
limit, the azimuthally averaged tangential distortion in each circular
annulus of radius $\theta$, $\langle g_+ \rangle(\theta)$ (see
Eq.~\ref{eq:1d_gt}), is related to the projected mass density (e.g.,
Bartelmann \& Schneider 2000) as
\begin{equation}
\langle g_+ \rangle(\theta) \simeq \langle \gamma_+\rangle(\theta) 
= \bar{\kappa}(<\theta)-\langle
 \kappa\rangle (\theta),
\label{eq:g-kappa} 
\end{equation}
where $\langle\cdots\rangle(\theta)$ denotes the azimuthally averaged
shear in the circular annulus, and $\bar{\kappa} $ is the mean
convergence within a circular aperture of radius $\theta$ defined as
$\bar{\kappa}(<\theta)\equiv (1/\pi\theta^2)\int_{|\bmf{\theta}'|\le
\theta}\!d^2\bmf{\theta}' \kappa(\bmf{\theta}')$. Note that the relation
(\ref{eq:g-kappa}) holds for an arbitrary mass distribution.

As implied by Eq.~(\ref{eq:g-kappa}), if the tangential distortion
profile $\langle g_+\rangle$ can be measured out to sufficiently large
radii from the cluster center, where the local convergence likely
drops down to $\kappa\approx0$, the measured tangential distortion at
a large radius gives a direct estimate on the two-dimensional mass
enclosed within the circular aperture: $\langle
\gamma_+\rangle(\theta)\simeq\bar\kappa(<\theta)=M_{\rm
2D}(<\theta)/\pi\theta^2$.  The large Suprime-Cam FoV is ideally
suited to such measurements because the single pointing observations
used in this study span cluster-centric radii of $\sim1-2r_{\rm vir}$.

In this paper we employ the so-called $\zeta_c$-statistics (slightly
modified version made in Clowe et al.\ 2000 from the original method
developed in Fahlman et al.\ 1994):
\begin{eqnarray}
\zeta_c(\theta_{ m},\theta_{ o1}, \theta_{ o2})&\equiv &
2\int_{\theta_{ m}}^{\theta_{ o1}}\!d\ln \theta~
\langle\gamma_+\rangle(\theta) \nonumber\\
&&+\frac{2}{1-\theta_{ o1}^2
/\theta_{ o2}^2}\int^{\theta_{ o2}}_{\theta_{ o1}}\!
d\ln\theta~ \langle\gamma_+\rangle(\theta)\nonumber\\
&=&\bar{\kappa}(<\theta_{ m})-\bar{\kappa}(\theta_{
 o1}\le\theta\le\theta_{o2}),
\label{eq:zeta}
\end{eqnarray}
where the radii $\theta_m$, $\theta_{o1}$ and $\theta_{o2}$ satisfy
$\theta_m<\theta_{o1}<\theta_{o2}$.  If the radius $\theta_{m}$ is
also taken to be sufficiently large so that the weak lensing limit
$g_+\approx\gamma_+$ holds, the quantity $\zeta_c$ can be directly
estimated from the measured tangential distortion profile, although
the discrete summation for the radial binned profile, instead of the
radial integration, needs to be employed.  The radius $\theta_{m}$ is
the {\em target} radius which encloses the projected mass we aim to
measure (see below).  On the other hand, the two outermost radii
$\theta_{o1}$ and $\theta_{o2}$ are taken to be sufficiently far from
the cluster center and are also, as suggested in Clowe et al.\ (2004),
chosen so that any prominent substructures in the annulus of
$\theta_{o1}\le\theta\le\theta_{o2}$, regardless of being associated
with the cluster or not, are absent in the reconstructed mass
map. Once these outer radii are set, we can safely consider
$\bar{\kappa}(\theta_{o1}\le\theta\le\theta_{o2})\sim0$ to be valid
in the second equality on the r.h.s. of Eq.~(\ref{eq:zeta}), and
therefore the enclosed mass can be estimated as
\begin{equation}
M_{\rm 2D}(<\theta_m) \simeq
\pi\theta_{m}^2\Sigma_{\rm cr}\zeta_c(\theta_m,\theta_{o1},\theta_{o2}).
\label{eqn:m_2D}
\end{equation}
More precisely, the estimated mass above, $M_{\rm 2D}(<\theta_m)$, gives
a lower limit on the true mass because there may be a non-vanishing mass
contribution from the annulus region of $\theta_{o1}\le\theta\le
\theta_{o2}$ as well as a constant mass-sheet contribution that does not
change the measured distortion signals at all.

The uncertainty in $\zeta_c$ is estimated as
\begin{eqnarray}
\sigma^2(\zeta_c)&=&4\sum_{i=N_{m}}^{N_{o1}}\left(
\frac{\Delta\theta_i}{\theta_i} \right)^2\sigma^2_{g_+}(\theta_i)
\nonumber\\
&&\hspace{0em}+\left(\frac{2}{1-\theta_{o1}^2/\theta^2_{o2}}\right)^2
\sum_{i=N_{o1}}^{N_{o2}}\left(\frac{\Delta\theta_i}{\theta_i}
\right)^2\sigma^2_{g_+}(\theta_i),
\end{eqnarray}
where we have again assumed that the lensing measurement uncertainty is
dominated by the intrinsic ellipticity noise, and $N_{m}$, $N_{o1}$ and
$N_{o2}$ are the indices of the discretized radial bins corresponding to
the radii, $\theta_m$, $\theta_{o1}$ and $\theta_{o2}$ in
Eq.~(\ref{eq:zeta}), respectively.

The weak lensing measurements thus offer a unique and powerful method
to estimate the projected mass of a cluster in a {\em
model-independent way}.

\section{RESULTS}
\label{sec:results}

\begin{table*}
 \caption{Summary of Cluster Subsamples}
\begin{center}
\begin{tabular}{l|l|l|l}\hline\hline
Name & $\#$ of clusters & Comments & Main Results \\ \hline
All & 30 clusters & As given in Table~\ref{tab:cog} & App.\ref{sec:massmap}
	     for the
mass map and shear
	     profile\\
Complex& 4 clusters & Complex mass maps & ZwCl0823, A689, A750 and A2345
	     (App.~\ref{sec:massmap})\\ 
Two filters & 25 clusters & Color used to correct for dilution & 
--
\\

&&(include ZwCl0823, A689 and A750)
\\
Shear profile & 22 clusters & Compared with spherical mass models
& Figs.~\ref{fig:Msis_vs_Mnfw}, \ref{fig:dM_to_M}, 
\ref{fig:mass_vs_2Dmass}, \ref{fig:compare_c+M} and \ref{fig:Mall_to_Mredblue}\\ 
NFW & 19 clusters & Well fitted by NFW model
& Figs.~\ref{fig:C-M}, \ref{fig:Cvir_hist} and \ref{fig:stack}\\ 
\hline  
\end{tabular}
\label{tab:subsample}
\end{center}
\textrm{NOTES $\singlebond$
Column (1): name of cluster subsamples studied in this paper; 
Column (2): the number of clusters contained in each subsample; 
Column (3): comments used to define each subsample;
Column (4): the figures and tables showing the main results derived from
 each subsample} 
\end{table*}

This section presents our main results, i.e.\ constraints on cluster
masses and density profile shapes based on the weak-lensing
measurements.

%MT
In the following we will often 
show the results using different subsamples of
clusters each of which is defined according to the available
information on color and lensing properties. Table~\ref{tab:subsample}
gives a brief  summary of the subsamples.

\subsection{Tangential Distortion Profiles}

All of our results are based on the tangential distortion profile of
galaxy images for each cluster, which are shown in
Appendix~\ref{sec:massmap} for all 30 clusters.  Our $X$-ray luminous
clusters at low redshifts ($z\simeq0.2$) typically show the lensing
distortion strength of $O(0.1)$ on small angular scales $\sim1'$. On
these small scales, the nonlinear correction to the lensing shear,
$g_+=\gamma_+/(1-\kappa)$ (see Eq.~[\ref{eq:model_gt}]), must be
included in the model fitting.  The distortion signals decrease 
down to a few per cent on large scales $\sim10'$ (a few
Mpc scales).  Impressively, a $1\%$ shear signal is detected at
$\simgt 2\sigma$ significance in most of our clusters (A209, A267,
A291, A383, A586, ZwCl0740, ZwCl0823, ZwCl0839 A611, A697, A750, A963,
A1835, A2219, A2261, A2345, RXJ2129, A2390 and A2631), thus
highlighting the unique capability of Subaru/Suprime-Cam data for
accurate weak lensing measurements thanks to its excellent image
quality and depth (Broadhurst et al.\ 2005; and also see 
Kneib et al. 2003 for the space-based lensing observation).  
Given the trade-off between
the radial dependence of the number  of background galaxies and
the distortion strengths of clusters at $z\simeq0.2$, the distortion
signals are most accurately measured around radii of $\sim5'$.

The figures in Appendix~\ref{sec:massmap} also show the radial profile
of the $g_\times$ distortion component for each cluster, providing a
monitor of the systematic errors inherent in the lensing measurements,
as described below Eq.~(\ref{eq:1d_gt}).  The $g_\times$ profiles are
consistent with a null signal in most of radial bins, confirming the
reliability of our lensing measurements.

\subsection{Two-dimensional Mass Reconstruction}

To understand the broad-brush features of the cluster mass
distributions, and thus understand the distortion profiles better, we use
the Kaiser \& Squires (1993) algorithm to reconstruct the projected
mass distribution in each cluster field, as shown in
Appendix~\ref{sec:massmap}.

Some clusters have depressions in the tangential distortion profiles
spanning a few radial bins.  When compared with the corresponding mass
map, it becomes apparent that these depressions correspond to
prominent structures in the annulus of the same radius -- e.g., A115
has a big depression in the distortion signals at $2-3'$; the mass map
contains three structures at this distance from the cluster center.
The mass maps therefore provide a useful cross-check on the distortion
profiles.  However in this paper we concentrate on a simple
one-dimensional (i.e.\ tangential distortion profile) analysis as the
first step in a series of papers on this sample.  Future papers will
employ more sophisticated modeling schemes, including substructures
and halo triaxiality in order to model more precisely the full two-
and three-dimensional structure of the cluster mass distributions
(e.g. Oguri et al. 2010).

\subsection{Parametrized Distortion Profile Models}

\begin{table*}
  \caption{Best-fit Mass Profile Parameters for SIS, CIS and NFW Models}
\label{tab:massprofile}
\begin{center}
\begin{tabular}{ccccccccc}
\hline
\hline\\
 Cluster          &  SIS
                  &  
                  &  CIS
                  &  
                  &  
                  &  NFW
                  &  
                  &  \\
                  &  $\sigma_{\rm SIS}$
                  &  $\chi^2_{\nu}$(d.o.f)
                  &  $\theta_E$
                  &  $\theta_C$
                  &  $\chi^2_{\nu}$(d.o.f)
                  &  $M_{\rm vir}$
                  &  $c_{\rm vir}$
                  &  $\chi^2_{\nu}$(d.o.f) \\
                  &  $({\rm km~ s^{-1}})$ 
                  &  
                  &  (arcmin)
                  &  (arcmin)
                  & 
                  &  $(10^{14}h^{-1}M_\odot)$ 
                  &
                  & \\                 
(1)               &  (2)
                  &  (3)
                  &  (4) 
                  &  (5)
                  &  (6)
                  &  (7)
                  &  (8)
                  &  (9) \\
\hline %%%%%%%%RED+BLUE SAMPLE
A68                &  $869.03^{+70.82}_{-75.14}$
                  &  $0.15(11)$
                  &  $0.30^{+0.13}_{-0.09}$
                  &  $0.11^{+0.37}_{-0.11}$
                  &  $0.13(10)$
                  &  $5.49^{+2.56}_{-1.81}$
                  &  $4.02^{+3.36}_{-1.82}$ 
                  &  $0.14(10)$ \\ 
A115                &  $818.02^{+86.85}_{-86.08}$
                  &  $0.66(12)$
                  &  $0.27^{+0.19}_{-0.08}$
                  &  $0.06^{+0.58}_{-0.06}$
                  &  $0.71(11)$
                  &  $5.36^{+4.08}_{-2.45}$
                  &  $3.69^{+5.03}_{-2.04}$ 
                  &  $0.75(11)$ \\ 
$[$ZwCl0104$]$                &  $665.85^{+42.71}_{-57.95}$
                  &  $1.29(12)$
                  &  $0.14^{+0.03}_{-0.14}$
                  &  $0.00^{+0.06}_{-0.00}$
                  &  $1.41(11)$
                  &  $1.73^{+0.58}_{-0.47}$
                  &  $8.08^{+8.20}_{-3.43}$ 
                  &  $1.35(11)$ \\ 
A209                &  $918.76^{+34.06}_{-40.37}$
                  &  $3.36(12)$
                  &  $0.70^{+0.13}_{-0.11}$
                  &  $0.65^{+0.30}_{-0.21}$
                  &  $0.89(11)$
                  &  $14.00^{+3.31}_{-2.60}$
                  &  $2.71^{+0.69}_{-0.60}$ 
                  &  $0.84(11)$ \\ 
RXJ0142              &  $886.80^{+43.55}_{-46.56}$
                  &  $0.56(12)$
                  &  $0.27^{+0.06}_{-0.27}$
                  &  $0.03^{+0.08}_{-0.03}$
                  &  $0.56(11)$
                  &  $4.49^{+1.23}_{-1.01}$
                  &  $7.12^{+2.71}_{-1.89}$ 
                  &  $0.49(11)$ \\ 
A267                &  $778.05^{+45.65}_{-37.28}$
                  &  $0.63(12)$
                  &  $0.26^{+0.05}_{-0.05}$
                  &  $0.07^{+0.09}_{-0.06}$
                  &  $0.54(11)$
                  &  $3.85^{+1.08}_{-0.88}$
                  &  $6.00^{+2.11}_{-1.58}$ 
                  &  $0.58(11)$ \\ 
A291                &  $801.74^{+53.89}_{-51.28}$
                  &  $1.17(12)$
                  &  $0.42^{+0.17}_{-0.11}$
                  &  $0.50^{+0.59}_{-0.33}$
                  &  $0.86(11)$
                  &  $7.02^{+3.10}_{-2.06}$
                  &  $2.36^{+1.34}_{-0.94}$ 
                  &  $0.87(11)$ \\ 
A383                &  $875.19^{+34.37}_{-41.47}$
                  &  $1.95(12)$
                  &  $0.27^{+0.03}_{-0.27}$
                  &  $<0.04$
                  &  $2.13(11)$
                  &  $3.62^{+1.15}_{-0.86}$
                  &  $8.87^{+5.22}_{-3.05}$ 
                  &  $2.78(11)$ \\ 
A521                &  $789.23^{+43.63}_{-43.87}$
                  &  $1.89(12)$
                  &  $0.33^{+0.08}_{-0.07}$
                  &  $0.28^{+0.23}_{-0.15}$
                  &  $1.50(11)$
                  &  $5.85^{+1.45}_{-1.22}$
                  &  $3.06^{+1.01}_{-0.79}$ 
                  &  $1.29(11)$ \\ 
A586                &  $1035.32^{+40.04}_{-67.58}$
                  &  $0.90(11)$
                  &  $0.46^{+0.11}_{-0.46}$
                  &  $0.07^{+0.12}_{-0.07}$
                  &  $0.87(10)$
                  &  $7.37^{+2.89}_{-2.08}$
                  &  $8.38^{+3.52}_{-2.52}$ 
                  &  $1.08(10)$ \\ 
ZwCl0740               &  $726.93^{+66.62}_{-58.09}$
                  &  $0.80(12)$
                  &  $0.47^{+0.29}_{-0.17}$
                  &  $0.94^{+1.34}_{-0.67}$
                  &  $0.54(11)$
                  &  $5.89^{+5.48}_{-2.39}$
                  &  $2.85^{+2.03}_{-1.37}$ 
                  &  $0.53(11)$ \\ 
$[$ZwCl0839$]$                &  $766.05^{+57.14}_{-47.89}$
                  &  $0.40(10)$
                  &  $0.20^{+0.06}_{-0.02}$
                  &  $<0.12$
                  &  $0.45(9)$
                  &  $2.91^{+1.08}_{-0.82}$
                  &  $7.24^{+5.04}_{-2.72}$ 
                  &  $0.49(9)$ \\ 
A611                &  $929.34^{+57.70}_{-45.26}$
                  &  $1.45(12)$
                  &  $0.33^{+0.08}_{-0.07}$
                  &  $0.11^{+0.19}_{-0.11}$
                  &  $1.47(11)$
                  &  $6.65^{+1.75}_{-1.42}$
                  &  $4.23^{+1.77}_{-1.23}$ 
                  &  $1.37(11)$ \\ 
A697                &  $1021.91^{+41.13}_{-45.14}$
                  &  $2.07(12)$
                  &  $0.56^{+0.11}_{-0.09}$
                  &  $0.38^{+0.23}_{-0.16}$
                  &  $1.13(11)$
                  &  $12.36^{+2.68}_{-2.21}$
                  &  $2.97^{+0.85}_{-0.69}$ 
                  &  $1.04(11)$ \\ 
$[$A963$]$                &  $816.53^{+37.85}_{-42.83}$
                  &  $2.25(13)$
                  &  $0.40^{+0.10}_{-0.08}$
                  &  $0.46^{+0.35}_{-0.23}$
                  &  $1.72(12)$
                  &  $6.96^{+2.17}_{-1.59}$
                  &  $2.57^{+1.00}_{-0.79}$ 
                  &  $1.76(12)$ \\ 
A1835                &  $1050.55^{+56.49}_{-41.65}$
                  &  $1.65(11)$
                  &  $0.61^{+0.14}_{-0.11}$
                  &  $0.46^{+0.27}_{-0.20}$
                  &  $0.71(10)$
                  &  $13.69^{+3.65}_{-2.86}$
                  &  $3.35^{+0.99}_{-0.79}$ 
                  &  $0.56(10)$ \\ 
ZwCl1454             &  $702.37^{+69.49}_{-67.89}$
                  &  $0.91(11)$
                  &  $0.20^{+0.10}_{-0.20}$
                  &  $0.09^{+0.28}_{-0.09}$
                  &  $0.94(10)$
                  &  $3.45^{+2.02}_{-1.36}$
                  &  $4.01^{+3.44}_{-1.96}$ 
                  &  $0.99(10)$ \\ 
$[$A2009$]$                &  $800.80^{+40.11}_{-49.15}$
                  &  $1.20(12)$
                  &  $0.31^{+0.08}_{-0.07}$
                  &  $0.13^{+0.18}_{-0.11}$
                  &  $1.14(11)$
                  &  $3.86^{+1.20}_{-0.93}$
                  &  $6.59^{+2.40}_{-1.71}$ 
                  &  $0.89(11)$ \\ 
ZwCl1459.4                &  $864.90^{+53.43}_{-71.78}$
                  &  $1.21(12)$
                  &  $0.28^{+0.08}_{-0.28}$
                  &  $0.04^{+0.12}_{-0.04}$
                  &  $1.28(11)$
                  &  $4.40^{+1.50}_{-1.20}$
                  &  $6.55^{+3.34}_{-2.18}$
                  &  $1.17(11)$ \\
RXJ1720               &  $879.13^{+61.09}_{-54.04}$
                  &  $0.52(12)$
                  &  $0.28^{+0.09}_{-0.03}$
                  &  $<0.14$
                  &  $0.57(11)$
                  &  $4.07^{+1.65}_{-1.22}$
                  &  $8.73^{+5.60}_{-3.08}$ 
                  &  $0.57(11)$ \\ 
A2219                &  $1132.87^{+43.65}_{-58.12}$
                  &  $1.73(12)$
                  &  $0.47^{+0.08}_{-0.47}$
                  &  $<0.07$
                  &  $1.89(11)$
                  &  $9.11^{+2.54}_{-2.06}$
                  &  $6.88^{+3.42}_{-2.16}$ 
                  &  $2.26(11)$ \\ 
A2261                &  $1078.32^{+54.66}_{-29.73}$
                  &  $0.77(12)$
                  &  $0.50^{+0.09}_{-0.08}$
                  &  $0.08^{+0.11}_{-0.07}$
                  &  $0.68(11)$
                  &  $9.49^{+2.01}_{-1.69}$
                  &  $6.04^{+1.71}_{-1.31}$ 
                  &  $0.67(11)$ \\ 
RXJ2129               &  $879.92^{+62.12}_{-52.16}$
                  &  $0.52(12)$
                  &  $0.33^{+0.12}_{-0.09}$
                  &  $0.15^{+0.33}_{-0.15}$
                  &  $0.48(11)$
                  &  $6.71^{+2.73}_{-1.96}$
                  &  $3.32^{+2.16}_{-1.34}$ 
                  &  $0.56(11)$ \\ 
A2390                &  $951.38^{+55.34}_{-31.23}$
                  &  $1.05(12)$
                  &  $0.49^{+0.08}_{-0.07}$
                  &  $0.13^{+0.09}_{-0.07}$
                  &  $0.53(11)$
                  &  $8.20^{+1.93}_{-1.63}$
                  &  $6.20^{+1.53}_{-1.28}$ 
                  &  $0.61(11)$ \\ 
A2485                &  $777.49^{+60.07}_{-64.03}$
                  &  $0.82(12)$
                  &  $0.28^{+0.11}_{-0.08}$
                  &  $0.17^{+0.34}_{-0.16}$
                  &  $0.78(11)$
                  &  $4.56^{+1.84}_{-1.38}$
                  &  $3.52^{+2.24}_{-1.44}$ 
                  &  $0.77(11)$ \\ 
A2631                &  $959.71^{+50.90}_{-35.71}$
                  &  $0.86(12)$
                  &  $0.30^{+0.04}_{-0.30}$
                  &  $<0.04$
                  &  $0.93(11)$
                  &  $5.24^{+1.15}_{-0.98}$
                  &  $7.84^{+3.54}_{-2.28}$ 
                  &  $1.09(11)$ \\ 
\hline
\end{tabular}
\end{center}
\textrm{ NOTES $\singlebond$ Column (1): cluster name (clusters with
name in brackets have only one filter data available, and the rest
has two filter data); Column (2): best-fit velocity dispersion for SIS
model (Eq.~[\ref{eq:sis}]);
Column (3): reduced $\chi^2$ for the best-fit SIS model, and the
 degrees-of-freedom in parenthesis; Column (4): the Einstein radius
 parameter for the CIS model (see Eq.~[\ref{eq:cis}]); Column (5): The core
 radius; Column (6): the reduced $\chi^2$; 
 Column (7): best-fit viral mass for the NFW model (Eq.[\ref{eq:M_nfw}]); 
 Column (8): the best-fit NFW concentration parameter; 
 Column (9): the reduced $\chi^2$.  }\\
\end{table*}
First we use the tangential distortion profile of each cluster to
constrain the spherical mass profile models discussed in
\S~\ref{sec:model}: NFW, SIS and CIS models.
Table~\ref{tab:massprofile} summarizes the best-fit parameters of
each model.  The clusters in parentheses have been observed through
just one filter, and the results are thus likely to be adversely
affected by the dilution effect discussed in \S\ref{sec:sample}.  Note
that the results are not shown for 4 clusters (ZwCl0823, A689, A750 and
A2345) because the complex mass distribution revealed by the mass maps
in Appendix~\ref{sec:massmap} suggest strongly that a spherically
symmetric model is wholly inappropriate for these systems. The 4
clusters are ``complex cluster subsample'' in Table~\ref{tab:subsample}.

We quantify the goodness-of-fit of each model by using the
significance probability $Q(\nu/2,\chi^2/2)$ that the data gives as
poor fit as the observed value of $\chi^2$ by chance (see \S~15.2 in
Press et al.\ 1992).   Specifically, $Q$ values greater than 0.1
  indicate a satisfactory agreement between the data and the model; if
  $Q\simgt0.001$, the fit may be acceptable, e.g.\ the measurement
  errors may be moderately underestimated; if $Q\simlt0.001$, the
  model may be called into question.  Note that the $Q$ value can be
computed from the chi-square value and the degrees-of-freedom given in
Table~\ref{tab:massprofile}.  For simplicity we adopt the threshold
$Q_{\rm th}=0.1$ as the dividing line between acceptable ($Q>Q_{\rm
  th}$) and unacceptable ($Q<Q_{\rm th}$) fits to the measured
profile.  By this criterion three clusters (A383, A2219 and A963) are
not well fitted by any of the three models.  Of the remaining 23
clusters, four (A209, A521, A697 and A1835) are not well fitted by an
SIS model having $Q\simeq6\times10^{-5}, 0.03, 0.02$, and $ 0.08$,
respectively, while either of CIS or NFW models gives an acceptable
fit.  In fact, as shown in Appendix~\ref{sec:massmap}, these 4
clusters display a clear radial curvature in the measured distortion
profile, which cannot be fitted by a 
%single power law of SIS.  
the unbroken power law of SIS. 
The
remaining 19 clusters are all well-fitted by the three models.
% MT
Note that we checked that, even if we include the `poor-fit' 3 clusters
in the analysis, the following results shown below are not largely
changed.  To be more precise, for example, the best-fit virial masses
obtained from the $\chi^2$-fitting of an NFW model to the stacked shear
profile (see below for the details) are changed by less than $5\%$.

With the exception of the 4 clusters noted above, we cannot discriminate
statistically between the three mass models -- i.e.\ we cannot make
statistically robust choice as to which model is a better description of
the observational data.  This is partly because the statistical
precision of the lensing measurements is insufficient, and partly
because the radial range of the data is not wide enough to discriminate
characteristic radial curvatures of CIS or NFW models from a single
power law of SIS model.  Specifically, the weak-lensing information 
%on
at radii smaller than a few arcminutes is limited by the smaller number
densities of background galaxies due to the small solid angle subtended
by annuli at these radii.  There are several ways to overcome this
limitation: (i) the statistical precision, especially on small scales,
can be boosted by stacking the  distortion profiles over cluster samples
(see below), and (ii) the weak-lensing information presented here can be
combined with strong-lensing constraints on small scales, allowing the
the cluster-by-cluster 
mass distribution to be measured to high precision over a wider
range of radii (e.g., Kneib et al.\ 2003; Broadhurst
et al.\ 2005).  Strong-lensing constraints are available for most of the
clusters from \emph{Hubble Space Telescope} observations (GO:11312; PI:
G.~P.~Smith) plus ground-based spectroscopic follow-up 
(Richard et al. 2009).
The improved constraints on the mass profile
parameters for the joint fitting to the strong and weak lensing
information will be presented elsewhere (Smith et al.\ in preparation).

\begin{figure}
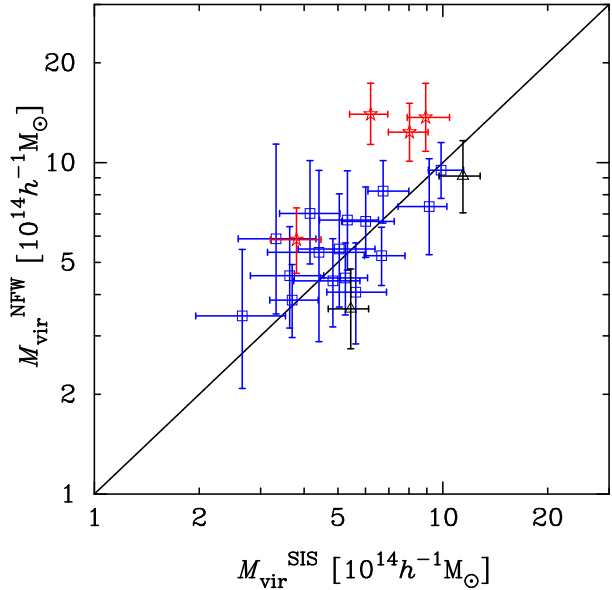

\begin{center}
%\FigureFile(80mm,80mm){compare_Msis_vs_Mnfw_photoz.ps}
\FigureFile(80mm,80mm){f4.ps}
\end{center}
\caption{Comparison of the virial mass estimates derived from the
 fitting of SIS and NFW models to the tangential distortion profile
 measured for each of the 22 clusters that have color information of
 galaxies (to define the red+blue galaxy sample). The clusters are
 classified into 3 different groups based on the results of
 Table~\ref{tab:massprofile}: the triangle symbols with error bars show
 the clusters for which any of SIS, CIS and NFW models does not give an
 acceptable fit (A383 and A2219); the star symbols show clusters for
 which an SIS model is disfavored compared with CIS and NFW models
 (A209, A521, A697 and A1835); the square symbols denote the other
 clusters for which all the three modes give an acceptable fit. While
 the star symbols show a significant smaller mass from SIS than that
 from NFW, an agreements within $1\sigma$ error bars can be found for
 other clusters, but the scatter around the relation $M_{\rm vir}^{\rm
 SIS}$-$M_{\rm vir}^{\rm NFW}$ is rather substantial.}
 \label{fig:Msis_vs_Mnfw}
\end{figure}
We now turn to constraints on the virial mass of each cluster, $M_{\rm
vir}$; from a theoretical perspective this is the most useful cluster
mass measurement.  In Table~\ref{tab:massprofile} we list the best-fit
virial mass and the $1\sigma$ statistical uncertainties obtained from
the NFW model fits.  The marginalized error on one parameter, obtained
by projecting the confidence region in a higher dimensional parameter
space onto one particular parameter axis, can be obtained by 
%monitoring
measuring
the range of the parameter that satisfies $\Delta \chi^2\le 1$ 
%with
while
varying other parameter(s) (e.g. see Section 15.3 in Numerical Recipes
in Press et al. 1992). 

In Fig.~\ref{fig:Msis_vs_Mnfw} we compare the
virial masses derived from the NFW models with the masses derived from
the SIS models, where the latter are estimated by inserting the
best-fit $\sigma_{\rm SIS}$ values from Table~\ref{tab:massprofile}
into Eq.~(\ref{eq:M_sis}).  Note that here we consider only clusters
with data available in two filters, which corresponds to the subsample
named ``shear profile'' in Table~\ref{tab:subsample} 
consisting of 22 clusters (25 clusters with two filter data minus 3
clusters showing the complex mass distribution).  
The two mass estimates agree
within the uncertainties for 13 out of 22 clusters.  Adopting a fixed
slope of unity, the relationship between the two model-dependent mass
measurements is found to be
 $M^{\rm NFW}_{\rm vir}/M_{\rm vir}^{\rm SIS}=1.20\pm0.25$,
 where the  quoted uncertainty is the scatter around the mean, and
   is dominated by the measurement errors.  Nevertheless, SIS mass is
 systematically smaller than the NFW mass by $\sim 20\%$, implying
 that model choice does influence the mass measurement.  We defer
 consideration of the origin of the difference between the SIS and NFW
 mass estimates to \S\ref{sec:stack} in which we study stacked
 distortion profiles.

\begin{figure}
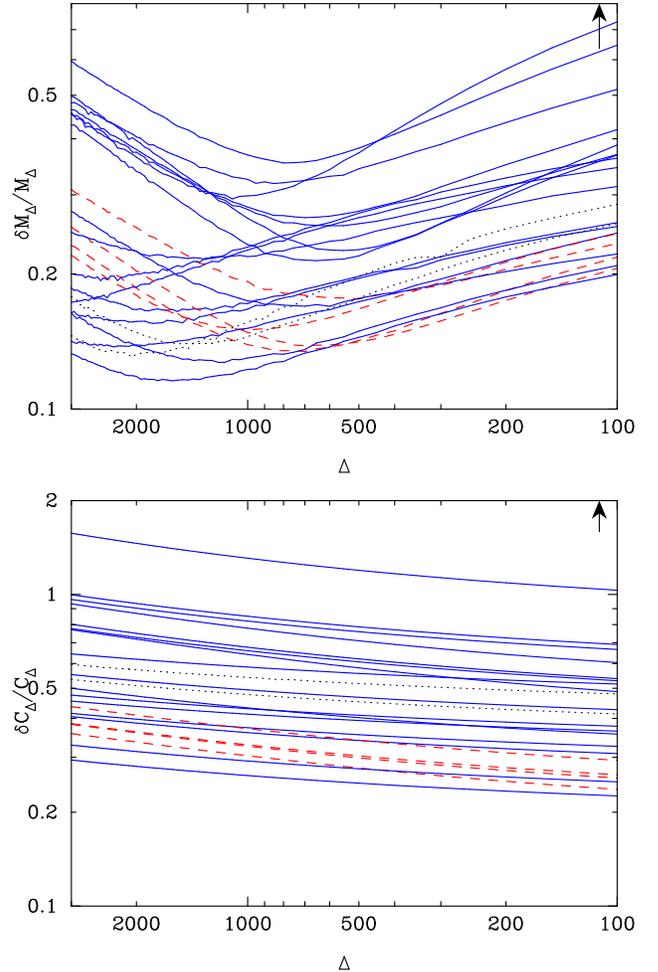

\begin{center}
%\FigureFile(83mm,83mm){dM_to_M_vs_delta_for_eachcluster_photoz.ps}
%\FigureFile(83mm,83mm){dC_to_C_vs_delta_for_eachcluster_photoz.ps}
\FigureFile(83mm,83mm){f5a.ps}
\FigureFile(83mm,83mm){f5b.ps}
\end{center}
\caption{{\em Upper panel}: Relative accuracies of the cluster mass
estimations, from the NFW model fitting, as a function of the average
over-density assumed, $\Delta$, by which the enclosed mass $M_\Delta $ is
defined based on Eq.~(\ref{eq:M_nfw}).  The solid, dashed and dotted
curves show the clusters that are marked with the square, star and
triangle symbols in Figure~\ref{fig:Msis_vs_Mnfw}, respectively. For
most clusters, the cluster mass can be estimated at a best precision
when assuming $\Delta\simeq 500$ -- $2000$. The arrow denotes the virial
over-density at $z\simeq 0.2$: $\Delta_{\rm vir}\simeq 110$. {\em Lower
panel}: The similar plot, but for the concentration parameter.  }
\label{fig:dM_to_M}
\end{figure}

The fractional error on virial masses in Table~\ref{tab:massprofile}
is typically $20-30\%$.  The precision to which cluster masses can be
measured is central to attempts to measure intrinsic scatter in
cosmological scaling relations.  We therefore explore whether
alternative definitions of cluster mass yield similar or, hopefully,
greater precision.

Despite its theoretical appeal, the virial mass is neither a unique
nor necessarily the most observationally appealing choice of cluster
mass measurement.  There are many alternative cluster mass
definitions, the use of which depends to a large extent on the nature
of observational data available (strong-lensing, weak-lensing,
$X$-ray, SZ) to constrain the cluster mass.  In Fig.~\ref{fig:dM_to_M}
we plot the variation of the fractional error on cluster mass and
concentration with the over-density $\Delta$ at which the parameters
are defined.  More precisely, for each $\Delta$, we first express
the NFW model in terms of the two parameters $(c_\Delta, M_\Delta)$,
instead of their virial counterparts, and then estimate the best-fit
parameters and statistical uncertainties from the model fitting. While
the best-fit NFW model is unchanged for any
$\Delta$, given the measured
distortion profile, the statistical uncertainties in the parameters
$M_{\Delta}$ and $c_{\Delta} $ change because the variations in NFW
profile are given with respect to 
%the pivot radius 
$r_\Delta$
corresponding to the enclosed over-density $\Delta$.

The upper panel of Figure~\ref{fig:dM_to_M} indeed shows that the
accuracies of cluster mass determination do vary with $\Delta$.
Interestingly, the optimal over-density is $\Delta\simeq500-2000$ for
majority of our clusters.  This result can be understood as follows.
These clusters are found to be well fitted by an NFW model with
concentration $c_{\rm vir}\simlt 5$, which roughly matches the
$\Lambda$CDM simulation predictions for cluster-scale halos (e.g.\
Bullock et al.\ 2001; Dolag et al. 2004; Neto et al. 2007).  
Given the cluster redshifts ($z\simeq0.2$) and
the number densities of background galaxies available from Subaru, the
weak lensing signals have a maximum signal-to-noise ratio over a range
of radii corresponding to the over-density $\Delta\simeq500-1000$. 

The lower panel shows the results for the concentration parameter. The
concentration parameter is not as tightly constrained as mass, with
the fractional error in excess of $20\%$ in every case, at all
$\Delta$.  The precision does increase slowly with decreasing $\Delta$
or increasing the pivot radius $r_\Delta$.  This reflects the fact
that the larger pivot radius $r_\Delta$ gives greater leverage when
measuring the curvature of an NFW profile with respect to the scale
radius, $r_s(\equiv r_\Delta/C_\Delta)$, yielding a superior precision
on the concentration parameter for smaller $\Delta$.

\subsection{The $M_{\rm vir}$--$c_{\rm vir}$ Relation}

Numerical simulations based on the CDM model have revealed that the
two parameters of NFW halos, e.g.\ $c_{\rm vir}$ and $M_{\rm vir}$,
are correlated, i.e.\ halo concentration is a weakly decreasing
function of mass (e.g.\ Bullock et al.\ 2001).  Such a correlation is
expected to naturally arise from the nature of hierarchical
clustering.  According to the CDM structure formation scenario, less
massive halos first form and then more massive halos form as a result
of mergers of smaller halos and/or mass accretion onto halos.  Hence,
since the progenitors of more massive halos should have formed at
lower redshifts at which the mean background mass density is lower,
more massive halos at a given observing redshift tend to possess a
less centrally concentrated profile, given the fact that the mean
over-density within the virial radius is fixed for all halos.  Thus the
properties of halo profile contain rich information on cosmological
models as well as mass assembly history of halos.

\begin{figure}
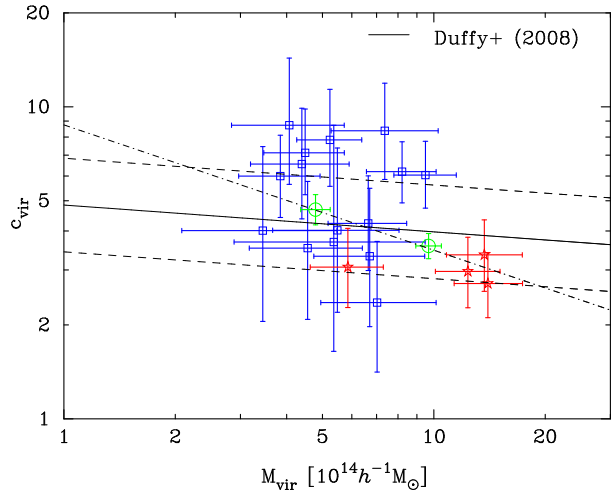

\begin{center}
\FigureFile(80mm,80mm){f6.ps}
%\FigureFile(80mm,80mm){redblue_photoz_Cvir_vs_Mvir_withBestfit.ps}
\end{center}
\caption{The observed distribution of the concentration parameters
 $c_{\rm vir}$ as a function of the cluster masses $M_{\rm vir}$, for 19
 clusters that are better fitted by NFW than SIS models.  The solid line
 indicates the median relation found from the CDM simulations for the
 WMAP5 cosmological model, while the region enclosed within the dashed
 lines correspond to the range of $\sigma(\log_{10}c_{\rm vir})=0.1$
 within which most of simulated clusters are distributed (Duffy et
 al.\ 2008).  The dotted-dashed line denotes the best-fit model of
 $c_{\rm vir}=c_N(M_{\rm vir}/10^{14}h^{-1}M_\odot)^{-\alpha}$ to the
 cluster distribution that is given by 
 $c_{N}=8.55$ and $\alpha=0.40$. The two circle symbols are the
 results for the stacked lensing signals obtained from the low- and
 high-mass samples that are divided with mass threshold, $M_{\rm vir,
   thresh}=6\times 10^{14}h^{-1}M_\odot$ (see \S~\ref{sec:stack} for
 the details). 
The star and square symbols are as in Fig.~\ref{fig:Msis_vs_Mnfw}.}
 \label{fig:C-M}
\end{figure}
We can therefore use our large cluster sample to explore whether such
a correlation between $c_{\rm vir}$ and $M_{\rm vir}$ is present in
real  clusters, concentrating on 19 clusters (the ``NFW'' subsample in
Table~\ref{tab:subsample})
-- i.e.\ we exclude 3
  clusters from the 22 with 2-filter data: A383 and A2219 are not well
  fit by an NFW profile, and ZwCl0740 because its redshift is
  estimated photometrically.  Figure~\ref{fig:C-M} shows how these 19
  clusters are distributed in the ($c_{\rm vir}, M_{\rm vir}$) plane.
Interestingly visual inspection of the data suggests a marginal trend
that the measured $c_{\rm vir}$ becomes smaller for more massive
halos, although the scatter is large.  It is also interesting to note
that none of our morphologically unbiased $X$-ray selected sample of
clusters, including those with the highest masses ($\simgt
10^{15}M_\odot$), show very high concentrations of $c_{\rm vir}\simgt
10$ as had been found for some strong lensing clusters (e.g., Gavazzi
et al.\ 2003; Kneib et al.\ 2003; Broadhurst et al.\ 2005, 2008).

\begin{table*}
\caption{Best-fit parameters for the mass-concentration relation of
$c(M)=c_N(M/10^{14}h^{-1}M_\odot)^{-\alpha}$ }
\begin{center}
\begin{tabular}{l|ccc}
\hline
\hline
                & $c_N$ 
                & $\alpha$
                & $\sigma(\log_{10}c)$ \\
\hline
$c_{\rm vir}(M_{\rm vir})$
                    & $8.75_{-2.89}^{+4.13}$
                    & $0.40\pm0.19$
                    & $0.17$\\
Duffy+08: $c_{\rm vir}(M_{\rm vir})$ & 4.96 & 0.086 & $\sim 0.15$\\ 
Buote+07: $c_{\rm vir}(M_{\rm vir})$ & $7.5\pm 0.33$ & $0.172\pm 0.026 $
& $\sim 0.1$\\ 
\hline
$c_{\rm 200}(M_{200})$
                    & $5.75_{-1.90}^{+2.47}$
                    & $0.37_{-0.21}^{+0.20}$
                    & $0.18$\\
Duffy+08: $c_{200}(M_{200})$&3.71&0.089&$\sim 0.15$ \\ 
\hline\hline
\end{tabular}
\end{center}
\textrm{NOTES: The row labeled as ``Duffy+08'' shows the results
 obtained from numerical simulations for the WMAP 5-year cosmological
 model in Duffy et al.\ (2008), corrected for clusters at $z=0.24$, the
 mean redshift of our sample clusters. The
 row labeled as ``Buote+07'' shows the results obtained from the $X$-ray
 data sets of 39 galaxy- and cluster-scale halos in Buote et
 al.\ (2007). } \label{tab:c-m}
\end{table*}
We quantify the possible trend by fitting the following function to the
$c_{\rm vir}$-$M_{\rm vir}$ data points:
\begin{eqnarray}
c_{\rm vir} = c_N 
\left(\frac{M_{\rm vir}}{10^{14}h^{-1}\MO}\right)^{-\alpha}.
%(1+z)^{-\beta}.  
\label{eq:c-m}
\end{eqnarray}
This form is motivated by the simulation based studies (e.g.\ Bullock
et al 2001) and specified by two free parameters: normalization
$c_{N}$ and mass slope $\alpha$.  The best-fit parameters and
$1\sigma$ uncertainties are: 
$c_N=8.75_{-2.89}^{+4.13}$ and $\alpha=0.40\pm 0.19$.  Thus the mass scaling of $c_{\rm vir}(M_{\rm
vir})$ is marginally detected at a $2\sigma$ level.  
Our results are more significant than the earlier weak lensing result
(Comerford \& Natarajan 2007; Mandelbaum et al. 2008). 

Note that the two parameters $M_{\rm vir}$ and $c_{\rm vir}$ are
correlated for each cluster: the measured shear profile can be
explained by NFW profiles with larger $M_{\rm vir}$ and smaller
$c_{\rm vir}$ than the best-fit values and vice versa.   We
therefore check whether this intrinsic correlation might be
exaggerating the significance of our result.  We randomly draw
$M_{\rm vir}$ and $c_{\rm vir}$ for each cluster from the respective
posterior distributions, and re-calculate the best-fit $M_{\rm
vir}$-$c_{\rm vir}$ 30000 times.  From the mean relation derived
from these samples, and the scatter around the mean, we estimate
that the significance of our detection of anti-correlation between
mass and concentration is unchanged, and conclude that
the intrinsic correlation has a negligible effect on our results.

We also checked whether the parameter fitting above causes a bias in the
best-fit slope $\alpha$ of the scaling relation $c_{\rm vir}(M_{\rm
vir})$ by using simulated data. First, we generated 3000 simulated
catalogs of the tangential shear profiles for 19 clusters including the
errors at each radial bins that are modeled to reproduce the measured
errors. In making this simulations the mass and concentration of each
cluster are randomly chosen from the observed ranges of $2\le M_{\rm
vir}/(10^{14} h^{-1}M_{\odot})\le 15$ and $2\le c_{\rm vir}\le 10$, and
the redshift of all clusters is fixed to a single redshift $z_{\rm
l}=0.23$ corresponding to the mean redshift of the 19 Subaru clusters.
Note that the simulated cluster catalogs have no intrinsic scaling
relation between $c_{\rm vir}$ and $M_{\rm vir}$ on average,
i.e. $\alpha=0$.  Then we estimated $M_{\rm vir}$ and $c_{\rm vir}$
parameters for each simulated cluster from the shear profile fitting to
an NFW model, and made a fitting of the distribution of $M_{\rm vir}$
and $c_{\rm vir}$ for 19 clusters to the model $c_{\rm vir}$-$M_{\rm
vir}$ relation (Eq.~[\ref{eq:c-m}]).  From the 3000 catalogs we found
that the estimated slope of $c_{\rm vir}(M_{\rm vir})\propto M_{\rm
vir}^{-\alpha}$ tends to be slightly biased as $\langle\alpha
\rangle=0.06 $ from the input value $\alpha=0$. The origin of this bias
can again be ascribed to the degeneracy between mass and concentration
for the shear profile fitting.  Nevertheless the amount of the bias is
smaller than the $1\sigma$ statistical errors of $\alpha$ estimation,
$\sigma(\alpha)=0.19$, therefore we conclude that this effect is also
insignificant.

The observed concentration-mass relation can be compared with the
theoretical predictions based on large $N$-body simulations.  The
solid line in Figure~\ref{fig:C-M} shows the median relation obtained
by Duffy et al.\ (2008, hereafter Duffy08), and given by
$(c_N,\alpha)=(4.96,0.086)$ in Eq.~(\ref{eq:c-m}), where the relation
was obtained assuming the concordance $\Lambda$CDM model that is
constrained by the WMAP 5-year data.  Note that the redshift
dependence of $c_{\rm vir}(M_{\rm vir})$ is corrected to match halos
at the mean redshift $z=0.23$, based on the fitting results in Table 1
of Duffy08.  The observed concentrations of $M_{\rm vir}\sim 5\times
10^{14}M_\odot$ clusters, i.e.\ $c_{\rm vir}\sim 5$ are consistent
with the prediction, however the observed slope is steeper than the
prediction, albeit at very modest statistical significance.  It is
also important to note that the clusters (star symbols) that are well
fitted by an NFW profile have low concentrations of $c_{\rm vir}\sim
3$, while the distribution of clusters (squares) for which we cannot
discriminate between CIS and NFW models extends to much large
concentrations.  

The region enclosed by the two dashed lines shows the range of
$\sigma(\log_{10}c_{\rm vir})=0.1$ 
in which simulated clusters are
typically distributed as shown in Duffy08 (also see Jing 2000). The
scatter for the observed concentrations is given by
$\sigma(\log_{10}c_{\rm vir})\simeq 0.17$, which is estimated by
weighting the cluster distribution with the inverse square of the
statistical error of each cluster concentration.  
The observed statistical errors are so large that it's not
possible to say whether there is any intrinsic scatter contribution.

Our results for the concentration distribution are summarized in
Table~\ref{tab:c-m}, together with the predictions of Duffy08 and
Buote et al.'s (2007) observational results based on $X$-ray data.
Note that the $X$-ray results are derived using a much wider range of
halo masses than our results -- from galaxy to galaxy cluster scales.
Both the lensing and $X$-ray observations imply a significantly higher
normalization $c_N$ than the simulations, and also a steeper
dependence (higher $\alpha$) on halo masses. Comparing the lensing and
$X$-ray results, the lensing results indicate a steeper dependence
than the $X$-ray results, but the discrepancy is not yet conclusive
due to the large statistical errors.  A further, careful study will be
needed to resolve these possible discrepancies.

\begin{figure}
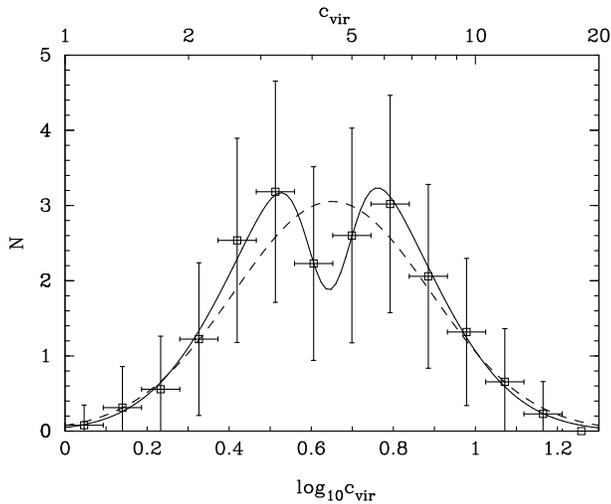

\begin{center}
%%\FigureFile(80mm,80mm){logCvir_hist_photoz.ps}
\FigureFile(80mm,80mm){f7.ps}
\end{center}
\caption{The one-dimensional distribution of the observed concentration
parameters for 19 clusters, obtained by projecting the cluster
distribution in Figure~\ref{fig:C-M} onto the axis of $\log_{10}c_{\rm
vir}$. The square symbols and error bars in each bin are computed from
the mean and variance of 3000 Monte Carlo redistributions of the $c_{\rm
vir}$ distribution, taking into account the uncertainties in $c_{\rm
vir}$ for each clusters. The solid and dashed curves show the
best-fit models of one- and two-lognormal distributions,
respectively.  }  \label{fig:Cvir_hist}
\end{figure}
In Figure~\ref{fig:Cvir_hist} we show the one-dimensional distribution
of the concentration parameters for the 19 clusters in
Figure~\ref{fig:C-M}.  The mean values and error bars in each bin are
computed from 3000 Monte Carlo redistributions of the clusters,
assuming that the halo concentration of each cluster obeys the
Gaussian distribution with width given by the measurement error
$\sigma(c_{\rm vir})$. 
% MT
Note that the data points in different bins are correlated. 
Interestingly the
observed distribution contains a dip at $c_{\rm vir}\simeq4$,
suggesting that a single log-normal model distribution may not fit the
distribution very well.  The solid and dashed curves show the results
of fitting one and two log-normal distributions, respectively.  Given
the large error bars, the two models both give an acceptable fit to
the data: the two log-normal distributions (additional two model
parameters compared to the one lognormal distribution) improve the
$\chi^2$ value only by 
%%NO
$\Delta\chi^2\simeq 0.9$.  Nevertheless it is interesting to note that
simulations have found similar structure in the distribution of
predicted concentrations (Jing 2000; Shaw et al.\ 2006; Neto et al.\
2007; Duffy08).  It is argued in these studies that the physical
origin of the structure lies in the dynamical status of clusters: more
relaxed clusters tend to have high concentrations and vice versa (also
see Smith \& Taylor  2008).
% for such an indication from the actual measurement).
It will therefore be
important to explore further the concentration distribution by
enlarging the sample size of clusters.

\subsection{Stacked Lensing Signal}
\label{sec:stack}

\begin{figure*}[t]
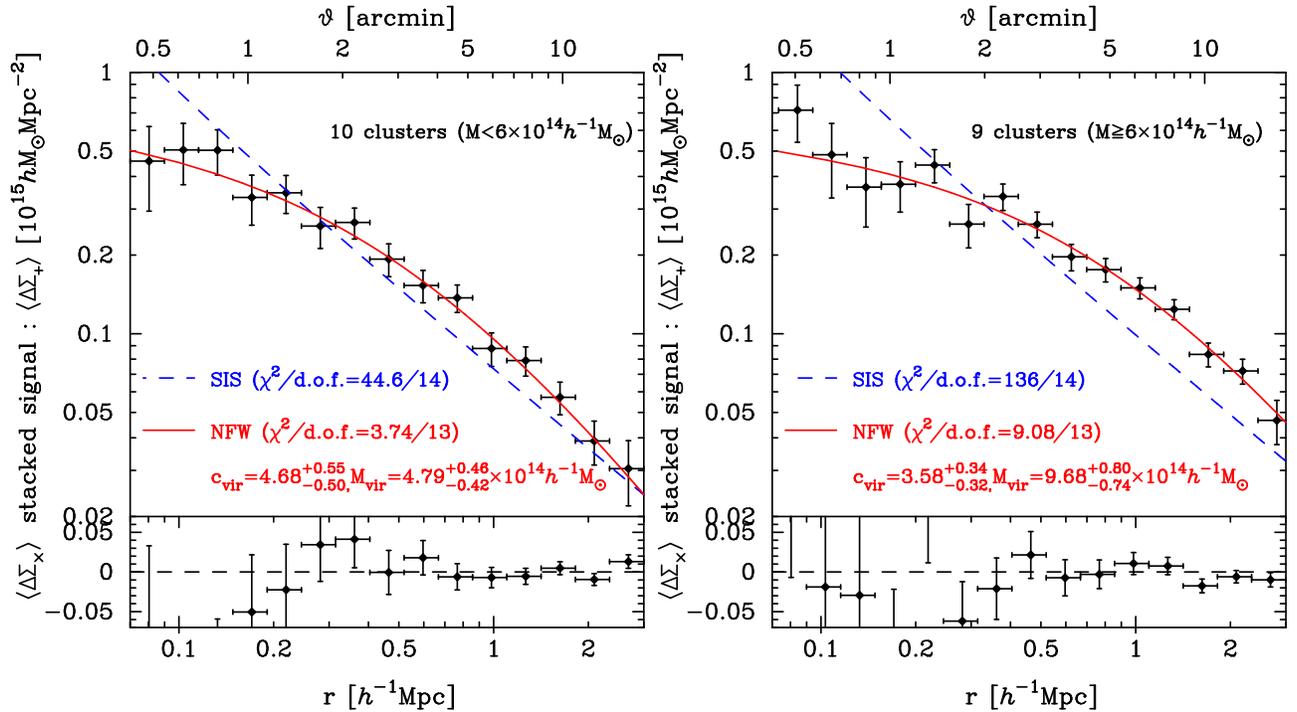

\begin{center}
%\FigureFile(170mm,170mm){stacked.ps}
\FigureFile(170mm,170mm){f8.ps}
\end{center}
\caption{{\em Left panel}: The mean distortion profile with $1\sigma$
 statistical uncertainties as a function of the projected radius, which
 is obtained by stacking the distortion signals for 10 clusters that are
 selected with the virial masses $M_{\rm vir}\le 6\times
 10^{14}h^{-1}M_{\odot}$ from 19 clusters in
 Figure~{\ref{fig:C-M}}. Note that the distortion profile is plotted in
 the unit of the projected mass density, and the projected radius is
 computed from the weighted mean redshift of clusters. The dashed and
 solid curves are the best-fit SIS and NFW models,
 respectively. {\em Right panel}: The similar plot, but for 9 halos with
 $M_{\rm vir}>6\times 10^{14}h^{-1}M_\odot$. For both the results, the
 SIS model is strongly disfavored: 
%%NO
$\Delta\chi^2\equiv \chi^2_{\rm SIS, min} -\chi^2_{\rm NFW, min}\simeq 41$ and 127 for
 the low- and high-mass cluster samples, respectively. The combined
 results also show a $2\sigma$-level evidence that the NFW concentration
 is greater for more massive halos, which is exactly consistent with the
 result in Figure~\ref{fig:C-M}. } \label{fig:stack}
\end{figure*}
In this section we study the stacked weak-lensing signal of 19
clusters in Figure~\ref{fig:C-M}.  This approach has several important
advantages.  First, the {\em average} distortion profile is less
sensitive to substructures within and asphericity of the individual
cluster mass distributions and also to uncorrelated large-scale
structure along the same line-of-sight.  This is because these
``contaminating signals'' are averaged out 
via the
stacking, under the assumption that the universe is {\em
statistically} homogeneous and isotropic.  Second, stacking should
boost the signal-to-noise ratio of the distortion signal at very small
and large radii.  The signal-to-noise ratio at small radii is limited
for a single cluster because the solid angle subtended by a radial bin
shrinks at small radii thus reducing the number of galaxies
over which the mean distortion signal is calculated.  Hence the
signal-to-noise ratio suffers despite the signal peaking in these
regions.  On the other hand, at large radii, the binned solid-angle is
much larger, helping to maintain signal-to-noise, however the signal
becomes very small, and thus the signal-to-noise ratio declines.  As
discussed in \S~5.3, the signal-to-noise ratio is optimized at
intermediate radii.  Therefore, stacking helps to improve
signal-to-noise as a function of radius, thus enabling a clearer
investigation of (i) the curvature of the density profile that is a
characteristic signature of the NFW prediction, helping us potentially
to address the nature of dark matter (e.g.\ Yoshida et al.\ 2000), and
(ii) the distribution of mass outside the virial radius to address
whether or not the outer slope of the NFW profile, $\rho\propto
r^{-3}$, continues outside the virial radius (e.g.\ Bertschinger 1985;
Busha et al.\ 2003).

To study the stacked lensing signal, we divide the 19 clusters into
two mass bins, based on whether the NFW model fits to individual
clusters yielded a virial mass estimate of greater than or less than
$M_{\rm vir}=6\times 10^{14}h^{-1}M_\odot$.  This results in two
sub-samples of 10 low-mass and 9 high-mass clusters.
Figure~\ref{fig:stack} shows the average distortion profile as a
function of the projected radius in the physical length scale. Note
that the effect of different cluster redshifts was taken into account
by using the weighting method in terms of the lensing efficiency
functions of averaging clusters (Mandelbaum et al.\ 2006; also see
Sheldon et al.\ 2007), and the projected radius is computed from the
weighted mean redshift of the sampled clusters.  However, we checked
that, even if we use the single lensing efficiency for the mean
cluster redshift, the results are almost unchanged due to the narrow
redshift coverage of our cluster samples.
% MT
Note that the mean lens redshifts are $\langle z_{\rm l}\rangle
=0.251$ and $0.236$ for the low- and high-mass samples, respectively.

First, unsurprisingly, the stacked profiles yield \emph{very}
significant detections: the total signal-to-noise ratios are
 $S/N=24$ and $30$ for the low- and high-mass samples respectively.  Second, the
lensing distortion signals are recovered over a wide range of radii,
from $ 70h^{-1}$kpc to 3$h^{-1}$Mpc scales, spanning 
a factor of 50 in 
radius. Note that the outer radial boundary corresponds to the size of
the Suprime-Cam's FoV for clusters at 
$z\simeq0.24$, and is a factor $\sim1.5-2$ beyond the cluster virial
radii determined from the individual NFW model fits.  Visual
inspection of the stacked profiles also reveals that they are clearly
not described by a single power law model, displaying very obvious
curvature, reminiscent of the NFW prediction.  We therefore fitted SIS
and NFW models to the stacked profiles, and show as solid and dashed
curves are the best-fit NFW and SIS models respectively.  The
non-linear corrections in the measured reduced shear are taken into
account in these fits following the method in Mandelbaum et al.\
(2006), however for simplicity we ignore uncertainties in the
  alignment of cluster halo centers; we will discuss this effect in
  detail in \S~\ref{sec:center}.  Now very clearly the SIS model is
strongly disfavored at
$6\sigma$ and $11\sigma$ significance  for the low- and high-mass
  samples, respectively, estimated from the difference between the
  $\chi^2$ values of the best-fit NFW and SIS models:
  $\Delta\chi^2\equiv\chi^2_{\rm SIS, min}-\chi^2_{\rm NFW,
    min}\simeq41$ and $127$, respectively. The NFW model gives an
acceptable fit to the data (the CIS model is also acceptable).  

The best-fit NFW parameters are $c_{\rm
vir}=4.68_{-0.50}^{+0.55}$, $M_{\rm vir}=4.79_{-0.42}^{+0.46}\times
10^{14}h^{-1}M_\odot$ for the low-mass sample, 
and $c_{\rm vir}=3.58_{-0.32}^{+0.34}$, $M_{\rm vir}= 9.68_{-0.74}^{+0.80}\times
10^{14}h^{-1}M_\odot$ for the high-mass sample,
i.e.\ relative accuracies of about $10\%$ for both $c_{\rm vir}$ and
$M_{\rm vir}$, an improvement by factor of 2-5 compared to the
individual cluster constraints in Figure~\ref{fig:dM_to_M}. Comparing
the two mass bins reveals that the concentration parameter appears to
be greater for the low-mass sample than for the high-mass one at
$2\sigma $ significance. It is re-assuring that this difference is
exactly consistent with the relation found from the individual cluster
analysis of 19 clusters in Figure~\ref{fig:C-M}, even though the
individual-cluster and stacked analyses involve non-trivial
differences in the averaging procedures that are not necessarily
equivalent for real clusters (e.g.\ due to non-spherical mass
distribution and substructures).  

The measured distortion profile outside the virial radius is
consistent with the outer slope of NFW profile, i.e.\ we could not
find any evidence that the mass distribution outside the virial
radius, which mostly contains gravitationally unbound mass, declines
more rapidly than is predicted by NFW.  This is in contrast to the
sharply truncated profile at the virial radius discussed by Busha et
al.\ (2005; see also Takada \& Jain 2004; Prada et al.\ 2006; Baltz et
al.\ 2007).  The stacked distortion profiles also do not show any
signature of associated large-scale structures such as filamentary
structures surrounding the clusters, unlike the SDSS stacked lensing
results (Johnston et al.\ 2007).  However the large-scale structure
lensing signals are only expected to dominate at projected radii
greater than $\sim10$Mpc.  Hence, by further extending the observed
fields to obtain more radial range covered, it would be interesting to
explore the lensing signals outside the virial radius to test the CDM
structure formation scenarios sitting more in the linear regime.

Finally, we note that the results presented in this section help to
explain the systematic difference between virial mass estimates
between SIS and NFW model fits to the individual clusters as found in 
Fig.~\ref{fig:Msis_vs_Mnfw}.
The virial mass estimates are dominated by the integral of
the density profile on large radii around the virial
radius.  Figure 7 reveals that when an SIS model is fitted to
distortion profile data from an NFW halo, the inability of the SIS
model to capture the curvature of the distortion profile causes it to
underestimate the amount of mass in the cluster on large radii.
This short-fall on large scales is compensated to some extent, but
not entirely by the overestimation of the cluster mass on small
scales.

\subsection{Results for Model-Independent Mass Estimates}
\label{sec:results_model_independent}

\begin{table*}
  \caption{Weak Lensing Mass Estimates for the 22 Clusters} 
\label{tab:2Dmass}
\begin{center}
\begin{tabular}{cccccccc}
\hline
\hline\\
 Cluster          &  $M_{\rm 2D}(<500h^{-1}{\rm kpc})$
                  &  $M_{\rm 2D}(<\theta_{500})$
                  &  $M_{\rm 2D}(<\theta_{\rm vir})$
                  &  $M_{2500}^{{\rm NFW}}$
                  &  $M_{500}^{{\rm NFW}}$
                  &  $M_{200}^{{\rm NFW}}$\\
(1)               &  (2)
                  &  (3)
                  &  (4)
                  &  (5)
                  &  (6)
                  &  (7)\\
\hline %%%%%%%%RED+BLUE SAMPLE : NFW 10^14
A68                  &  $2.62\pm0.69$
                  &  $4.17\pm1.13$
                  &  $7.87\pm3.02$
                  &  $1.00^{+0.42}_{-0.43}$
                  &  $2.92^{+0.86}_{-0.75}$
                  &  $4.45^{+1.75}_{-1.35}$  \\
A115                  &  $3.23\pm1.00$
                  &  $5.20\pm1.82$
                  &  $8.79\pm6.61$
                  &  $0.86^{+0.46}_{-0.47}$
                  &  $2.70^{+1.15}_{-0.93}$
                  &  $4.24^{+2.60}_{-1.79}$  \\
A209                  &  $4.71\pm0.49$
                  &  $8.40\pm1.11$
                  &  $13.16\pm4.00$
                  &  $1.53^{+0.33}_{-0.33}$
                  &  $6.19^{+0.95}_{-0.86}$
                  &  $10.62^{+2.17}_{-1.81}$  \\
RXJ0142                 &  $2.36\pm0.62$
                  &  $3.97\pm0.98$
                  &  $5.60\pm2.91$
                  &  $1.37^{+0.22}_{-0.22}$
                  &  $2.85^{+0.60}_{-0.53}$
                  &  $3.86^{+0.98}_{-0.82}$  \\
A267                  &  $1.87\pm0.45$
                  &  $3.14\pm0.68$
                  &  $3.94\pm1.93$
                  &  $1.01^{+0.18}_{-0.18}$
                  &  $2.31^{+0.48}_{-0.43}$
                  &  $3.23^{+0.82}_{-0.69}$  \\
A291                  &  $2.55\pm0.48$
                  &  $3.79\pm0.84$
                  &  $5.23\pm3.12$
                  &  $0.63^{+0.30}_{-0.27}$
                  &  $2.88^{+0.70}_{-0.62}$
                  &  $5.19^{+1.80}_{-1.34}$  \\
A383                  &  $2.54\pm0.45$
                  &  $3.72\pm0.79$
                  &  $8.69\pm2.53$
                  &  $1.23^{+0.17}_{-0.17}$
                  &  $2.37^{+0.51}_{-0.43}$
                  &  $3.11^{+0.88}_{-0.69}$  \\
A521                  &  $3.85\pm0.61$
                  &  $5.35\pm1.15$
                  &  $9.29\pm4.58$
                  &  $0.77^{+0.22}_{-0.22}$
                  &  $2.78^{+0.51}_{-0.48}$
                  &  $4.58^{+1.00}_{-0.88}$  \\
A586                  &  $3.75\pm0.99$
                  &  $7.54\pm2.53$
                  &  $12.69\pm8.57$
                  &  $2.41^{+0.45}_{-0.42}$
                  &  $4.74^{+1.40}_{-1.14}$
                  &  $6.29^{+2.26}_{-1.69}$  \\
ZwCl0740                &  $2.25\pm0.48$
                  &  $2.77\pm0.85$
                  &  $6.31\pm3.66$
                  &  $0.64^{+0.27}_{-0.26}$
                  &  $2.55^{+1.11}_{-0.75}$
                  &  $4.36^{+3.14}_{-1.60}$  \\
A611                  &  $3.86\pm0.59$
                  &  $5.78\pm1.11$
                  &  $8.77\pm3.52$
                  &  $1.30^{+0.33}_{-0.34}$
                  &  $3.63^{+0.70}_{-0.64}$
                  &  $5.47^{+1.31}_{-1.11}$  \\
A697                  &  $3.86\pm0.56$
                  &  $7.74\pm1.12$
                  &  $11.09\pm3.62$
                  &  $1.60^{+0.38}_{-0.38}$
                  &  $5.87^{+0.89}_{-0.82}$
                  &  $9.73^{+1.86}_{-1.61}$  \\
A1835                  &  $5.53\pm0.82$
                  &  $9.15\pm2.53$
                  &  $16.39\pm10.02$
                  &  $2.03^{+0.40}_{-0.41}$
                  &  $6.78^{+1.20}_{-1.07}$
                  &  $10.86^{+2.53}_{-2.08}$  \\
ZwCl1454                  &  $2.90\pm0.82$
                  &  $3.12\pm1.17$
                  &  $5.42\pm4.04$
                  &  $0.63^{+0.27}_{-0.29}$
                  &  $1.83^{+0.69}_{-0.57}$
                  &  $2.80^{+1.39}_{-1.03}$  \\
ZwCl1459                 &  $3.24\pm0.66$
                  &  $3.92\pm1.08$
                  &  $3.25\pm2.83$
                  &  $1.26^{+0.30}_{-0.30}$
                  &  $2.74^{+0.71}_{-0.63}$
                  &  $3.77^{+1.17}_{-0.98}$  \\
RXJ1720                  &  $2.17\pm0.64$
                  &  $3.13\pm1.05$
                  &  $6.31\pm3.11$
                  &  $1.36^{+0.28}_{-0.26}$
                  &  $2.64^{+0.78}_{-0.66}$
                  &  $3.48^{+1.28}_{-0.99}$  \\
A2219                  &  $4.54\pm0.71$
                  &  $7.68\pm1.62$
                  &  $12.45\pm4.92$
                  &  $2.65^{+0.41}_{-0.44}$
                  &  $5.67^{+1.05}_{-0.95}$
                  &  $7.75^{+1.89}_{-1.60}$  \\
A2261                  &  $4.32\pm0.61$
                  &  $7.94\pm1.44$
                  &  $10.64\pm4.75$
                  &  $2.49^{+0.31}_{-0.31}$
                  &  $5.70^{+0.86}_{-0.78}$
                  &  $7.97^{+1.51}_{-1.31}$  \\
RXJ2129                  &  $2.53\pm0.57$
                  &  $4.78\pm1.02$
                  &  $8.17\pm3.36$
                  &  $0.97^{+0.37}_{-0.38}$
                  &  $3.28^{+0.77}_{-0.69}$
                  &  $5.29^{+1.76}_{-1.38}$  \\
A2390                  &  $4.69\pm0.68$
                  &  $8.84\pm1.31$
                  &  $18.32\pm3.74$
                  &  $2.21^{+0.31}_{-0.30}$
                  &  $4.97^{+0.90}_{-0.82}$
                  &  $6.92^{+1.50}_{-1.29}$  \\
A2485                  &  $2.84\pm0.72$
                  &  $3.36\pm1.11$
                  &  $8.04\pm3.97$
                  &  $0.71^{+0.30}_{-0.30}$
                  &  $2.30^{+0.63}_{-0.56}$
                  &  $3.63^{+1.26}_{-1.02}$  \\
A2631                  &  $3.13\pm0.49$
                  &  $3.97\pm0.87$
                  &  $8.16\pm2.49$
                  &  $1.70^{+0.25}_{-0.26}$
                  &  $3.40^{+0.53}_{-0.49}$
                  &  $4.54^{+0.89}_{-0.78}$  \\
\hline
\end{tabular}
\end{center}
\textrm{NOTES $\singlebond$
 Column (1): cluster name;
 Column (2): the aperture mass within the projected radius of
 $500h^{-1}$kpc at the cluster redshift,  
in the unit of  $10^{14}h^{-1}M_\odot$;
 Columns (3,4): the aperture masses within the radius 
 corresponding to the over-density $\Delta=500$ and the virial radius,
 respectively, where the radii are computed from the
 best-fit NFW model to the tangential distortion profile;
 Columns (5-7): the three-dimensional masses estimated from 
the NFW model fitting,  
 $M_{2500}$, $M_{500}$ and $M_{200}$, for the over-densities
 $\Delta=2500$, 500 and $200$, respectively.
}
\end{table*}
\begin{figure}
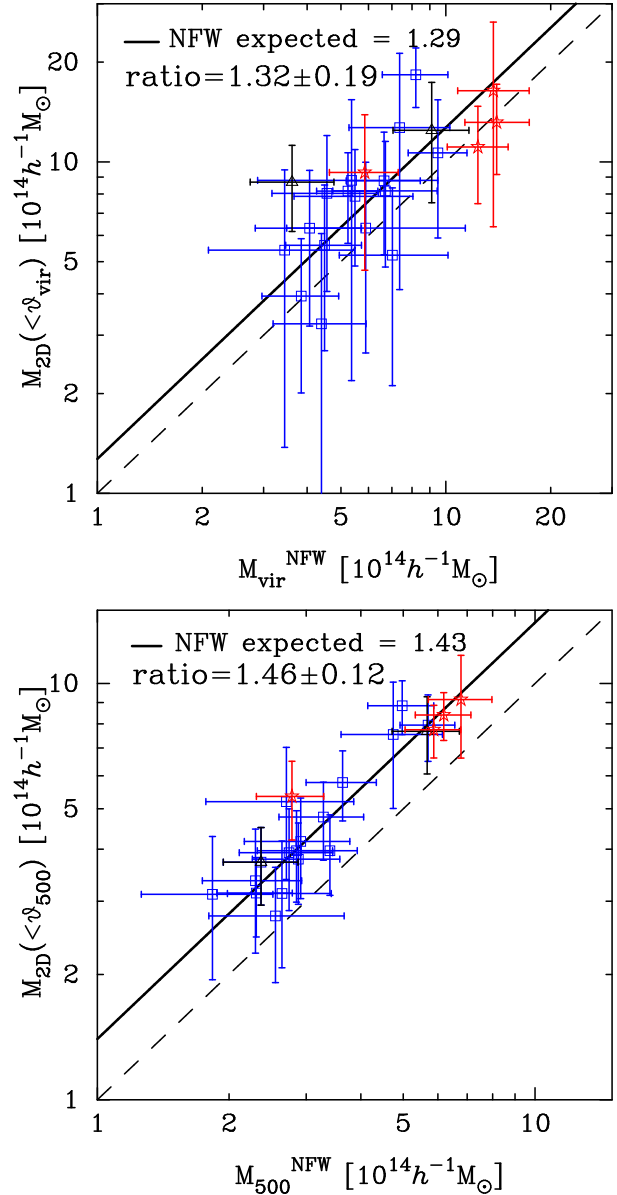

\begin{center}
%\FigureFile(80mm,80mm){compare_Map_vs_Mnfw_photoz.ps}
%\FigureFile(80mm,80mm){compare_Map_vs_M500_photoz.ps}
\FigureFile(80mm,80mm){f9a.ps}
\FigureFile(80mm,80mm){f9b.ps}
\end{center}
\caption{Comparing the lensing aperture mass with the
  three-dimensional mass that is obtained from the NFW model fitting,
  for each of 22 clusters as in Figure~\ref{fig:Msis_vs_Mnfw}. 
The
  upper panel shows the results obtained when the virial radius of the
  best-fit NFW model is assumed for the aperture radius, while the
  lower panel shows the results for the radius of the over-density
  $\Delta=500$. The 2D aperture masses are systematically greater than
  the 3D mass for both the cases.  In each panel the numbers labeled
  as ``ratio'' are the ratio of the 3D and 2D masses and the
  dispersion over all the clusters. For comparison, the solid line
  denotes the mass difference expected from a cluster-scale NFW
  profile with concentration parameters $\langle c_{\rm
    vir}\rangle=3.6$ and $\langle c_{\rm 500}\rangle=1.7$, computed
  using Eq.~(\ref{eqn:nfw2Dvs3D}): $M^{\rm NFW}_{2D}/M^{\rm
    NFW}_{3D}\simeq 1.29$ and $1.43$ for the radii with
  $\Delta=\Delta_{\rm vir}$ and $500$, respectively, which are in good
  agreement with the actual measurements. Note that the dashed line
  denotes $M_{2D}=M_{3D}$. 
The different symbols are as in Fig.~\ref{fig:Msis_vs_Mnfw}.
%{\bf ZZ what are the points marked by red?
%    ZZ}
} \label{fig:mass_vs_2Dmass}
\end{figure}
%%%NO re-calculate M2d/M3d using new <cvir> and <c500> 

We now turn to model-independent estimates of the projected mass of
each cluster, using the $\zeta_c$-method described in
\S~\ref{sec:model-independ-mass}.

The first three columns of Table~\ref{tab:2Dmass} list, for the 22
clusters in Table~\ref{tab:massprofile}, the aperture masses within
several different radii.  Note again that these 22 clusters
have color information -- the results in Table~6 are therefore based
on the red+blue background galaxy samples.  The statistical accuracy
of the aperture mass within a given aperture radius $\theta_m$ is
determined by the measurement accuracy of the $\zeta_c$-statistics
(see Eq.~[\ref{eq:zeta}]) that is computed by integrating the
measured distortion profile over the annulus taken outside the
aperture radius $\theta_m$.  Therefore, the aperture mass accuracy
decreases with increasing aperture radius, because at larger radii the
cluster lensing signal become weaker and thus noisier.
Table~\ref{tab:2Dmass} shows that, at the viral radius and $r_{500}$,
the typical accuracies are $\sigma(M_{2D})/M_{2D}\sim 50\%$ or $25\%$,
respectively.  Note that the aperture mass at the virial radius is
somewhat sensitive to the choice of the control annulus
($\theta_{o1}\le \theta\le\theta_{o2}$ -- see Eq.~[\ref{eq:zeta}]).
However the $M_{2D}$ estimates vary within the $1\sigma$ statistical
errors quoted in Table~6 when the control annulus is varied -- this is
therefore not a dominant source of errors.  The second column shows the
results for a fixed projected radius, $r=500h^{-1}$kpc.
%, which is computed from each cluster redshift.

For comparison we also list the model-dependent results for the
three-dimensional masses obtained from the NFW model fitting, at
several over-densities: 
%$M_{2500}, M_{500}$ and $M_{200}$ for the over-densities 
$\Delta=2500$, 500 and $200$ (the virial
mass and the errors were already given in
Table~\ref{tab:massprofile}).  The masses $M_{2500}$ and $M_{500}$ are
often used when estimating cluster masses based on $X$-ray
observations (e.g.\ Vikhlinin et al.\ 2008).

Figure~\ref{fig:mass_vs_2Dmass} compares the aperture masses with the
three-dimensional best-fit NFW messes for 22 clusters.  The upper panel
shows the comparison at the virial radius -- on this scale the mass
estimates agree within the error bars, the scatter around the equality
line being dominated by measurement error.  A formal fit to the data
points, holding the slope of the line fixed at unity, gives a best fit
ratio of
$M_{\rm 2D}(<\theta_{\rm vir})/M_{\rm vir}^{\rm NFW}=1.32\pm0.19$. Note that the fit is done in the linear scale of
masses, rather than the log space. On average the aperture masses are
therefore $\sim32\%$ higher than the 3D NFW masses, at $\sim2\sigma$
significance.  The comparison at $r_{500}$ is shown in the lower panel.
In this case a systematic excess of aperture masses over 3D NFW masses
is immediately obvious for most of clusters.  Repeating the fit
described above to the data at $r_{500}$ gives a ratio of 
$M_{\rm 2D}(<\theta_{500})/M_{500}^{\rm NFW}=1.46\pm0.12$, i.e.\ a $46\%$
difference at $\sim 4\sigma$ significance.

These results are naturally expected as follows.  Recalling that the
two-dimensional projected mass includes all the mass contributions
contained in the cylinder from the observer to the source galaxies
along the line-of-sight, the aperture mass has an additional mass
contributions to the three-dimensional spherical mass within the same
radius.  The main contribution arises from integration of the cluster
mass distribution itself along the line of sight to calculate the mass
within a cylinder of the same radius on the sky as the sphere used in
the calculation of the 3D NFW mass.  Aperture masses are therefore
always expected to be larger than the 3D NFW mass.
For example, the amplitude of the mass biases calculated above
is well explained by a cluster-scale NFW profile.
As described explicitly in
Appendix~\ref{app:NFWmass}, the ratio of the projected 2D and 3D
masses of such an NFW halo are calculated analytically to be: $M^{\rm
NFW}_{2D}(<\theta_{m})/M^{\rm NFW}_{3D}(<r=D_l\theta_m)\simeq 1.29$
and $1.43$ for $\Delta=\Delta_{\rm vir}$ and $500$, respectively,
assuming the concentration parameter $\langle 
c_{\rm vir}\rangle=3.6$, the mean
concentration for all the clusters. 
These
biases are shown by the solid lines in
Figure~\ref{fig:mass_vs_2Dmass}, showing nice agreement with
the measured biases.  In other words the three-dimensional spherical
mass can be estimated from the aperture mass by correcting for the
mass bias, assuming an NFW profile (see Mahdavi et al.\ 2008 for such
an example). 
Note that the correction factor is not so sensitive to the
assumed concentration parameter, because the aperture mass does not
measure shear signals at inner radii, which are sensitive to halo 
concentration. 
Even if $c_{\rm vir}=8$ is assumed, the
correction factor becomes smaller only by about $10\%$.  

\subsection{Discussion of Systematic Errors}

There are several sources of systematic errors involved in the weak
lensing measurements. In this subsection we discuss possible
effects of the systematic errors on our results.

\vspace{0.5em}
\subsubsection{Dilution contamination} \label{subsec:dil}

\begin{figure*}
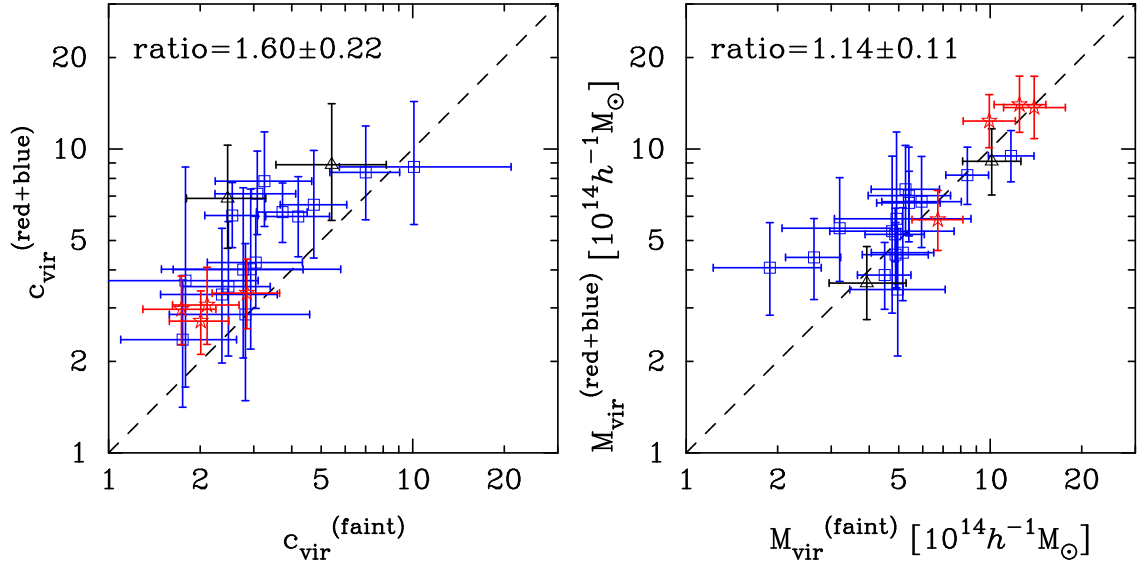

\begin{center}
%%\FigureFile(150mm,150mm){compare_Cvir+Mvir_photoz.ps}
\FigureFile(150mm,150mm){f10.ps}
\end{center}
\caption{Comparison of the best-fit parameters of NFW model obtained
using the ``faint'' galaxy sample and the ``red+blue'' galaxy sample in
the weak lensing analysis for the 22 clusters as in
Figure~\ref{fig:Msis_vs_Mnfw}, where the faint galaxy sample is likely
to be more contaminated by unlensed member galaxies and therefore suffer
from the dilution effect (see \S~\ref{sec:sample}).  The left panel
shows the results for the concentration parameter, and the right panel
for the virial mass. The concentration parameter is systematically
underestimated by the dilution effect, while the virial mass is little
affected. This is because the dilution effect is indeed caused mainly by
member galaxies, which reduces the measured distortion signals on small
radii, but preserves the signals at large radii to which the virial mass
is sensitive.}  \label{fig:compare_c+M}
\end{figure*}
One of the most important systematic errors to which we have paid
particular attention is the dilution of the weak-lensing signal due to
contamination of background galaxy catalog by faint cluster galaxies.

As described in \S~\ref{sec:sample}, we defined several samples of
background galaxies according to different color/magnitude selection
criteria: the magnitude-selected faint galaxy sample that is often
used in the literature and a more secure ``red+blue'' galaxy sample,
defined as faint galaxies redder and bluer than the cluster
red-sequence by a minimum color offset.  Figure~\ref{fig:compare_c+M}
demonstrates the impact of dilution on estimates of the cluster
parameters, comparing the best-fit NFW parameters obtained when using
the faint and red+blue galaxy samples.  It is clear that the
concentration parameter for the faint galaxy sample is systematically
smaller than for the red+blue sample for most of the clusters, i.e.\
underestimated due to the dilution effect inherent in the faint
sample.  The bias is measured to be
%%NO
 $c_{\rm vir}^{\rm
(red+blue)}/c^{\rm (faint)}_{\rm vir}\simeq 1.60\pm 0.22$.
%%NO
On the other hand, the virial mass constraints are consistent between the two
samples within the error bars:
%%NO
 $M_{\rm vir}^{\rm (red+blue)}/M_{\rm vir}^{\rm (faint)}\simeq 1.14\pm 0.11$. 
%%NO
 This is because the virial
mass is mainly sensitive to the overall shear amplitudes at large
radii ($\simgt 10'$), and relatively insensitive to the distortion
signals at small radii to which the concentration parameter is
particularly sensitive.  It is important to remember here that the
dilution effect increases as cluster-centric distance decreases
because the number density of faint cluster galaxies that contaminate
the faint galaxy catalog is expected to roughly follow the underlying
density profile of the cluster.  Thus our results indicate that
correcting for the dilution effect is important to obtain unbiased,
accurate constraints on cluster parameters, especially on the
concentration parameter.

It is nevertheless worth noting that, due to the limited information
on galaxy colors and redshifts, the red+blue galaxy sample we have
used may still be contaminated by member and foreground
galaxies.  According to the results in Figure~\ref{fig:compare_c+M},
we should also bear in mind that the virial mass estimates are
relatively unbiased, but the best-fit concentration parameters given
in Table~\ref{tab:massprofile} may still underestimate the true value
(if an NFW profile represents the true mass distribution).

\begin{figure}
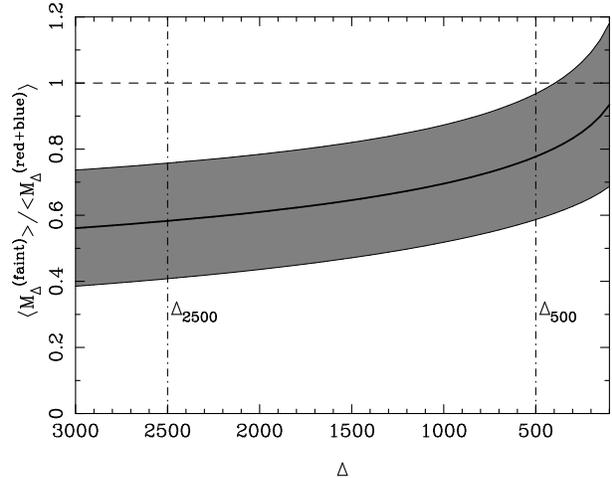

\begin{center}
%%\FigureFile(80mm,80mm){Mall_to_Mredblue_vs_delta.ps}
\FigureFile(80mm,80mm){f11.ps}
\end{center}
\caption{The solid curve shows the ratio of NFW mass estimates for the
 red+blue galaxy sample and for the faint galaxy sample, as a function
 of the over-density used to define cluster mass.  The shaded, gray
 region around the solid curve is the dispersion of 19 clusters. The
 dilution effect causes cluster masses to be more significantly
 underestimated with increasing the over-density.  }
 \label{fig:Mall_to_Mredblue}
\end{figure}
However, unsurprisingly given the expected variation of dilution as a
function of radius, the amplitude of the bias in mass measurements
depends on the chosen aperture radius within which mass is measured.
Figure~\ref{fig:Mall_to_Mredblue} shows the variation of the ratio of
mass estimates from the faint and red+blue galaxy samples change as a
function of the over-density used to define the cluster mass.  
As $\Delta$ increases the cluster masses become
progressively underestimated due to  more significant 
dilution of the weak-lensing signal by cluster members.  This
is an important result when considering studies in which lensing-based
mass estimates are compared with cluster observables at other
wavelengths that are conventionally measured at over-densities
exceeding $\Delta_{\rm vir}$. 
%virial one.  
For example, $X$-ray observations, especially with
\emph{Chandra}, are typically sensitive out to
$\Delta=2500$.  Figure~\ref{fig:Mall_to_Mredblue} shows that, in this
case, weak lensing may underestimate $M_{2500}$ by a factor of 2 if
the faint galaxy sample, based solely on the magnitude selection, is
employed. Therefore, the dilution effect should be carefully corrected
for if weak lensing is used to estimate cluster masses with higher
over-densities.

\vspace{0.5em}
\subsubsection{Source redshift uncertainty}\label{subsec:zs}

%\begin{figure}
%\begin{center}
%%\FigureFile(80mm,80mm){f11.ps}
%\end{center}
%\caption{The effect of uncertainties in the mean source redshift on the
% best-fit NFW parameters, the concentration parameter (upper panel)
% and the virial mass (lower). The triangle and square symbols show the
% results when assuming the mean redshift $\langle z_s\rangle=0.8$ and
% 1.2, respectively, instead of our fiducial choice of $\langle
% z_s\rangle=1$. The data points are plotted as a function of cluster
% redshift for 21 clusters (the clusters shown in
% Figure~\ref{fig:Msis_vs_Mnfw} minus ZwCl0740 whose redshift is
% unknown). The concentration parameter is little affected by the unknown
% source redshifts, while the virial mass may be biased by up to $\sim
% 15\%$, a more significant bias for clusters at higher redshifts.  }
% \label{fig:zs_error}
%\end{figure}
%

%MT
As described in \S~\ref{ssec:zs}, we estimated redshifts of source
galaxies using the well-calibrated COSMOS photo-$z$
catalog. However, our analysis includes faint galaxies sometimes down to
$i=26$, while the COSMOS galaxies are available only down to
$i=25$. Hence our lensing results may be affected by a residual
uncertainty in the source redshift, although such faint galaxies are
generally assigned a smaller weight.
% in estimate the lensing distortion.

A $5\%$ or $10\%$ change in the average distance ratio, which controls
the overall amplitude of distortion signal, corresponds to $\simeq 10\%
$ or $20\%$ in the mean source redshift for a cluster at $z\simeq 0.2$
or 0.3, respectively. A typical uncertainty in the mean source
redshift, inferred from the photo-$z$ errors in the COSMOS catalog, is a
few $\%$ at most, therefore a 10\% level change in the mean redshift is
unlikely.
%unexpected. 
Recall that a bias in the average distance ratio is linearly
propagated into a bias in cluster mass estimates (a 10$\%$ change in
$\langle D_{\rm ls}/D_{\rm s}\rangle$ yields a $10\%$ change in the
best-fit mass parameter). On the other hand, the concentration
parameter is less affected by the bias in the distance ratio about by a
factor of 2, because the concentration is constrained by the shape of
distortion profile. Therefore we believe that a residual uncertainty in
source redshifts is insignificant for our results.

\vspace{0.5em}
\subsubsection{Misalignment of the BCG position and halo center}
\label{sec:center}

\begin{figure}
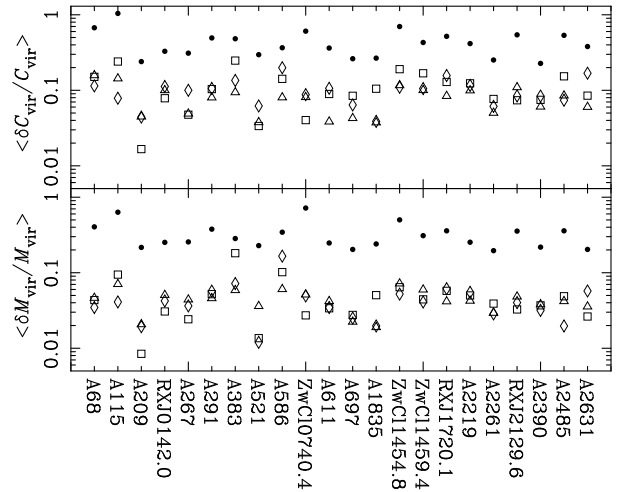

\begin{center}
%%\FigureFile(80mm,80mm){M+C_accuracies.ps}
\FigureFile(80mm,80mm){f12.ps}
\end{center}
\caption{
The diamond symbols show typical biases in the best-fit parameters,
 $c_{\rm vir}$ (upper panel) and $M_{\rm vir}$ (lower), when the cluster
 center is randomly taken from arbitrary point in the vicinity of BCG
 within $10^{\prime\prime} $ in radius. Note that the $y$-axis is
 plotted in a logarithmic scale.  The squares show typical biases in the
 best-fit parameters obtained when shifting the innermost radial bin
 by $\Delta\theta_{\rm min}= \pm0\farcm2$ in the tangential distortion
 profile, while the triangles show the biases in the parameters when
 changing the number of radial bins in the range of $N_{\rm
 rad}=[7,16]$, instead of their fiducial choices. For comparison, the
 filled symbols show the statistical accuracies of these parameter
 estimations given in Table~\ref{tab:massprofile}.  } \label{fig:center}
\end{figure}

Our analysis has so far adopted the angular position of the BCG as the
cluster center around which to measure the tangential distortion
profile.
 However the BCG might be offset from the
true center of dark matter halo hosting the cluster.  Such a
misalignment may cause a bias in measuring the tangential distortion
profile and thus cluster model parameters.  One advantage of our weak
lensing analysis is we can measure variations in the goodness-of-fit
of the NFW model fitting to the distortion profile by varying the
cluster center, on an individual cluster basis; this can be contrasted
to the the stacked cluster-galaxy
lensing where the cluster center of each cluster has to be {\em a
priori} assumed before stacking
%the stacking average  
(Johnston et al.\ 2007).
In our case, if the BCG position
is close to the true center, the $\chi^2$ value should be close to its
true minimum when the BCG is taken as the cluster center in the
analysis.  On the other hand, if we adopt the BCG as the cluster
center in a cluster in which the BCG is significantly offset from the
true center, then the resulting $\chi^2$ value may become
significantly degraded.

In Figure~\ref{fig:center} we examine the impact of the uncertainty in
the assumed cluster center on the NFW model parameters. For each of 22
clusters shown in Figure~\ref{fig:Msis_vs_Mnfw}, the open diamond
symbols show typical variations in the best-fit parameters when taking
a random point as the cluster center that is away from the BCG
position by within 10$^{\prime\prime}$ in radius. More precisely, the
results are computed from the variance of the best-fit parameters
obtained from 100 Monte Carlo realizations of random cluster center
identifications.  The range of 10$^{\prime\prime}$ radius is based on
the fact that the $\chi^2$ value for the best-fit model significantly
degrades for most of our clusters if the cluster center is taken to be
offset from the BCG position by more than $10^{\prime\prime}$,
and is also consistent with the distribution of
   offsets between BCG centres and X-ray centroids (Sanderson et al.
   2009).
Comparing the results with the filled symbols, one finds that
possible variations in the true cluster center around the BCG position
cause negligible biases in the parameters, typically smaller by almost
one order of magnitude than the statistical errors on
%accuracies for 
our
fiducial analysis (i.e the BCG is taken as the cluster center).
Physically, the cluster parameters we are interested in are sensitive
to the weak lensing distortions at larger radii compared to the size
of cluster center variations.  Therefore, the relative inaccuracy in
the cluster center position is negligible.  These results are also
consistent with numerous strong lensing studies (e.g.\ Kneib et
al. 1996; Smith et al. 2001, 2002, 2005; Sharon et al. 2005; Richard
et al., 2007) in which negligible BCG-cluster center offsets were
found.  We are also testing this more thoroughly with our new HST data
(SNAP:10881; Hamilton-Morris et al., in prep., and GO:11312; 
  Richard et al., 0911.3302) and the new Bayesian version of lenstool
(Jullo et al., 2008).

\subsubsection{Radial binning}

Our fiducial analysis did not use the distortion signals at very small
radii to avoid the effect of noisy measurements in bins that subtend
small solid angles on the sky, in addition to seeking to minimize the
impact of any mis-identification of the cluster center on the model
fitting (see the tangential distortion plots for each cluster field in
Appendix~\ref{sec:massmap} to find the range of angular scales used).
The square symbols in Figure~12 show the mean variation in the NFW
parameters obtained when shifting the innermost radial bin used in the
analysis by $\Delta\theta_{\rm min}=\pm 0.\!\!^{\prime}2$ with the
cluster center being fixed to the BCG position.  This uncertainty has
a similar-level impact on the model parameters to the diamonds, and is
again considered as an insignificant effect compared to the current
statistical precision.

Finally we also study the effect of the radial binning scheme on
the model fitting.  {We typically use $13$ bins in the tangential
distortion profiles;} finer or coarser binning may change the results,
because the intrinsic ellipticity noise contribution to the
measurement errors is sensitive to the radial binning that determines
the number of background galaxies contained in each radial annulus.
The triangles show typical variations in the parameters when varying
the bin number in the range $N_{\rm rad}=[7,16]$, confirming that the
best-fit parameters do not change significantly.  This is partly
because the effect of substructures on the azimuthally averaged
tangential profile do not largely change with the radial bin
variations.  The possible biases are again small compared to the
statistical errors.

\vspace{0.5em}
\subsubsection{Projection effect}

A chance projection of foreground/background mass structures can
potentially affect the cluster parameter determination based on the
``non-local'' distortion profile, which is sensitive to the total
interior mass in projection.  It can locally boost the surface mass
density, and hence can affect the tangential distortion measurement if
this physically unassociated mass structure is contained within the
measurement radius.  For the determination of the NFW concentration
parameter, it can lead to either an under- or over-estimation of the
concentration depending on the apparent position of the projected mass
structure.  One way to overcome this is to utilize the convergence
profile to examine the cluster mass profile, by locally masking out the
contribution of the known foreground/background structure in the
reconstructed mass map (Appendix~\ref{sec:massmap}; also see Umetsu et
al.\ 2008 for the case of A2261). It should be again worth noting that
these projection effects are averaged out in the stacked lensing
signals. Since our results for the individual clusters are consistent
with the stacked lensing results (see Figure~\ref{fig:C-M}), the
projection effect does not seem to cause significant biases in our
results. 
%MT
The projection effect is studied in more detail in our
subsequent paper, confirming an insignificant projection effect for the
current measurements (Oguri et al. 2010).

\subsubsection{Shape measurement}

The shape measurement method may involve systematic errors.  As
studied in detail by the STEP project (Massey et al. 2007;
Heymans et al.\ 2007), the various shape measurement methods developed to date
differ in galaxy ellipticity measurements by up to a multiplicative
bias of $\sim10\%$.  It is important to note 
It is important to note that STEP was conceived to
%that these results were
%designed to 
inform analysis strategies for cosmic shear experiments,
and thus concentrated on weak lensing lensing signals of $\simlt5\%$ in
contrast to cluster signals that typically reach $\simgt10\%$.  STEP also
used exclusively synthetic data.  Nevertheless, possible
method-dependent systematic biases in galaxy shape measurement are
also relevant for cluster lensing studies.  We therefore repeated the
galaxy shape measurement steps of our analysis for a representative
sub-set of our sample using the im2shape method (Bridle et al., 2002)
as implemented by Smith et al.\ (2009, in prep.).  The resulting
distortion profiles were identical within the measurement errors to
those based on the KSB methods described earlier in this
paper.  In summary, whilst further careful tests are required to
validate the shape measurement methods on both synthetic and real
cluster lensing data, we found no evidence for shape measurement
systematic biases in our analysis, and do not expect them to be a
dominant source of errors.

\subsection{Characteristics of Mass Maps vs. $X$-ray and 
Radio Information}

Two-dimensional maps of projected mass density can be reconstructed
from the measured ellipticity distribution of background galaxy shapes
(e.g.\ Kaiser \& Squires 1993).  The mass maps of individual clusters
are shown in Appendix~\ref{sec:massmap}.  Since the shear and mass
density fields are equivalent in the weak lensing regime, the mass
maps do not carry any additional information on cluster parameters.
Also, in practice uncertainties in reconstructed mass maps are highly
correlated between different pixels -- it is therefore important to
include the error covariance in order to properly propagate the
measurement uncertainties into accuracies of parameter estimations
from mass maps (see Oguri et al.\ 2006 and Umetsu \& Broadhurst 2008
for such studies).  Nevertheless mass maps are useful when comparing
the total matter distribution with cluster properties obtained from
other wavelengths (optical, $X$-ray, etc.), in order to study the
evolutionary processes and dynamical stages of each cluster (e.g.\
Clowe et al.\ 2006; Okabe \& Umetsu 2008).  Here we comment on
features in the mass maps from a multi-wavelength perspective.

Our cluster samples contain 2 {\em cold-front} clusters that have
sharp discontinuities of $X$-ray cores observed in the $X$-ray surface
brightness: ZwCl1454 (also known as MS1455.0+2232: Mazzotta et al.\
2001b) and RXJ1720 (Mazzotta et al.\ 2001a).  The formation of cold
fronts is one of the outstanding problems in cluster physics.  In fact
the mass maps of these two clusters suggest a bi-modal mass
distribution in the core of each cluster.  In both clusters, one
sub-clump of the bi-modal mass distribution appears to be the
``counterpart'' of hot intra-cluster gas at a similar position, while
the other does not have any clear counterpart (see Okabe, Mazzotta et
al.\ in preparation for a more quantitative study).  This bi-modal
structure is consistent with results on the other three cold-front
clusters studied to date, including the bullet cluster, A2034 and
A2142 (Clowe et al.\ 2006; Okabe \& Umetsu 2008).

The origin of diffuse radio emission within clusters, emanating from
synchrotron radiation of relativistic non-thermal electrons, remains
an unsolved mystery.  One possible scenario discussed in the
literature is that the non-thermal electrons are produced by
hierarchical mergers that every cluster universally experiences in the
CDM scenario.  Weak-lensing mass maps are useful tools with which to
test this picture because they allow to search for direct merging
signatures, e.g.\ prominent substructures in the mass maps due to
cluster-cluster mergers.  An important advantage of this approach is
that the collisionless nature of dark matter should result in the
merger signatures surviving longer in the dark matter distribution
that dominates weak-lensing maps compared to the intra-cluster hot gas
(e.g., see Okabe \& Umetsu 2008 and Tormen et al.\ 2004 for the
observational and theoretical studies, respectively).  On the other
hand, $X$-ray substructures may not be a good tracer of mass
substructures, indeed sometimes they are not associated with the
lensing substructures, depending on the stage that the merger has
reached (Okabe \& Umetsu 2008; see also Smith et al. 2005 for a
strong-lensing/X-ray comparison).

Our cluster sample contains 8 clusters in which diffuse radio emissions
have been found to date: A209 \citep{gio06}, A697 \citep{kem01},
RXJ1720 \citep{maz08}, ZwCl1454 \citep{ven08}, A115 \citep{gio99},
A2345 \citep{gio99}, A521 \citep{fer03}, and A2219 \citep{kem01}.  These
clusters appear to show the substructures that are seen more prominent
than those in other clusters, and the
substructure locations generally match well the morphology of
the radio emission.  This trend was also reported for other clusters
with diffuse radio sources (Clowe et al.\ 2006; Okabe \& Umetsu 2008).
A more quantitative comparison between the mass map and the radio
sources, further including the $X$-ray information, will be presented
elsewhere (Okabe et al.\ in preparation).

\section{SUMMARY AND DISCUSSION}
\label{sec:discussion}

In this paper we have presented a systematic weak-lensing study of 30
$X$-ray luminous clusters at $0.15<z<0.3$ as part of the Local Cluster
Substructure Survey (LoCuSS), based on high-quality Subaru/Suprime-Cam
data.  Our findings are summarized as follows:
\begin{itemize}
\item The high-quality Subaru data allowed a significant detection of
      the individual cluster lensing signals
      (Table~{\ref{tab:wl_para}}).  The total signal-to-noise ($S/N$)
      ratios for the tangential distortion profile, integrated over
      the range of radii probed, are $5\ltsim S/N\ltsim13$ for all
      30 clusters.
\item We made a detailed comparison of the measured distortion profile
      with mass profile models (Table~\ref{tab:massprofile} and
      Figure~\ref{fig:Msis_vs_Mnfw}) -- among the secure 22 clusters
      (with color information and suitable for the spherical model
      fitting), 3 clusters favor an NFW profile compared to an SIS
      model, 2 clusters cannot be well fitted by either model, and the
      other clusters are well-fitted by either model.
\item The virial mass estimates from NFW and SIS models are in good
      agreement, albeit with large measurement errors.  However, the
      best-fit mass tends to be underestimated if an SIS model is
      employed.  We understand this, in the context of the stacked
      analysis discussed below, to be caused by the SIS model
      under-predicting the amplitude of the gravitational distortion
      on large scales due to its inability to describe the curvature
      of the distortion profile of an NFW halo.
%    \item The fractional error on cluster  masses derived from NFW
%        model fits is minimized at larger over-densities, typically
%      in the range $\Delta\simeq 500-2000$ as opposed to the commonly
%      used virial over-density of $\Delta_{\rm vir}\simeq 110$.  
%        This is the result of the trade-off between decreasing signal
%      and decreasing noise per bin as over-density decreases and
%      cluster-centric radius increases (Figure~\ref{fig:dM_to_M}).
    \item  We detect anti-correlation between mass and
        concentration at $2\sigma$ significance: $c_{\rm vir}(M_{\rm
        vir})=8.75^{+4.13}_{-2.89}\times (M_{\rm vir}/10^{14}
      h^{-1}M_\odot)^{\alpha}$ with $\alpha\approx -0.40\pm 0.19$.
       This is in qualitative agreement with predictions from
        numerical simulations, but with a tentative detection of a
        steeper slope than predicted (Figure~\ref{fig:C-M}).
    \item The distribution of $c_{\rm vir}$ for our morphologically-
      and strong-lensing-unbiased sample does not contain any clusters
      with extremely high concentrations as have been reported in the
      literature for spectacular strong-lensing clusters. 
% MT
	  More precisely, our best-fit $c_{\rm vir}$-$M_{\rm vir}$
	  scaling predicts $c_{\rm vir}\simeq 3.48^{+1.65}_{-1.15}$ for
	  massive clusters with $M_{\rm
	  vir}=10^{15}h^{-1}M_\odot$. 
Therefore the high
   concentrations of $c_{\rm vir}\sim 10$ 
inferred from strong-lensing-selected clusters are
   inconsistent with our statistical analysis of X-ray selected clusters at
   $4\sigma$ significance.
%Therefore such a high concentration
%	  $c_{\rm vir}\sim 10$ inferred from the strong lensing clusters
%	  is not consistent with our result at $4\sigma$ level. 
\item The stacked distortion signals, for the two sub-samples of 19
      clusters binned into mass bins, show a pronounced radial
      curvature over radii ranging from 70$h^{-1}$kpc to 3$h^{-1}$Mpc
      (Figure~\ref{fig:stack}).  The profiles are well-fitted by a
 curved (cored isothermal or NFW) density profile,
%NFW profile, 
supporting the individual cluster lensing results,
      and strongly rule out the SIS model at $6\sigma$ and $11\sigma$
      for low ($M_{\rm vir}<6\times10^{14}h^{-1}M_\odot$) and high
      ($M_{\rm vir}>6\times10^{14}h^{-1}M_\odot$) mass bins respectively.
\item The projected 2D mass within the cylinder enclosed within a
      given projected radius, estimated from the model-independent
      aperture mass method, tends to be greater than the 3D spherical
      masses enclosed within the same radius in 3D, obtained from the
      NFW model fitting (Table~\ref{tab:2Dmass} and
      Figure~\ref{fig:mass_vs_2Dmass}).  The ratio of 2D to 3D mass is
      $\simeq 1.32$ and $\sim1.46$ at $\Delta=500$ and
      $\Delta=\Delta_{\rm vir}$ respectively, which can be well
      explained by the projected mass contribution of 
      a cluster-scale NFW halo with $c_{\rm vir}\simeq 4$.
\end{itemize}

Our results are an important step towards a more thorough empirical
understanding of the mass distribution 
in galaxy clusters, and thus
towards testing the nature of dark matter and dark energy (through the
cluster mass function for the latter). 
However the results are limited by (i) the
modest statistical precision available from a sample of $\sim20$
clusters, (ii) the limited color information available on the
background galaxy samples, (iii) the simplistic spherical mass
modeling approach applied to the data, and (iv) we have ignored other
data available to constrain the cluster mass distributions, most
notably strong-lensing arcs in the cluster cores.

For example, the detection of a slope in the observed $c_{\rm
vir}(M_{\rm vir})$ relation is significant at just $2\sigma$.  Simply
doubling or quadrupling the sample size would improve this to a $3$ or
$4\sigma$ result respectively.  Measurements of concentration
parameters appear to be more sensitive to systematic errors than
measurements of cluster mass.  We therefore plan to combine the Subaru
weak-lensing constraints with strong-lensing constraints available
from our \emph{HST} and Keck data (Richard et al., 2009)  to
build joint strong/weak-lensing models of the clusters, from which to
obtain more robust concentration measurements (Smith et al., 2009, in
prep.).  An important feature of these models will be the use of
pseudo-elliptical NFW models (Golse et al., 2001) and inclusion of
multiple halos in the models to capture the full two-dimensional
structure of the clusters in the plane of the sky.  Jing \& Suto
(2002) have also used numerical simulations to show that CDM halos are
better fitted by a triaxial mass distribution than a spherical NFW
model even in the statistical average sense, as naturally expected
from collision-less nature of CDM particles.  This is a very
interesting possibility that has been explored recently by Oguri et
al.\ (2005) and Corless \& King (2008), and can be
explored in a straightforward manner using the same data sets used in
this paper (Oguri et al. 2010).

It is also interesting to compare our results on the distribution of
cluster concentrations with the high concentration results obtained
for several well-known strong lensing clusters, notably A\,1689,
Cl\,0024 and MS\,2137 (Gavazzi et al.\ 2003; Kneib et al.\ 2003;
Broadhurst et al.\ 2005; Limousin et al.\ 2007; Broadhurst et al.\
2008; Oguri et al.\ 2009).  The important difference, beyond sample
size, between these detailed single-object studies and our statistical
study is that our cluster sample is unbiased 
with respect to 
the prevalence of strong-lensing arcs
in the cluster cores (Figure~\ref{fig:lx_vs_z}).  As shown in
Figure~\ref{fig:C-M}, the massive clusters in our sample generally
have the lowest statistical errors, and indeed have low
concentrations, $c_{\rm vir}\sim3$; i.e.\ consistent with the
simulation results.  On the other hand, there are clusters displaying
relatively high concentrations $c_{\rm vir}\sim 8$.  An important test
of the joint interpretation of our statistical results with those of
single-object studies will be whether the presence of strong-lensing
arcs in clusters is correlated with the high concentration of the cluster.
Increased sample size and joint strong/weak-lens modeling will both be
central to this investigation.

Vikhlinin et al.\ (2008) recently claimed very tight cosmological
constraints based on the cluster mass functions at $\Delta=500$
derived from Chandra observations under the assumption of hydrostatic
equilibrium.  The relationship between $X$-ray observables and mass
was calibrated using numerical simulations (Kravtsov et al.\ 2006;
Nagai et al.\ 2007; Vikhlinin et al.\ 2008), and the level of residual
uncertainty in the absolute mass calibration was assessed by comparing
the $X$-ray derived masses with the lensing mass estimates of Hoekstra
(2007), claiming possible $5\%$-level residual uncertainties in the
mass estimate.  However, our results indicate that the lensing masses
estimated at $\Delta=500$ are sensitive to dilution of the
weak-lensing signal by faint cluster galaxies, cluster masses being
underestimated by $\simgt20\%$ at $\Delta=500$ if dilution is not
properly corrected for.  Therefore, if the absolute mass calibration
primarily rests on the comparison with the lensing masses, the $X$-ray
derived masses may still involve additional biases.  In this sense, a
further large detailed comparison of $X$-ray and lensing masses for
joint $X$-ray and lensing cluster samples is crucial.  In particular
detailed \emph{cluster-by-cluster} comparison will be very important
to pin down the sources of systematic errors due to physical
differences between the clusters.  The mass maps shown in
Appendix~\ref{sec:massmap} will be useful for this purpose because the
mass distribution directly reflects the dynamical stages of a cluster
(relaxed, merging, etc.).  These studies will be presented elsewhere
(Okabe et al.\ in preparation).

\section*{Acknowledgments}

We are very grateful to A.~Finoguenov, C.~Haines, H.~Hoekstra,
Y.~Itoh, A.~Leauthaud, M.~Oguri, Y.~Y.~Zhang, and the members of
LoCuSS collaboration for invaluable discussions and comments.  We also
thank annonymous referee for useful comments which led to improvement on
the manuscript. Finally we also thank 
Subaru Support Astronomers for the Subaru/Suprime-CAM and 
N.\ Kaiser for developing the IMCAT package publicly available.  NO,
MT, and GPS  acknowledge warm hospitality at the Kavli Institute
  for Cosmological Physics at U.\ Chicago, where some of this work was
  carried out.  GPS also acknowledges warm hospitality at Tohoku
  University.  NO, MT and TF are in part supported by a Grant-in-Aid
from the Ministry of Education, Culture, Sports, Science, and
Technology of Japan (NO: 20740099; MT: 20740119; TF: 20540245).  MT is
also in part supported by the World Premier International Research
Center Initiative of MEXT of Japan.  KU is partially supported by the
National Science Council of Taiwan under the grant
NSC97-2112-M-001-020-MY3.
%NSC95-2112-M-001-074-MY2. 
GPS acknowledges support from the Royal Society and the Science and
Technology Facilities Council.  This work is also supported by a
Grant-in-Aid for the 21st Century COE Program ``Exploring New Science
by Bridging Particle-Matter Hierarchy'' and the GCOE Program ``Weaving
Science Web beyond Particle-matter Hierarchy'' at Tohoku University as
well as by a Grant-in-Aid for Science Research in a Priority Area
"Probing the Dark Energy through an Extremely Wide and Deep Survey
with Subaru Telescope" (No. 18072001).  This work was supported in
part by the Kavli Institute for Cosmological Physics at the University
of Chicago through grants NSF PHY-0114422 and NSF PHY-0551142 and an
endowment from the Kavli Foundation and its founder Fred Kavli.

\appendix

\section{Defining Galaxy Samples}
\label{app:samp}

 We have concentrated on clusters for which Suprime-Cam data are
available in two filters, and used the following galaxy samples to
select background galaxies robustly for our weak-lensing analysis:
{\em member galaxy sample}, {\em faint galaxy sample}, {\em red galaxy
  sample} and {\em blue galaxy sample}.  In this appendix, we describe
how the four galaxy samples are defined based on the color-magnitude
diagram of each cluster.

\subsection{Color-Magnitude Diagram}

We typically used the color-magnitude information, e.g.  the
$(V-i')$-$i'$ information, to separate cluster members from
non-members.  Note that because we focus on relatively low-redshift
clusters, most non-member galaxies are very likely background galaxies
thanks to the deep imaging data and the limited volume that lies
between us and each cluster.  To define the galaxy samples, we first
analyze the data using SExtractor (Bertin \& Arnouts 1996) in the
dual-image mode, using the redder passband (typically $i'$-band) for
source detection.  We extract all objects with isophotal areas larger
than 10 contiguous pixels 
%MT
where each pixel 
($2.\!\!^{\prime\prime}02$) needs to be 
%each
$\ge3\sigma\,{\rm pixel}^{-1}$ of the local sky background.  We
calculate for each source the total magnitude in the AB-magnitude
system using the MAG\_AUTO parameter.  Colors are calculated using the
MAG\_APER parameters with the aperture size being set to 10 pixels in
diameter.

\subsection{Member Galaxy Sample}
\label{sec:member}

Early-type cluster galaxies occupy a narrow well-defined locus, the
so-called red sequence, in the color-magnitude diagram.   Red
  sequence galaxies were selected as follows.  First, point-sources
were removed from the object catalog, and then the following relation
(or its equivalent in the case that different filters were available)
was fitted to galaxies brighter than 22nd magnitude in the redder
filter:
\begin{equation}
(V-i')_{\rm E/S0}=a\,i' + b. 
\label{eq:r-s}
\end{equation}
The best-fit values of $a$ and $b$ were determined such that the
number of galaxies contained in the red-sequence is maximized allowing
the red-sequence to have a finite width such as $\delta(V-i')\simeq
\pm0.1 \!\mbox{ mag}$ depending on the tightness
of the observed color-magnitude relation.  For example, the
green points in Figure~\ref{fig:cmr} show the member galaxy sample for
A68.

In a few cases, multiple combinations of the parameters $a$ and $b$
were found to fit the data.  In such cases we identified the sequence
that is most likely one inferred from the cluster redshift based on a
passive evolution model of galaxy color and magnitude.  Interestingly,
as discussed in Appendix~\ref{sec:massmap}, galaxies sitting in other
red-sequences generally coincide with peaks in the weak-lensing mass
maps, suggesting that they correspond to over-densities at other
redshifts.

We also identify the brightest cluster galaxy (hereafter BCG) in each
cluster, and defined the nominal center of each cluster as the angular
position of the BCG in each cluster.  Note that in some clusters the
BCG does not sit on the red-sequence -- we therefore visually checked
such clusters to ensure correct identification of BCGs.

The BCGs and the galaxies contained in the red-sequence with a finite
width,
which are all brighter than 22 mag (AB), gives our member galaxy
sample.  This member galaxy sample is used to estimate the number
density field as well as the luminosity density field of cluster
galaxies for comparison with the lensing mass maps in
Appendix~\ref{sec:massmap}.

\subsection{Faint Galaxy Sample}

Magnitude-selected background galaxy samples have often been used in
previous studies of cluster weak lensing.  Although our main results
are based on color-selected galaxies, we first define here our
magnitude-selected, or ``faint'' galaxy samples.  These samples are
mainly used as a suite of reference samples against which our more
sophisticated color-selection methods can be compared.

To ensure that the shape of galaxies can be measured reliably, the
``background'' galaxies used for weak-lensing analyses are required to
be both well-resolved and have a sufficiently large integrated
signal-to-noise ratio.  On the latter point, we restrict our attention
to galaxies with signal-to-noise ratios of $\nu\ge 10\sigma$ as calculated
with the IMCAT software.  We also select galaxies with a half-light
radius, $r_h$, in the range $\bar{r}_h^*+\sigma(r_h^\ast)<r_h<10\,{\rm
pixels}$, where $\bar{r}_h^*$ and $\sigma(r_h^*)$ are the median and
rms of the half-light radii of stellar objects selected over the
entire Suprime-Cam FoV.  Note that the upper limit of $r_h=10\,{\rm
pixels}$ is chosen based on trial and error 
to avoid galaxies with
saturated pixels and/or strange shapes typically originating from
superpositions of two or more galaxies  (e.g.\ Okabe \& Umetsu 
2008)\footnote{For the clusters A115
and A2345, we impose more restrict conditions on the half-light radius
due to poor seeing as listed in Table~\ref{tab:wl_para}.}.  Then the
faint galaxy sample is defined from the resolved, high
signal-to-noise galaxies as those lying in the apparent magnitude
ranges listed in Table~\ref{tab:wl_para} -- typically $22\le i'\le
26$.  The bright magnitude limit is designed to minimize contamination
of this sample by bright cluster members, and corresponds to the
apparent magnitude of $\sim\,i'^\star+3.5$ for an early-type galaxy at
the median redshift of a cluster in our sample.  The faint limit is a
consequence of the signal-to-noise and size cuts discussed above.

\subsection{Background Red/Blue Galaxy Samples} \label{subsec:red+blue}

Several authors have shown that faint galaxy samples such as those
described above suffer contamination by faint cluster galaxies, and
therefore weak-shear measurements based on such samples are diluted by
cluster and foreground (and thus unlensed) galaxies (e.g.\ Broadhurst
et al.\ 2005; Limousin et al.\ 2007).  In this paper we employ the
method described by Medezinski et al.\ (2007) and Umetsu \& Broadhurst
(2008). First, to quantify the dilution effect we calculate the mean
distortion strength of each cluster by averaging the tangential
distortion profile (Eq.~[\ref{eq:1d_gt}]) over a range of radial bins:
\begin{equation}
\langle\langle g_+\rangle\rangle\equiv 
\frac{1}{N_{\rm rad}}\sum_{n; 1'\le \theta_n\simlt 20'}
\langle g_+\rangle(\theta_n), 
\label{eqn:aagt}
\end{equation}
where $n$ runs over the radial bin labels, $\theta_n$, in
Eq.~(\ref{eq:1d_gt}) and $N_{\rm rad}$ is the total number of the
radial bins used.  
The cluster lensing signals are greater with decreasing cluster-centric
radius, and therefore this calculation assigns a greater weight to the
lensing signals closer to the cluster center than those towards the edge
of the FoV.
This is useful when
quantifying the effect of dilution, because dilution is expected to be
more significant at smaller cluster-centric radii as it roughly 
traces the cluster mass distribution, while the lensing
distortion signals are non-local (non-vanishing even beyond the virial
radius) and slowly decreasing with increasing radius. Note that we do not
include the lensing signals at very small radii $\theta_n\le 1'$,
because on these scales the signals are very noisy due to the small
numbers of galaxies in these radial bins, in addition to the impact of
uncertainties in the cluster center position.  

As described above, SExtractor was used to build the photometric
catalogs, while IMCAT was used to measure galaxy shapes.  Therefore,
before varying the color-selection criteria, it was necessary to match
the SExtractor and IMCAT catalogs.  In doing so, we define the
following matching criteria. For each object in the IMCAT catalog, the
closest neighbor on the sky in the SExtractor catalog is identified;
if positional difference between the two catalogs is less than 2
pixels ($0.\!\!^{\prime\prime}404$), then the two objects are regarded
as the same object, and otherwise are rejected.  If we find multiple
candidates in this matching procedure, although very rare, we take the
one with the closest total magnitude as the corresponding object.

\begin{figure}
\begin{center}
%\FigureFile(90mm,90mm){a68_dilution.eps}
\FigureFile(90mm,90mm){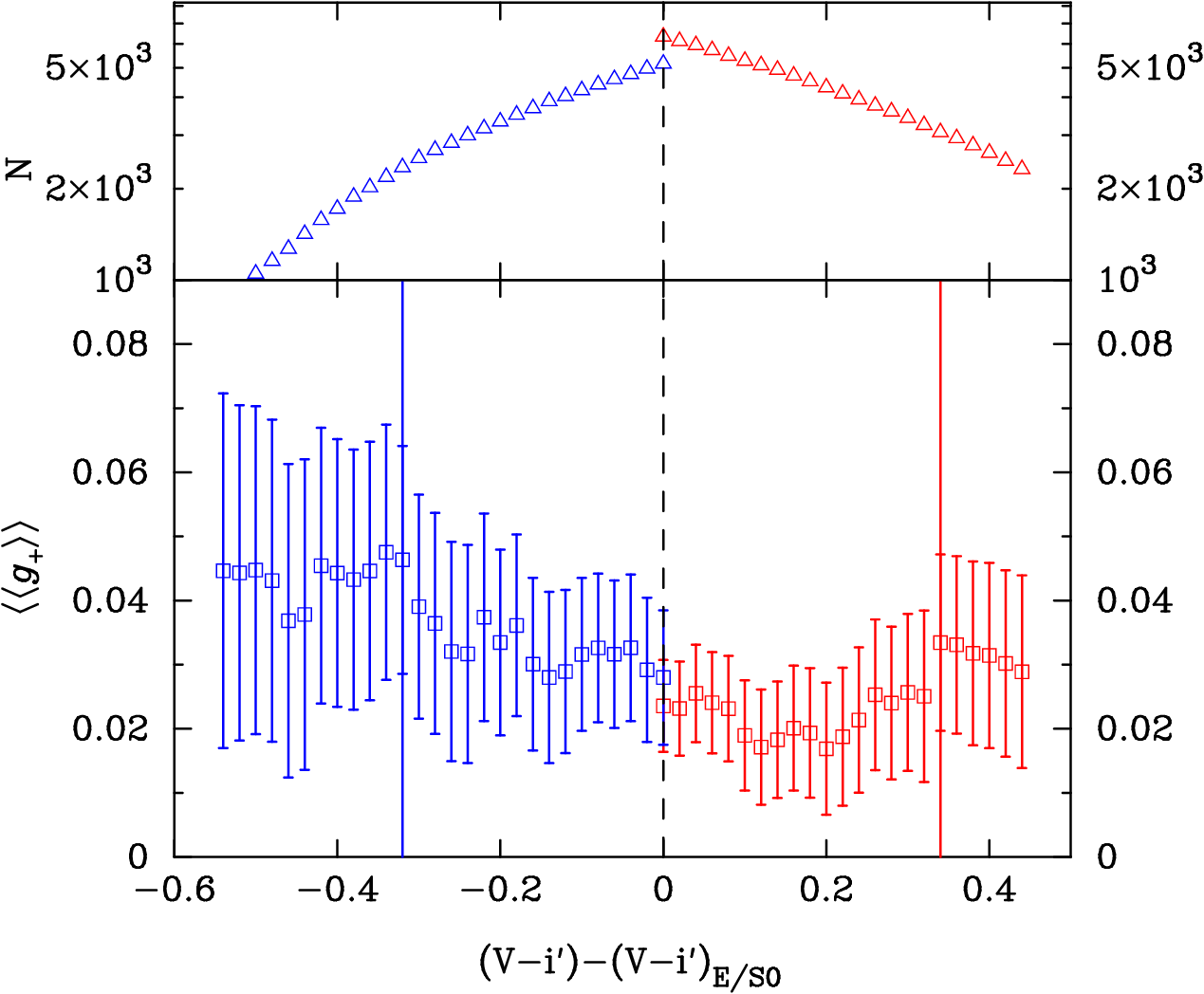}
\end{center}
\caption{The total number of galaxies
    ({\em upper panel}) and the mean tangential distortion strength
    ({\em lower}; also see Eq.\ref{eqn:aagt}) over the radii of
    $1'\le r\le 17'$, as a function of the varying background galaxy
    samples, for A68.  The background samples are defined with
    galaxies redder or bluer than the red-sequence at least by the
    color offsets given by $x$-label. The distortion strength is
    changed due to the dilution by cluster members and also partly due
    to the change in average source redshift. The two solid lines in
    the lower panel denote our choices of the color cuts used to
    define the red/blue background galaxy samples shown in
    Figure~\ref{fig:cmr} (see text for the details). }
    \label{fig:dilution}
\end{figure}
Figure~\ref{fig:dilution} shows, for A68 as a typical example, the
mean distortion strengths as a function of the varying background
galaxy samples, where each galaxy sample is selected from the faint
galaxy sample by further requiring that galaxies are redder or bluer
than the red-sequence (the vertical dashed line) by a given color
offset in the horizontal axis\footnote{The errors on the mean
distortion strengths are estimated as $\sigma^2_{\langle\langle
g_+\rangle\rangle}=(1/N_{\rm rad}^2)\sum_{n}\sigma^2_{g_+}(\theta_n)$
from Eq.~(\ref{eq:sig_g+}).}.  Note that the data points in the
different color bins are highly
   correlated because each data point includes all galaxies at larger color
   offsets than the offset at which the point is plotted.
%highly correlated because some galaxies 
%used for the $\langle\langle g_+\rangle\rangle$ calculation are
%overlapping between different galaxy samples -- i.e. the different data
%points contain the same information on galaxy shapes. 

First let us consider the results for galaxy samples redder than the
red-sequence -- i.e. right-ward of the vertical dashed line.  The
distortion strength changes as the color-cut becomes progressively
redder due to both reduced cluster member contamination and to the
change in average redshift of galaxies.  
All other things being equal the distortion strength
should, in principle, become insensitive to color-offset when the
color-cut is sufficiently large so as to render contamination and thus
dilution negligible.  In the case of A68, we therefore adopt a color
cut of $\Delta{\rm color}\equiv(V-i')-(V-i')_{\rm E/S0}=0.34$, as
shown by the vertical red line.  The background
galaxy redshift distribution is expected to vary slowly with color
offset, suggesting that the relatively abrupt jump in distortion
strength either side of the vertical red line is
contamination-related\footnote{However, note that, for fewer galaxies defined
by the larger color-cut, additional large scatters may be caused by
violation of the single source redshift assumption.}.  
Similar red-side color-cuts are adopted for
the other clusters, with values lying in the range $\Delta{\rm
color}\simeq [0.1,0.35]$.  Following the same logic on the blue-side we
adopt a color-cut of $\Delta{\rm color}=-0.32$ and mark this with
a vertical blue line; in this case the insensitivity of distortion
strength at the blue-side cut is more obvious than the red-side
discussed above.  The blue-side color-cuts are in the range of
$\Delta {\rm color}\simeq [-0.4,-0.1]$ for the entire cluster sample.

We then use the combined red$+$blue galaxy samples in our lensing
analysis throughout this paper.  Despite the care that we have taken
over the color-selection of background galaxies, the rather limited
color information that we have used here will inevitably allow some
unlensed galaxies to leak into the red$+$blue galaxy catalogs.
Nevertheless, we are able to prove that our red$+$blue samples are
less affected by the contamination than the faint galaxy sample. The
effectiveness of our color selection methods is demonstrated in
Fig.~\ref{fig:Nprofile}.

Our method may be compared to alternative method where the dilution
effect is corrected for by multiplying the measured distortion signal at
a given radial bin with a correction factor inferred from the increased
number density of faint galaxies at the radius (e.g., see Kneib et al.\
2003; Hoekstra 2007).  In this method, the stacked number density
profile, as shown in Figure~\ref{fig:Nprofile}, are usually used to
infer the correction factor, because a measurement of the number
density profile is noisy for individual cluster field -- we have also
found that we cannot necessarily find a clear increase in the number
density of faint galaxies at small cluster-centric radii for a
single cluster field\footnote{This is probably because of the intrinsic
clustering contamination of galaxies and of another lensing effect,
magnification bias, that affects the number counts of galaxies in
complex, different ways for blue and red background galaxies
(e.g. Broadhurst et al. 2005).}.  Therefore this alternative method does
not allow the cluster-by-cluster correction of dilution.  Our method
using a color-selected galaxy sample can thus be recognized as a more
direct, unbiased way in a sense that our method purely rests on the
lensing shape measurements and do not employ any correction factor to
obtain cleaner distortion signals. We are planning to further improve
the dilution correction with more accurate photometric redshifts
obtained by adding more passband data, which is also invaluable to
calibrate the source redshift uncertainties.

\begin{figure}
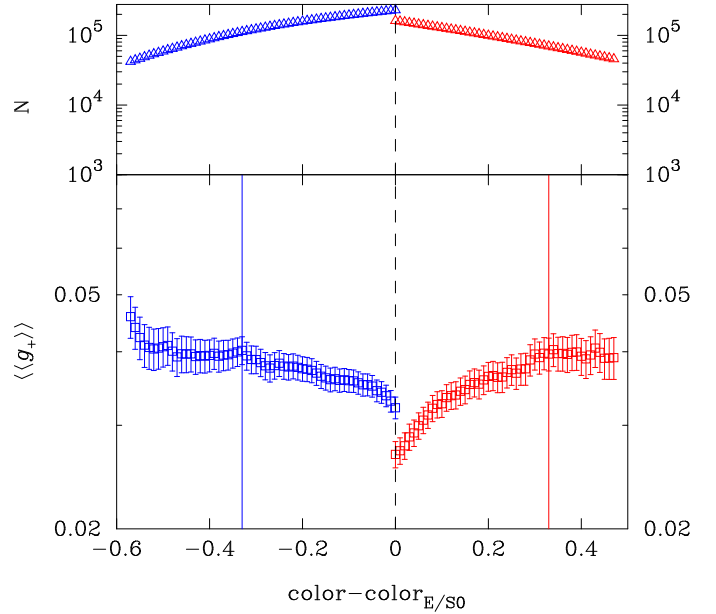

\begin{center}
%\FigureFile(90mm,90mm){stacked_g+_vs_color.ps}
\FigureFile(90mm,90mm){f14.ps}
\end{center}
\caption{As in Figure~\ref{fig:dilution}, but stacked for 21 clusters
 (22 clusters with color information minus ZwCl0740). A significant
 dilution of lensing signals can be found for small color offsets,
 i.e. if including faint galaxies with color closer to that of cluster
 red-sequence. The distortion strength becomes almost constant for color
 offsets, $|\Delta{\rm color}|\simgt 0.3$, at both red- and blue-sides. 
% The total number of galaxies for 21 clusters
%    ({\em upper panel}) and the stacked tangential distortion strength
%    ({\em lower}; over the radii of 
%$1\farcm\ler\le18\farcm$
}
    \label{fig:stacked_dilution}
\end{figure}

Since a selection of background galaxies is important, we also made
another test as follows. Figure~\ref{fig:stacked_dilution} shows the
{\em stacked} distortion strength for 21 clusters (22 clusters with
color information minus ZwCl0740) against different background galaxy
samples 
as in Figure~\ref{fig:dilution},
but selected with a single color offset for all the clusters. 
With the help of stacking the
distortion strength is smoothly varying against color, and a significant
dilution of lensing signals is clearly seen if including faint galaxies
with color similar to color of red-sequence galaxies. Also evident is
the dilution strength stays constant for the color offset $|\Delta {\rm
color}|>0.3$ at both red- and blue-sides. Note that this color offset is
comparable with the color cut employed for each cluster region as listed
in Table~\ref{tab:wl_para}, giving another confirmation that our
background galaxy selection is considered secure.

\begin{figure*}
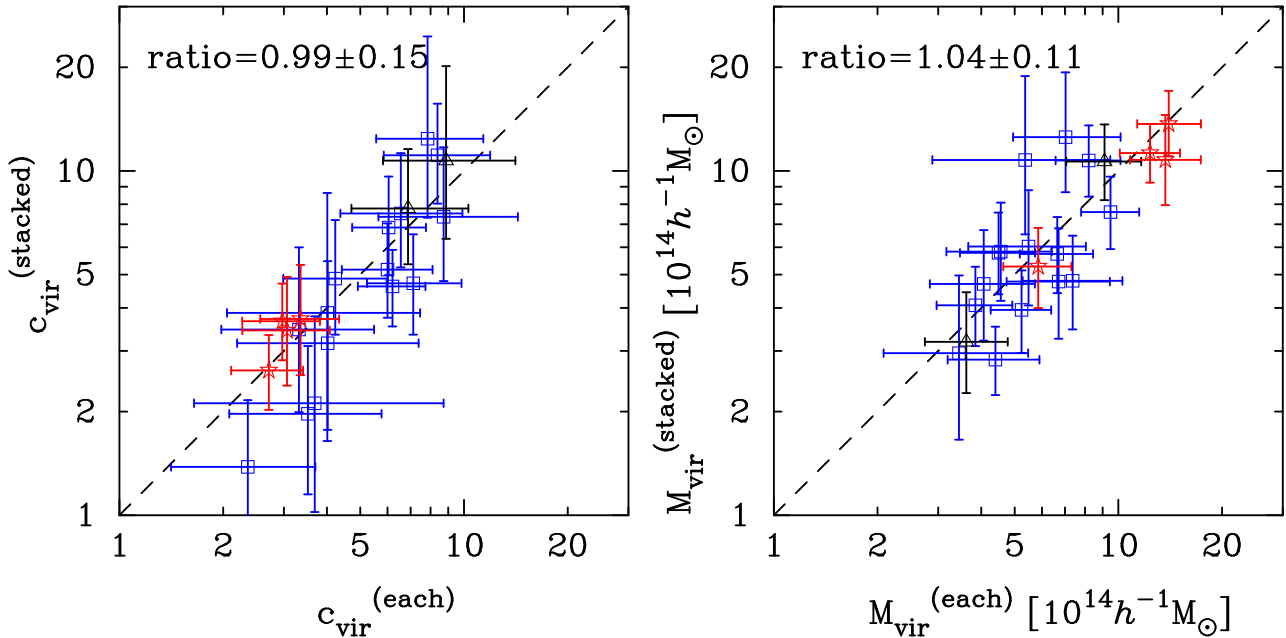

\begin{center}
%\FigureFile(170mm,170mm){compare_Cvir+Mvir_colorcut.ps}
\FigureFile(170mm,170mm){f15.ps}
\end{center}
\caption{Comparing the best-fit parameters, $c_{\rm vir}$ (left panel)
 and $M_{\rm vir}$, for each cluster field when using different samples
 of blue+red galaxies. The vertical axis in each panel shows the result
 obtained by using background galaxy sample defined with a single color
 cut $|\Delta {\rm color}|=0.3$, while the horizontal axis shows the
 result for our fiducial background galaxy sample. The two results agree
 well within the statistical errors.}
 \label{fig:compare_dil}
\end{figure*}

Given the results in Figure~\ref{fig:stacked_dilution}, 
Figure~\ref{fig:compare_dil} studies how best-fit parameters, $M_{\rm
vir}$ and $c_{\rm vir}$, change for each cluster if the background
galaxy sample defined with the single color cut $|\Delta {\rm
color}|=0.3$ is used, compared to the results of our fiducial red+blue
galaxy samples. It can be found that the results for two different
samples are consistent within the statistical errors.  Thus our
background galaxy selection is again considered robust.  Even so, we
believe that it is more secure to define background galaxy catalog by
setting the color cut on cluster-by-cluster basis, because the slope and
normalization of red-sequence is different for each cluster, and the
populations of member galaxies may also significantly differ for
different clusters.  More color information is needed to further refine
background galaxy selection based on improved photo-$z$ information,
which is our future project.

\section{2D and 3D Aperture Masses for an NFW Model}
\label{app:NFWmass}

The 3D mass enclosed within a sphere of a given radius $r_{\Delta}$ (see
Eq.~[\ref{eq:M_Delta}] for the definition of $r_\Delta$ in terms of the
mean over-density $\Delta $) is an important parameter to characterize
the cluster mass. The lensing fields at the projected radius
$\theta_\Delta=r_\Delta/D_l $ from the cluster center ($D_l$ is the
angular diameter distance up to the cluster) are sensitive to the 2D
mass enclosed within a cylinder of the radius $\theta_{\Delta}$ between
an observer and source galaxies. For an isolated NFW halo, the 2D and 3D
masses are found to be related as
\begin{equation}
\frac{M^{\rm NFW}_{2D}(<\theta_\Delta=r_{\Delta}/D_{l})}
{M^{\rm NFW}_{3D}(<r_{\Delta})}
=   f(c_\Delta) g(x=c_\Delta), 
\label{eqn:nfw2Dvs3D}
\end{equation}
where $f(c)\equiv 1/[\log(1+c)-c/(1+c)]$ and the function $g(x)$ is
defined below Eq.~(5) in Golse \& Kneib (2002). By using the equation
above, the 3D mass can be inferred from the 2D mass that is directly
estimated from the lensing observables in a model-independent way.
This inversion holds valid if the cluster mass distribution is well
represented by an NFW profile (see \S\ref{sec:model-independ-mass} for
the detailed discussion).

\section{Mass Maps}
\label{sec:massmap}

The coherent distortion pattern measured from background galaxy images
also allows to directly reconstruct the two-dimensional map of the
(projected) total matter distribution (Kaiser \& Squires 1993). The mass
density fields between different pixels in the mass map are highly
correlated, so the correlations need to be properly taken into account
when extracting some useful information from the mass map (e.g., Umetsu
\& Broadhurst 2008). Even so, the mass map is sometimes useful:
comparing the mass map with other wavelength information (member galaxy
distribution, $X$-ray and/or SZ maps, and so on), and inferring the
dynamical stages of a cluster from the mass distribution (the presence
of substructures and asphericity). For this reason, we show in this
Appendix the mass maps for the individual clusters of our sample,
comparing with the number density and luminosity density maps of member
galaxies.

In the following mass maps we also show the significance contours of the
mass density, relative to the $1\sigma$ noise level expected from the
intrinsic ellipticity noise. Following the method developed in Van
Waerbeke (2000) we use the Gaussian smoothing function to quantify the
noise level at an arbitrary angular position in the mass map, which is
given as
\begin{equation}
\sigma_\kappa^2=\frac{\sigma_g^2}{2}\frac{1}{2\pi\theta_g^2\bar{n}_g}, 
\label{eqn:sig_massmap}
\end{equation}
where $\sigma_g^2$ is the intrinsic ellipticity noise computed in the
similar manner as in Eq.~(\ref{eq:sig_g+}): $\sigma_g^2= \sum_i
w_{(i)}^2 \sigma_{g(i)}^2/[\sum_i w_{(i)}]^2$ using all the galaxies
used in the mass map reconstruction. The angular scale $\theta_g$ is the
width of the Gaussian smoothing function, $W(\bmf{\theta})=1/(\pi
\theta_g^2 )\exp(-|\bmf{\theta}|^2/\theta_g^2)$, and $\bar{n}_g$ is the
mean number density of galaxies over the field. Thus the noise level in
the mass map varies for each cluster, depending on the number density of
background galaxies and the smoothing scale used.

{\em For the following mass maps, we use the faint background galaxy
sample}, because some of our cluster samples do not have color
information, so only the background galaxy selection is available for
the whole sample, yielding fair comparisons between the mass maps of
different clusters.  Note that we also show the measured radial profiles
of tangential and its 45$^\circ$ rotated components of the galaxy images
for the clusters listed in Table~{\ref{tab:massprofile}}: the clusters 
whose lensing distortion profiles are used to constrain cluster
parameters (mass profiles and cluster masses). 

\end{document}